\documentclass[aps,prd,onecolumn,groupedaddress,showpacs,nofootinbib,amssymb]{revtex4}
\usepackage[latin1]{inputenc}
\usepackage{graphicx}
\usepackage[english]{babel}
\usepackage{amsmath}
\usepackage{amssymb}
\usepackage{amsfonts}
\usepackage{color}

\allowdisplaybreaks[4] 

\begin{document}

\newcommand{\be}{\begin{equation}}
\newcommand{\ee}{\end{equation}}
\newcommand{\bea}{\begin{eqnarray}}
\newcommand{\eea}{\end{eqnarray}}
\newcommand{\beaa}{\begin{eqnarray*}}
\newcommand{\eeaa}{\end{eqnarray*}}
\newcommand{\Lhat}{\widehat{\mathcal{L}}}
\newcommand{\nn}{\nonumber \\}
\newcommand{\e}{\mathrm{e}}
\newcommand{\tr}{\mathrm{tr}\,}

\tolerance=5000

\title{Generalized Galileon Model \\
-- cosmological reconstruction and the Vainshtein mechanism --}

\author{Norihito Shirai$^1$, Kazuharu Bamba$^2$, 
Shota Kumekawa$^1$, 
Jiro Matsumoto$^1$, and 
Shin'ichi Nojiri$^{1,2}$}

\affiliation{
$^1$ Department of Physics, Nagoya University, Nagoya
464-8602, Japan \\
$^2$ Kobayashi-Maskawa Institute for the Origin of Particles and
the Universe, Nagoya University, Nagoya 464-8602, Japan 
}

\begin{abstract}

Explicit formulae of the equations in the generalized Galileon models are given. 
We also develop the formulation of the reconstruction. 
By using the formulation, we can explicitly construct an action   
which reproduces an arbitrary development of the expansion of the universe. 
We show that we can completely and explicitly separate the action to the part 
relevant for the expansion and the irrelevant part. 
The irrelevant part are related with the stability of the reconstructed solution and 
we can further separate the part to the part relevant for the stability and the part 
irrelevant for the stability. 
The conditions how the reconstructed solution becomes stable and 
therefore it becomes an attractor solution are also given. 
Working in the static and spherically symmetric space-time, we 
investigate how the Vainshtein mechanism works in the generalized Galileon 
model and the correction to the Newton law becomes small. 
It is also shown that any spherically symmetric and static geometry 
can be realized by properly choosing the form of the action, 
which may tell that the solution could have fourth hair corresponding 
to the scalar field. 
We again separate the action to the part relevant for the reconstruction 
for the spherically symmetric and static geometry and the irrelevant part. 
We show that by choosing the relevant and irrelevant parts appropriately, 
we can obtain an action which admits both 
the solution corresponding to an arbitrarily given spherically symmetric and static 
geometry and the solution an arbitrarily given expansion history of the universe. 

\end{abstract}


\maketitle

\section{Introduction \label{In}}

The observation of the type Ia supernovae at the end of the last century 
tells that the expansion of the present universe is
accelerating \cite{Perlmutter:1998np,Perlmutter:1997zf,Riess:1998cb}.
In order that the accelerating expansion could occur in the Einstein gravity, 
we need cosmological fluid with the negative pressure and we call the fluid as 
dark energy (for review, see \cite{Silvestri:2009hh,Li:2011sd,Caldwell:2009ix}). 
The simplest model of dark energy is the cosmological term in the Einstein 
gravity, which is called $\Lambda$CDM model (CDM is cold dark matter). 
The $\Lambda$CDM model, however, suffers the so-called fine-tuning problem and/or coincidence problem. 
In order to avoid these problems, many kinds of dynamical models have been 
proposed. Especially there are dynamical models using the scalar field(s) 
like quintessence model 
\cite{Peebles:1987ek,Ratra:1987rm,Chiba:1997ej,Zlatev:1998tr,Peebles:1998qn}, 
$k$-essence models 
\cite{Chiba:1999ka,ArmendarizPicon:2000dh,ArmendarizPicon:2000ah}, 
or ghost condensation models 
\cite{ArkaniHamed:2003uy,ArkaniHamed:2003uz}. 
Such scalar models, however, generate large correction to the Newton law in 
general by the propagation of the scalar field. 
In order to make the correction decrease, so-called Chameleon mechanism has 
been proposed in \cite{Khoury:2003aq,Khoury:2003rn}. In the Chameleon 
mechanism, the mass of the scalar field becomes large when the mass density is large 
and therefore the force propagated by the scalar field becomes very short 
range, 
so that 
the correction can become very small not to conflict with the observational 
or experimental results. 
Another mechanism to suppress the contribution from the scalar field is the Vainshtein 
mechanism \cite{Vainshtein:1972sx}, where the scalar field is suppressed by 
the non-linear structure of the scalar field equation. Originally the Vainshtein mechanism was a mechanism 
for the decoupling of the longitudinal mode in the massive gravity. 
After that, it was found that similar mechanism works \cite{Luty:2003vm,Nicolis:2004qq} 
for the bending mode of the DGP model \cite{Dvali:2000hr,Deffayet:2000uy,Deffayet:2001pu}. 
Then the scalar field models where the Vainshtein mechanism works have been proposed. 
The actions of the original models have a symmetry called Galilean symmetry 
and hence 
the scalar field is called as the Galileon field 
\cite{Nicolis:2008in,Deffayet:2009mn,Deffayet:2010zh}. 
The equation of motion for the Galileon field does not include the derivative higher than two, 
which may prevent the existence of the ghost although the condition is not necessary nor 
sufficient condition for no ghost. We also note that the structure of the equation 
does not have a direct relation with the Galilean symmetry. 
Until now there are a number of 
activities related with the Galileon scalar field 
\cite{Galileon,Deffayet:2011gz,Kobayashi:2011nu}. 

In generalized models of the Galileon field, the field equations of motion is very complicated 
but it could be useful for future researches if the explicit forms of the equations are given. 
In this paper, we explain how the field equations of motion do not include the derivatives higher 
than two. As pointed in \cite{Deffayet:2011gz}, the Levi-Civita symbol 
$\epsilon^{\mu\nu\rho\sigma}$ 
plays the crucial role in the structure. 
In \cite{Kobayashi:2011nu}, it has been shown that the actions obtained in 
\cite{Deffayet:2011gz} are equivalent to those in \cite{Horndeski}. 
When we consider the curved space-time, in order to preserve the structure of the field equation, 
there appear the correction terms including the curvatures in the action, which also guarantees that 
the Einstein equations do not include the derivatives higher than two, either. 
We also develop the formulation of the reconstruction, which tells the explicit form of the action reproducing 
an arbitrary development of the expansion of the universe. 
We show the conditions how the reconstructed solution becomes stable and thus 
it becomes an attractor solution. 
We also investigate how the Vainshtein mechanism works in the curved space-time for the generalized Galileon model. 
We also show that any spherically symmetric and static geometry 
can be realized by properly choosing the form of the action, 
which may tell that the solution could have fourth hair corresponding 
to the scalar field. 

We give explicit forms of the equations as far 
as we can although they looks very complicated. 
We believe the explicit formulae could be necessary for later applications. 

In the next section, we give a general formulation of the Galileon field and show how the derivatives 
higher than two do not appear in the field equations and the Einstein equation. 
In Section \ref{III}, we consider the FRW dynamics and give formulae for the reconstruction. 
The stability of the reconstructed solution is also investigated. 
In Section \ref{IV}, we examine 
the spherically symmetric and static space-time and investigate how the 
Vainshtein mechanism works for the generalized Galileon model. 
We also realize any spherically symmetric and static geometry in the framework of the Galileon model. 
The last section is devoted to the summary and discussion. 

\section{Formulation of Galileon scalar \label{II}}

In this section, we give a general formulation of the Galileon models by using 
the Levi-Civita symbol $\epsilon^{\mu\nu\rho\sigma}$ in flat space-time and curved one. 

We first consider the flat space-time. 
We now find the contributions from the Galileon scalar $\pi$ to the equations of motion 
are expressed as \cite{Nicolis:2008in,Deffayet:2009mn,Deffayet:2010zh}
\bea
\label{Gl1}
\mathcal{E}_1 &=& 1 
= \frac{1}{4!} \epsilon_{\mu\nu\rho\sigma} \epsilon^{\mu\nu\rho\sigma} \, , \nn
\mathcal{E}_2 &=& - 2 \tr \Pi 
= - \frac{1}{3} \epsilon_{\mu\nu\rho}^{\ \ \ \delta} \epsilon^{\mu\nu\rho\sigma} 
\partial_\delta \partial_\sigma \pi \, , \nn
\mathcal{E}_3 &=& - 3 \left\{ \left( \tr \Pi \right)^2 - \tr \Pi^2 \right\} 
= - \frac{3}{2} \epsilon_{\mu\nu}^{\ \ \gamma\delta} \epsilon^{\mu\nu\rho\sigma} 
\partial_\gamma \partial_\rho \pi \partial_\delta \partial_\sigma \pi \, , \nn
\mathcal{E}_4 &=& - 2 \left\{ \left( \tr \Pi \right)^3 - 3 \tr \Pi \tr \Pi^2 +2 \tr \Pi^3 \right\} 
= - 2 \epsilon_\mu^{\ \beta\gamma\delta} \epsilon^{\mu\nu\rho\sigma} 
\partial_\beta \partial_\nu \pi \partial_\gamma \partial_\rho \pi 
\partial_\delta \partial_\sigma \pi \, , \nn
\mathcal{E}_5 &=& - \frac{5}{6}\left\{ \left( \tr \Pi \right)^4 - 6 \left(\tr \Pi\right)^2 \tr \Pi^2 
+ 8 \left(\tr \Pi\right) \tr \Pi^3 + 3 \left(\tr \Pi^2 \right)^2 -6 \tr \Pi^4 \right\} \nn
&=& - \frac{5}{6} \epsilon^{\alpha\beta\gamma\delta} \epsilon^{\mu\nu\rho\sigma} 
\partial_\alpha \partial_\mu \pi \partial_\beta \partial_\nu \pi \partial_\gamma \partial_\rho \pi 
\partial_\delta \partial_\sigma \pi \, .
\eea
Here $\Pi^\mu_{\ \nu} \equiv \partial^\mu \partial_\nu \pi$ and $\tr \Pi = \Pi^\mu_{\ \mu}$. 
We have also used the following formula:
\bea
\label{GG1}
&& \epsilon^{\alpha\beta\gamma\delta} \epsilon^{\mu\nu\rho\sigma}
= \eta^{\alpha\mu} \eta^{\beta\nu} \eta^{\gamma\rho} \eta^{\delta\sigma} 
 - \eta^{\alpha\mu} \eta^{\beta\nu} \eta^{\gamma\sigma} \eta^{\delta\rho}
+ \eta^{\alpha\mu} \eta^{\beta\rho} \eta^{\gamma\sigma} \eta^{\delta\nu} 
 - \eta^{\alpha\mu} \eta^{\beta\rho} \eta^{\gamma\nu} \eta^{\delta\sigma} 
+ \eta^{\alpha\mu} \eta^{\beta\sigma} \eta^{\gamma\nu} \eta^{\delta\rho} 
 - \eta^{\alpha\mu} \eta^{\beta\sigma} \eta^{\gamma\rho} \eta^{\delta\nu} \nn
&& \qquad - \eta^{\alpha\nu} \eta^{\beta\rho} \eta^{\gamma\sigma} \eta^{\delta\mu}
+ \eta^{\alpha\nu} \eta^{\beta\rho} \eta^{\gamma\mu} \eta^{\delta\sigma}
 - \eta^{\alpha\nu} \eta^{\beta\sigma} \eta^{\gamma\mu} \eta^{\delta\rho}
+ \eta^{\alpha\nu} \eta^{\beta\sigma} \eta^{\gamma\rho} \eta^{\delta\mu}
 - \eta^{\alpha\nu} \eta^{\beta\mu} \eta^{\gamma\rho} \eta^{\delta\sigma}
+ \eta^{\alpha\nu} \eta^{\beta\mu} \eta^{\gamma\sigma} \eta^{\delta\rho} \nn
&& \qquad + \eta^{\alpha\rho} \eta^{\beta\sigma} \eta^{\gamma\mu} \eta^{\delta\nu} 
 - \eta^{\alpha\rho} \eta^{\beta\sigma} \eta^{\gamma\nu} \eta^{\delta\mu} 
+ \eta^{\alpha\rho} \eta^{\beta\mu} \eta^{\gamma\nu} \eta^{\delta\sigma} 
 - \eta^{\alpha\rho} \eta^{\beta\mu} \eta^{\gamma\sigma} \eta^{\delta\nu} 
+ \eta^{\alpha\rho} \eta^{\beta\nu} \eta^{\gamma\sigma} \eta^{\delta\mu} 
 - \eta^{\alpha\rho} \eta^{\beta\nu} \eta^{\gamma\mu} \eta^{\delta\sigma} \nn
&& \qquad - \eta^{\alpha\sigma} \eta^{\beta\mu} \eta^{\gamma\nu} \eta^{\delta\rho} 
+ \eta^{\alpha\sigma} \eta^{\beta\mu} \eta^{\gamma\rho} \eta^{\delta\nu} 
 - \eta^{\alpha\sigma} \eta^{\beta\nu} \eta^{\gamma\rho} \eta^{\delta\mu} 
+ \eta^{\alpha\sigma} \eta^{\beta\nu} \eta^{\gamma\mu} \eta^{\delta\rho} 
 - \eta^{\alpha\sigma} \eta^{\beta\rho} \eta^{\gamma\mu} \eta^{\delta\nu} 
+ \eta^{\alpha\sigma} \eta^{\beta\rho} \eta^{\gamma\nu} \eta^{\delta\mu}\, . 
\eea
Then it is easy to find the corresponding Lagrangian densities as follows
\bea
\label{Gl2}
\mathcal{L}_1 &=& \pi\, , \nn
\mathcal{L}_2 &=& \frac{1}{6} \epsilon_{\mu\nu\rho}^{\ \ \ \delta} \epsilon^{\mu\nu\rho\sigma} 
\partial_\delta \pi \partial_\sigma \pi 
= \partial^\mu \pi \partial_\mu \pi \, , \nn
\mathcal{L}_3 &=& \frac{1}{2} \epsilon_{\mu\nu}^{\ \ \gamma\delta} \epsilon^{\mu\nu\rho\sigma} 
\partial_\gamma \pi \partial_\rho \pi \partial_\delta \partial_\sigma \pi 
=  \partial^\mu \pi \partial_\mu \pi \Box \pi
 - \partial^\mu \pi \partial^\nu \pi \partial_\mu \partial_\nu \pi  \, , \nn
\mathcal{L}_4 &=& \frac{1}{2}\epsilon_\mu^{\ \beta\gamma\delta} \epsilon^{\mu\nu\rho\sigma} 
\partial_\beta \pi \partial_\nu \pi \partial_\gamma \partial_\rho \pi 
\partial_\delta \partial_\sigma \pi \nn
&=& \frac{1}{2} \left( \partial_\mu \pi \partial^\mu \pi \left( \Box \pi \right)^2 
 - \partial_\mu \pi \partial^\mu \pi \partial^\sigma \partial_\rho \pi \partial^\rho \partial_\sigma \pi
 - 2 \partial^\mu \pi \partial^\nu \pi \partial_\mu \partial_\nu \pi \Box \pi 
+ 2 \partial^\nu \pi \partial^\mu \pi \partial^\rho \partial_\nu \pi \partial_\mu \partial_\rho \pi \right) 
\, , \nn
\mathcal{L}_5 &=& \frac{1}{6}\epsilon^{\alpha\beta\gamma\delta} \epsilon^{\mu\nu\rho\sigma} 
\partial_\alpha \pi \partial_\mu \pi \partial_\beta \partial_\nu \pi \partial_\gamma \partial_\rho \pi 
\partial_\delta \partial_\sigma \pi \nn
&=& \frac{1}{6} \left( \partial_\mu \pi \partial^\mu \pi \left( \Box \pi \right)^3 
 - 3 \partial_\mu \pi \partial^\mu \pi \Box \pi \partial^\nu \partial_\rho \pi \partial^\rho \partial_\nu \pi 
+ 2 \partial_\mu \pi \partial^\mu \pi \partial^\rho \partial_\nu \pi 
\partial^\sigma \partial_\rho \pi \partial^\nu \partial_\sigma \pi 
 - 3 \partial^\mu \pi \partial^\nu \pi \partial_\mu \partial_\nu \pi \left( \Box \pi \right)^2 \right. \nn
&& \left. + 3 \partial^\mu \pi \partial^\nu \pi \partial_\mu \partial_\nu \pi 
\partial^\sigma \partial_\rho \pi \partial^\rho \partial_\sigma \pi
+ 6 \partial^\nu \pi \partial^\mu \pi \partial^\rho \partial_\nu \pi \partial_\mu \partial_\rho \pi \Box \pi 
 - 6 \partial^\nu \pi \partial^\mu \pi \partial^\sigma \partial_\nu \pi \partial_\mu \partial_\rho \pi 
\partial^\rho \partial_\sigma \pi \right) \, .
\eea

In the curved space-time, instead of the above Lagrangian densities (\ref{Gl2}), 
we may consider the following ones:
\bea
\label{GG2}
\mathcal{L}_1 &=& \pi\, ,\nn
\mathcal{L}_2 &=& \frac{1}{6} \epsilon_{\mu\nu\rho}^{\ \ \ \delta} \epsilon^{\mu\nu\rho\sigma} 
\partial_\delta \pi \partial_\sigma \pi 
= \partial^\mu \pi \partial_\mu \pi \, , \nn
\mathcal{L}_3 &=& \frac{1}{2}
\epsilon_{\mu\nu}^{\ \ \gamma\delta} \epsilon^{\mu\nu\rho\sigma} 
\partial_\gamma \pi \partial_\rho \pi \nabla_\delta \partial_\sigma \pi 
= \partial^\mu \pi \partial_\mu \pi \Box \pi
 - \partial^\mu \pi \partial^\nu \pi \nabla_\mu \partial_\nu \pi \, , \nn
\mathcal{L}^{(0)}_4 &=& \frac{1}{2}\epsilon_\mu^{\ \beta\gamma\delta} \epsilon^{\mu\nu\rho\sigma} 
\partial_\beta \pi \partial_\nu \pi \nabla_\gamma \partial_\rho \pi 
\nabla_\delta \partial_\sigma \pi \, , \nn
&=& \frac{1}{2} \left( \partial_\mu \pi \partial^\mu \pi \left( \Box \pi \right)^2 
 - \partial_\mu \pi \partial^\mu \pi \nabla^\sigma \partial_\rho \pi \nabla^\rho \partial_\sigma \pi
 - 2 \partial^\mu \pi \partial^\nu \pi \nabla_\mu \partial_\nu \pi \Box \pi 
+ 2 \partial^\nu \pi \partial^\mu \pi \nabla^\rho \partial_\nu \pi \nabla_\mu \partial_\rho \pi \right) 
\, , \nn
\mathcal{L}^{(0)}_5 &=& 
\frac{1}{6} \epsilon^{\alpha\beta\gamma\delta} \epsilon^{\mu\nu\rho\sigma} 
\partial_\alpha \pi \partial_\mu \pi \nabla_\beta \partial_\nu \pi \nabla_\gamma \partial_\rho \pi 
\nabla_\delta \partial_\sigma \pi \, , \nn
&=& \frac{1}{6} \left( \partial_\mu \pi \partial^\mu \pi \left( \Box \pi \right)^3 
 - 3 \partial_\mu \pi \partial^\mu \pi \Box \pi \nabla^\nu \partial_\rho \pi \nabla^\rho \partial_\nu \pi 
+ 2 \partial_\mu \pi \partial^\mu \pi \nabla^\rho \partial_\nu \pi 
\nabla^\sigma \partial_\rho \pi \nabla^\nu \partial_\sigma \pi 
 - 3 \partial^\mu \pi \partial^\nu \pi \nabla_\mu \partial_\nu \pi \left( \Box \pi \right)^2 \right. \nn
&& \left. + 3 \partial^\mu \pi \partial^\nu \pi \nabla_\mu \partial_\nu \pi 
\nabla^\sigma \partial_\rho \pi \nabla^\rho \partial_\sigma \pi
+ 6 \partial^\nu \pi \partial^\mu \pi \nabla^\rho \partial_\nu \pi \nabla_\mu \partial_\rho \pi \Box \pi 
 - 6 \partial^\nu \pi \partial^\mu \pi \nabla^\sigma \partial_\nu \pi \nabla_\mu \partial_\rho \pi 
\nabla^\rho \partial_\sigma \pi \right) \, .
\eea
Here $\nabla_\mu$ is the covariant derivative and 
the meaning of the suffix ``$^{(0)}$'' in $\mathcal{L}^{(0)}_5$ and $\mathcal{L}^{(0)}_4$ 
will be clarified soon. 
In (\ref{GG2}), $\epsilon^{\alpha\beta\gamma\delta} \epsilon^{\mu\nu\rho\sigma}$ is {\it defined} by 
\bea
\label{GG3}
&& \epsilon^{\alpha\beta\gamma\delta} \epsilon^{\mu\nu\rho\sigma}
= g^{\alpha\mu} g^{\beta\nu} g^{\gamma\rho} g^{\delta\sigma} 
 - g^{\alpha\mu} g^{\beta\nu} g^{\gamma\sigma} g^{\delta\rho}
+ g^{\alpha\mu} g^{\beta\rho} g^{\gamma\sigma} g^{\delta\nu} 
 - g^{\alpha\mu} g^{\beta\rho} g^{\gamma\nu} g^{\delta\sigma} 
+ g^{\alpha\mu} g^{\beta\sigma} g^{\gamma\nu} g^{\delta\rho} 
 - g^{\alpha\mu} g^{\beta\sigma} g^{\gamma\rho} g^{\delta\nu} \nn
&& \qquad - g^{\alpha\nu} g^{\beta\rho} g^{\gamma\sigma} g^{\delta\mu}
+ g^{\alpha\nu} g^{\beta\rho} g^{\gamma\mu} g^{\delta\sigma}
 - g^{\alpha\nu} g^{\beta\sigma} g^{\gamma\mu} g^{\delta\rho}
+ g^{\alpha\nu} g^{\beta\sigma} g^{\gamma\rho} g^{\delta\mu}
 - g^{\alpha\nu} g^{\beta\mu} g^{\gamma\rho} g^{\delta\sigma}
+ g^{\alpha\nu} g^{\beta\mu} g^{\gamma\sigma} g^{\delta\rho} \nn
&& \qquad + g^{\alpha\rho} g^{\beta\sigma} g^{\gamma\mu} g^{\delta\nu} 
 - g^{\alpha\rho} g^{\beta\sigma} g^{\gamma\nu} g^{\delta\mu} 
+ g^{\alpha\rho} g^{\beta\mu} g^{\gamma\nu} g^{\delta\sigma} 
 - g^{\alpha\rho} g^{\beta\mu} g^{\gamma\sigma} g^{\delta\nu} 
+ g^{\alpha\rho} g^{\beta\nu} g^{\gamma\sigma} g^{\delta\mu} 
 - g^{\alpha\rho} g^{\beta\nu} g^{\gamma\mu} g^{\delta\sigma} \nn
&& \qquad - g^{\alpha\sigma} g^{\beta\mu} g^{\gamma\nu} g^{\delta\rho} 
+ g^{\alpha\sigma} g^{\beta\mu} g^{\gamma\rho} g^{\delta\nu} 
 - g^{\alpha\sigma} g^{\beta\nu} g^{\gamma\rho} g^{\delta\mu} 
+ g^{\alpha\sigma} g^{\beta\nu} g^{\gamma\mu} g^{\delta\rho} 
 - g^{\alpha\sigma} g^{\beta\rho} g^{\gamma\mu} g^{\delta\nu} 
+ g^{\alpha\sigma} g^{\beta\rho} g^{\gamma\nu} g^{\delta\mu}\, . 
\eea
which gives
\bea
\label{GG3B}
\epsilon_\mu^{\ \beta\gamma\delta} \epsilon^{\mu\nu\rho\sigma}
&=& g^{\beta\nu} g^{\gamma\rho} g^{\delta\sigma} 
 - g^{\beta\nu} g^{\gamma\sigma} g^{\delta\rho}
+ g^{\beta\rho} g^{\gamma\sigma} g^{\delta\nu} 
 - g^{\beta\rho} g^{\gamma\nu} g^{\delta\sigma} 
+ g^{\beta\sigma} g^{\gamma\nu} g^{\delta\rho} 
 - g^{\beta\sigma} g^{\gamma\rho} g^{\delta\nu} \, ,\nn
\epsilon_{\mu\nu}^{\ \ \ \gamma\delta} \epsilon^{\mu\nu\rho\sigma}
&=& 2\left( g^{\gamma\rho} g^{\delta\sigma} - g^{\gamma\sigma} g^{\delta\rho} \right) \, ,\nn
\epsilon_{\mu\nu\rho}^{\ \ \ \ \delta} \epsilon^{\mu\nu\rho\sigma}
&=& 6 g^{\delta\sigma} \, .
\eea
Since $\nabla_\rho g_{\mu\nu}=0$, we find 
$\nabla_\lambda \left( \epsilon^{\alpha\beta\gamma\delta} \epsilon^{\mu\nu\rho\sigma}
\right) =0$. 
By the variation of $\pi$, instead of (\ref{Gl1}), we obtain
\bea
\label{GG4}
\mathcal{E}_1 &=& 1 \, , \nn
\mathcal{E}_2 &=& - 2 \nabla^\mu \partial_\mu \pi \, , \nn
\mathcal{E}_3 &=& \epsilon_{\mu\nu}^{\ \ \gamma\delta} \epsilon^{\mu\nu\rho\sigma} \left\{ - \frac{3}{2} 
\nabla_\gamma \partial_\rho \pi \nabla_\delta \partial_\sigma \pi 
+ \frac{3}{4} R^\lambda_{\ \sigma\gamma\delta} \partial_\rho \pi \partial_\lambda \pi \right\} \, , \nn
\mathcal{E}^{(0)}_4 &=& \epsilon_\mu^{\ \beta\gamma\delta} \epsilon^{\mu\nu\rho\sigma} \left\{ - 2 
\nabla_\beta \partial_\nu \pi \nabla_\gamma \partial_\rho \pi 
\nabla_\delta \partial_\sigma \pi 
+ \frac{5}{2} R^\lambda_{\ \rho\beta\gamma} \partial_\nu \pi \partial_\lambda \pi 
\nabla_\delta \partial_\sigma \pi \right. \nn
&& \left.  - \frac{1}{2} \partial_\beta \pi \partial_\nu \pi 
\left( \nabla_\rho R^\lambda_{\ \sigma\gamma\delta} \right)
\partial_\lambda \pi 
 - \frac{1}{2} \partial_\beta \pi \partial_\nu \pi R^\lambda_{\ \sigma\gamma\delta} 
\nabla_\rho \partial_\lambda \pi \right\} \, , \nn
\mathcal{E}^{(0)}_5 &=& \epsilon^{\alpha\beta\gamma\delta} \epsilon^{\mu\nu\rho\sigma} \left\{ - \frac{5}{6} 
\nabla_\alpha \partial_\mu \pi \nabla_\beta \partial_\nu \pi \nabla_\gamma \partial_\rho \pi 
\nabla_\delta \partial_\sigma \pi 
+ \frac{7}{4} R^\lambda_{\ \nu\alpha\beta}
\partial_\mu \pi \partial_\lambda \pi \nabla_\gamma \partial_\rho \pi 
\nabla_\delta \partial_\sigma \pi \right. \nn
&& \left. + \frac{1}{4} \partial_\alpha \pi \partial_\mu \pi 
R^\lambda_{\ \rho\beta\gamma} \partial_\lambda \pi 
R^\tau_{\ \delta\nu\sigma} \partial_\tau \pi - \frac{1}{2} \partial_\alpha \pi 
\partial_\mu \pi \left(\nabla_\nu R^\lambda_{\ \rho\beta\gamma} \right)
\partial_\lambda \pi \nabla_\delta \partial_\sigma \pi 
 - \frac{1}{2} \partial_\alpha \pi \partial_\mu \pi R^\lambda_{\ \rho\beta\gamma} 
\nabla_\nu \partial_\lambda \pi \nabla_\delta \partial_\sigma \pi \right\}\, .
\eea
Here we have used the Bianchi identity
\be
\label{GG8}
0 = \nabla_\lambda R^\alpha_{\ \beta\mu\nu} + \nabla_\mu R^\alpha_{\ \beta\nu\lambda} 
+ \nabla_\nu R^\alpha_{\ \beta\lambda\mu}\, ,
\ee 
and the definition of the Riemann curvature:
\be
\label{GG5}
\left[ \nabla_\mu, \nabla_\nu \right] V_\rho = - R^\lambda_{\ \rho\mu\nu} V_\lambda\, .
\ee
Here $V_\lambda$ is a covariant vector. 
Note that in $\mathcal{E}^{(0)}_5$ and $\mathcal{E}^{(0)}_4$, there appear the derivatives 
of the Riemann curvature, which include the third derivatives. 
By using the Bianchi identity (\ref{GG8}), we find 
\bea
\label{Delta1}
 - \frac{1}{2} \epsilon^{\alpha\beta\gamma\delta} \epsilon^{\mu\nu\rho\sigma} 
\partial_\alpha \pi 
\partial_\mu \pi \left(\nabla_\nu R^\lambda_{\ \rho\beta\gamma} \right)
\partial_\lambda \pi \nabla_\delta \partial_\sigma \pi 
&=& \frac{1}{4} \epsilon^{\alpha\beta\gamma\delta} \epsilon^{\mu\nu\rho\sigma}
\partial_\alpha \pi \partial_\mu \pi \nabla^\lambda \left( R_{\rho\nu\beta\gamma} \right)
\partial_\lambda \pi \nabla_\delta \partial_\sigma \pi \, , \nn
 - \frac{1}{2} \epsilon_\mu^{\ \beta\gamma\delta} \epsilon^{\mu\nu\rho\sigma} 
\partial_\beta \pi \partial_\nu \pi 
\left( \nabla_\rho R^\lambda_{\ \sigma\gamma\delta} \right)
\partial_\lambda \pi 
&=& \frac{1}{4} \epsilon_\mu^{\ \beta\gamma\delta} \epsilon^{\mu\nu\rho\sigma} 
\partial_\beta \pi \partial_\nu \pi \nabla^\lambda 
\left( R_{\sigma\rho \gamma\delta}\right) \partial_\lambda \pi \, .
\eea
Then if we consider the following Lagrangian densities:
\bea
\label{Delta2}
\Delta\mathcal{L}_4 &=& \frac{1}{8} \epsilon_\mu^{\ \beta\gamma\delta} \epsilon^{\mu\nu\rho\sigma} 
\partial_\beta \pi \partial_\nu \pi 
R_{\sigma\rho \gamma\delta} \partial_\lambda \pi \partial^\lambda \pi\, , \nn
\Delta\mathcal{L}_5 &=& \frac{1}{8} \epsilon^{\alpha\beta\gamma\delta} \epsilon^{\mu\nu\rho\sigma}
\partial_\alpha \pi \partial_\mu \pi R_{\rho\nu\beta\gamma} 
\nabla_\delta \partial_\sigma \pi \partial_\lambda \pi \partial^\lambda \pi \, ,
\eea
by the variation of $\pi$, we have 
\bea
\label{Delta3}
\Delta\mathcal{E}_4 &=& - \frac{1}{8} \epsilon_\mu^{\ \beta\gamma\delta} \epsilon^{\mu\nu\rho\sigma} 
\left\{ 2 \nabla_\beta \partial_\nu \pi R_{\sigma\rho\gamma\delta} \partial_\lambda \pi \partial^\lambda \pi 
+ 8 \partial_\nu \pi R_{\sigma\rho\gamma\delta} \nabla_\beta \partial_\lambda \pi \partial^\lambda \pi \right. \nn
&& \left. + 2 \partial_\beta \pi \partial_\nu \pi R_{\sigma\rho\gamma\delta} \Box \pi 
+ 2 \partial_\beta \pi \partial_\nu \pi \nabla_\lambda R_{\sigma\rho\gamma\delta}
\partial^\lambda \pi \right\}\, , \nn
\Delta\mathcal{E}_5 &=& \frac{1}{8} \epsilon^{\alpha\beta\gamma\delta} \epsilon^{\mu\nu\rho\sigma}
\left\{ -3 \nabla_\alpha \partial_\mu \pi \nabla_\delta \partial_\sigma \pi R_{\rho\nu\beta\gamma} 
\partial_\lambda \pi \partial^\lambda \pi 
+ \frac{3}{2} \partial_\mu \pi R_{\rho\nu\beta\gamma} R^\tau_{\ \sigma\alpha\delta} \partial_\tau \pi 
\partial_\lambda \pi \partial^\lambda \pi \right. \nn
&& - 12 \partial_\mu \pi R_{\rho\nu\beta\gamma} \nabla_\delta \partial_\sigma \pi 
\nabla_\alpha \partial_\lambda \pi \partial^\lambda \pi 
+ 2 \partial_\alpha \pi \partial_\mu \pi R_{\rho\nu\beta\gamma} 
\nabla_\sigma \partial_\lambda \pi \nabla_\delta \partial^\lambda \pi 
 - 2 \partial_\alpha \pi \partial_\mu \pi R_{\rho\nu\beta\gamma} R^\tau_{\ \delta\sigma\lambda} 
\partial_\tau \pi \partial^\lambda \pi \nn
&& \left.  - 2 \partial_\alpha \pi \partial_\mu \pi R_{\rho\nu\beta\gamma} \nabla_\delta \partial_\sigma \pi 
\Box \pi 
 - 2 \partial_\alpha \pi \partial_\mu \pi \nabla_\lambda R_{\rho\nu\beta\gamma} 
\nabla_\delta \partial_\sigma \pi \partial^\lambda \pi \right\} \, .
\eea
Hence, if we define
\be
\label{Delta4}
\mathcal{L}_4 \equiv \mathcal{L}^{(0)}_4 + \Delta\mathcal{L}_4 \, , \quad
\mathcal{L}_5 \equiv \mathcal{L}^{(0)}_5 + \Delta\mathcal{L}_5 \, ,
\ee
by the variation of $\pi$, we acquire 
\bea
\label{Delta5}
\mathcal{E}_4  &=& \mathcal{E}^{(0)}_4 + \Delta\mathcal{E}_4 \nn
&=& \epsilon_\mu^{\ \beta\gamma\delta} \epsilon^{\mu\nu\rho\sigma} \left\{ - 2 
\nabla_\beta \partial_\nu \pi \nabla_\gamma \partial_\rho \pi 
\nabla_\delta \partial_\sigma \pi 
+ \frac{5}{2} R^\lambda_{\ \rho\beta\gamma} \partial_\nu \pi \partial_\lambda \pi 
\nabla_\delta \partial_\sigma \pi 
 - \frac{1}{2} \partial_\beta \pi \partial_\nu \pi R^\lambda_{\ \sigma\gamma\delta} 
\nabla_\rho \partial_\lambda \pi \right. \nn
&& \left. - \frac{1}{4} \nabla_\beta \partial_\nu \pi R_{\sigma\rho\gamma\delta} 
\partial_\lambda \pi \partial^\lambda \pi 
 - \partial_\nu \pi R_{\sigma\rho\gamma\delta} \nabla_\beta \partial_\lambda \pi \partial^\lambda \pi 
 - \frac{1}{4} \partial_\beta \pi \partial_\nu \pi R_{\sigma\rho\gamma\delta} \Box \pi 
\right\}\, , \nn
\mathcal{E}_5 &=& \mathcal{E}^{(0)}_5 + \Delta\mathcal{E}_5 \nn
&=& \epsilon^{\alpha\beta\gamma\delta} \epsilon^{\mu\nu\rho\sigma} \left\{ - \frac{5}{6} 
\nabla_\alpha \partial_\mu \pi \nabla_\beta \partial_\nu \pi \nabla_\gamma \partial_\rho \pi 
\nabla_\delta \partial_\sigma \pi 
+ \frac{7}{4} R^\lambda_{\ \nu\alpha\beta}
\partial_\mu \pi \partial_\lambda \pi \nabla_\gamma \partial_\rho \pi 
\nabla_\delta \partial_\sigma \pi \right. \nn
&& + \frac{1}{4} \partial_\alpha \pi \partial_\mu \pi 
R^\lambda_{\ \rho\beta\gamma} \partial_\lambda \pi 
R^\tau_{\ \delta\nu\sigma} \partial_\tau \pi 
 - \frac{1}{2} \partial_\alpha \pi \partial_\mu \pi R^\lambda_{\ \rho\beta\gamma} 
\nabla_\nu \partial_\lambda \pi \nabla_\delta \partial_\sigma \pi 
 - \frac{3}{8} \nabla_\alpha \partial_\mu \pi \nabla_\delta \partial_\sigma \pi R_{\rho\nu\beta\gamma} 
\partial_\lambda \pi \partial^\lambda \pi \nn
&& + \frac{3}{16} \partial_\mu \pi R_{\rho\nu\beta\gamma} R^\tau_{\ \sigma\alpha\delta} \partial_\tau \pi 
\partial_\lambda \pi \partial^\lambda \pi 
 - \frac{3}{2} \partial_\mu R_{\rho\nu\beta\gamma} \nabla_\delta \partial_\sigma \pi 
\nabla_\alpha \partial_\lambda \pi \partial^\lambda \pi 
+ \frac{1}{4} \partial_\alpha \pi \partial_\mu \pi R_{\rho\nu\beta\gamma} 
\nabla_\sigma \partial_\lambda \pi \nabla_\delta \partial^\lambda \pi \nn
&& \left.  - \frac{1}{4} \partial_\alpha \pi \partial_\mu \pi R_{\rho\nu\beta\gamma} R^\tau_{\ \delta\sigma\lambda} 
\partial_\tau \pi \partial^\lambda \pi 
 - \frac{1}{4} \partial_\alpha \pi \partial_\mu \pi R_{\rho\nu\beta\gamma} 
\nabla_\delta \partial_\sigma \pi \Box \pi \right\} \, .
\eea
Note that in the expressions in (\ref{Delta5}), there does not appear the derivatives higher than two. 

In the curved space-time, we may also consider the following Lagrangian densities
\bea
\label{GG6}
\mathcal{L}_6 &=& \epsilon_\mu^{\ \beta\gamma\delta} \epsilon^{\mu\nu\rho\sigma} 
\partial_\beta \pi \partial_\nu \pi R_{\gamma\delta\rho\sigma} \nn
&=& -4 \left( R_{\mu\nu} - \frac{1}{2} g_{\mu\nu} R \right) \partial^\mu \pi \partial^\nu \pi \, , \nn
\mathcal{L}_7 &=& \epsilon^{\alpha\beta\gamma\delta} \epsilon^{\mu\nu\rho\sigma} 
\partial_\alpha \pi \partial_\mu \pi \nabla_\beta \partial_\nu \pi 
R_{\gamma\delta\rho\sigma} \, , \nn
&=& 2 \partial_\mu \pi \partial^\mu \pi \Box \pi R 
 - 4 \partial_\mu \pi \partial^\mu \pi \nabla^\nu \nabla^\rho \pi R_{\nu\rho} 
 - 2 \partial^\nu \pi \partial^\mu \pi \nabla_\nu \partial_\mu \pi R \nn
&& + 8 \partial^\nu \pi \partial^\mu \pi \nabla_\nu \partial^\rho \pi R_{\mu\rho} 
 - 4 \partial^\rho \pi \partial^\mu \pi \Box \pi R_{\mu\rho} 
+ 4 \partial^\rho \pi \partial^\mu \pi \nabla^\sigma \partial^\nu \pi R_{\mu\nu\rho\sigma} \, .
\eea
By the variation $\pi$, we obtain
\bea
\label{GG7}
\mathcal{E}_6 &=& -2 \epsilon_\mu^{\ \beta\gamma\delta} \epsilon^{\mu\nu\rho\sigma} 
\nabla_\beta \partial_\nu \pi R_{\gamma\delta\rho\sigma} \, , \nn
\mathcal{E}_7 &=& - 3 \epsilon^{\alpha\beta\gamma\delta} \epsilon^{\mu\nu\rho\sigma} 
\nabla_\alpha \partial_\mu \pi \nabla_\beta \partial_\nu \pi 
R_{\gamma\delta\rho\sigma} 
+ \frac{3}{2} \epsilon^{\alpha\beta\gamma\delta} \epsilon^{\mu\nu\rho\sigma} 
\partial_\mu \pi \partial_\lambda \pi R^\lambda_{\ \nu\alpha\beta} 
R_{\gamma\delta\rho\sigma} \, .
\eea
Even in the curved space-time, $\mathcal{E}_i$'s $\left(i=1,2,\cdots,7\right)$ do not include the derivatives higher than the second ones\footnote{
In addition to the Lagrangian densities in (\ref{GG6}), the Brans-Dicke type Lagrangian density 
\[
\mathcal{L}_\mathrm{BD} = f(\pi) R\, ,
\]
does not generate the field equation and the Einstein equation including the derivatives higher than the second ones. 
Here $f(\pi)$ is an arbitrary function of the scalar field $\pi$.  
}. 
We should also note that 
\be
\label{Delta8}
\Delta\mathcal{L}_4 = - \frac{1}{8}\mathcal{L}_6 \partial_\lambda \pi \partial^\lambda \pi \, , \quad
\Delta\mathcal{L}_5 = - \frac{1}{8}\mathcal{L}_7 \partial_\lambda \pi \partial^\lambda \pi \, .
\ee
Now we find that the general Lagrangian has the following form
\bea
\label{DeltaGG8}
\mathcal{L} &=& \frac{R}{2\kappa^2} + G_2 (\pi, X) + G_3 (\pi, X) \mathcal{L}_3 
+ \frac{\partial G_4 (\pi, X)}{\partial X} \mathcal{L}^{(0)}_4 - \frac{1}{8} G_4 (\pi, X) \mathcal{L}_6 \nn
&& + \frac{\partial G_5 (\pi, X)}{\partial X} \mathcal{L}^{(0)}_5 - \frac{1}{8} G_5 (\pi, X) \mathcal{L}_7
+ G_6 (\pi) \mathcal{L}_6 + G_7 (\pi) \mathcal{L}_7 + G_8 (\pi) \mathcal{L}_8 \, .
\eea
Here $X=\partial_\mu \pi \partial^\mu \pi$ and 
$\mathcal{L}_8$ is four times the Gauss-Bonnet invariant:
\be
\label{GB}
\mathcal{L}_8 = \epsilon^{\alpha\beta\gamma\delta} \epsilon^{\mu\nu\rho\sigma} 
R_{\alpha\beta\mu\nu} R_{\gamma\delta\rho\sigma}
= 4 \left( R^2 - 4 R_{\mu\nu} R^{\mu\nu} + R_{\mu\nu\rho\sigma} R^{\mu\nu\rho\sigma} \right)\, .
\ee


We now show that even in the generalized action (\ref{DeltaGG8}), there do not appear the higher derivative (higher than second order) in the equations given by the variation of 
$\pi$. In fact, we find
\begin{eqnarray}
\label{V2}
\mathcal{E}^{(G)}_2 &=& \frac{\partial G_2}{\partial \pi} 
 -2 \partial_{\alpha} \frac{\partial G_2}{\partial X} \partial^{\alpha} \pi
 -2 \frac{\partial G_2}{\partial X} \Box \pi\, . \\
\label{V3}
\mathcal{E}^{(G)}_3 &=&
\epsilon_{\mu \nu} ^{ \ \  \ \gamma \delta} \epsilon^{\mu \nu \rho \sigma}
\left[ 
2 \frac{\partial^2 G_3}{\partial X^2} \partial_{\beta} \pi \partial_{\alpha} \pi \partial_{\gamma}
\pi \partial_{\rho} \pi 
( \nabla_{\sigma} \partial^{\beta} \pi \nabla_{\delta} \partial^{\alpha} \pi
 - \nabla^{\alpha} \partial^{\beta} \pi \nabla_{\delta} \partial_{\sigma} \pi)
\right. \nn
&& + \frac{\partial^2 G_3}{\partial \pi \partial X}
\partial_{\gamma} \pi \partial_{\rho} \pi 
( 2 \partial_{\sigma} \pi \partial_{\alpha} \pi \nabla_{\delta} \partial^{\alpha} \pi
 - \partial^{\alpha} \pi \partial_{\alpha} \pi \nabla_{\delta} \partial_{\sigma} \pi )
+ \frac{1}{2} \frac{\partial^2 G_3}{\partial \pi^2} \partial_{\sigma} \pi \partial_{\delta} \pi
\partial_{\gamma} \pi \partial_{\rho} \pi \nn
&& + \frac{\partial G_3}{\partial X}
( \nabla_{\sigma} \partial_{\alpha} \pi \partial_{\gamma} \pi \partial_{\rho} \pi \nabla_{\delta}
\partial^{\alpha} \pi
+ 3 \partial_{\alpha} \pi \nabla_{\sigma} \partial_{\gamma} \pi \partial_{\rho} \pi
\nabla_{\delta} \partial^{\alpha} \pi
 -4 \partial_{\alpha} \pi \nabla^{\alpha} \partial_{\gamma} \pi \partial_{\rho} \pi
\nabla_{\delta} \partial_{\sigma} \pi
 - \Box \pi \partial_{\gamma} \pi \partial_{\rho} \pi \nabla_{\delta} \partial_{\sigma} \pi  
) \nn
&& + \frac{\partial G_3}{\partial \pi} \left( \partial_{\sigma} \pi \partial_{\gamma} \pi
\nabla_{\delta} \partial_{\rho} \pi - \frac{3}{2} \partial_{\gamma} \pi \partial_{\rho} \pi
\nabla_{\delta} \partial_{\sigma} \pi \right) 
 - \frac{3}{2} G_3 \nabla_{\gamma} \partial_{\rho} \pi \nabla_{\delta} \partial_{\sigma} \pi \\
&& \left. + \frac{\partial G_3}{\partial X} \partial^{\alpha} \pi \partial_{\gamma} \pi \partial_{\rho} \pi
R_{\sigma \alpha \delta}^{ \ \ \ \ \beta} \partial_{\beta} \pi
 - \frac{3}{4} G_3 R_{\gamma \delta \sigma}^{ \ \ \ \ \alpha} \partial_{\alpha} \pi
\partial_{\rho} \pi \right]\, , \\
\label{V4}
\mathcal{E}^{(G,0)}_4 &=& \epsilon_{\mu} ^{ \ \  \beta \gamma \delta} \epsilon^{\mu \nu \rho \sigma}
\left[ - \partial_{\alpha} \frac{\partial^2 G_4}{\partial X^2} \partial^{\alpha} \pi \partial_{\beta} \pi
\partial_{\nu} \pi \nabla_{\gamma} \partial_{\rho} \pi \nabla_{\delta} \partial_{\sigma} \pi 
 - \frac{\partial^2 G_4}{\partial X^2} \partial_{\nu} \pi \nabla_{\gamma} \partial_{\rho} \pi
\nabla_{\delta} \partial_{\sigma} \pi (\Box \pi \partial_{\beta} \pi +2 \partial^{\alpha} \pi
\nabla_{\alpha} \partial_{\beta} \pi )
\right. \nn 
&& + 2 \partial_{\rho} \frac{\partial^2 G_4}{\partial X^2} \partial_{\alpha} \pi 
\nabla_{\gamma} \partial^{\alpha} \pi \partial_{\beta} \pi \partial_{\nu} \pi 
\nabla_{\delta} \partial_{\sigma} \pi
+  2 \frac{\partial^2 G_4}{\partial X^2} \nabla_{\rho} \partial_{\alpha} \pi \nabla_{\gamma} 
\partial^{\alpha} \pi \partial_{\beta} \pi \partial_{\nu} \pi \nabla_{\delta} \partial_{\sigma} \pi \nn
&& + \partial_{\rho} \frac{\partial^2 G_4}{\partial \pi \partial X} 
\partial_{\gamma} \pi \partial_{\beta} \pi \partial_{\nu} \pi \nabla_{\delta} \partial_{\sigma} \pi \nn 
&& + \frac{\partial G_4}{\partial X} \left( 
 -2 \nabla_{\beta} \partial_{\nu} \pi \nabla_{\gamma} \partial_{\rho} \pi \nabla_{\delta} \partial_{\sigma} \pi
+ \frac{5}{2} R^{\lambda}_{ \ \rho \beta \gamma} \partial_{\nu} \pi 
\partial_{\lambda} \pi \nabla_{\delta} \partial_{\sigma} \pi
 - \frac{1}{2} \partial_{\beta} \pi \partial_{\nu} \pi R^{\lambda}_{ \ \sigma \gamma \delta} 
\nabla_{\rho} \partial_{\lambda} \pi \right) \nn
&& + 2 \frac{\partial^2 G_4}{\partial X^2} 
\left( \partial_{\alpha} \pi \nabla_{\gamma} \partial^{\alpha} \pi
\partial_{\beta} \pi \partial_{\nu} \pi R_{\rho \sigma \delta}^{ \ \ \ \ \lambda} \partial_{\lambda} \pi
+ \partial^{\alpha} \pi R_{\rho \alpha \gamma}^{ \ \ \ \ \lambda} \partial_{\lambda} \pi 
\partial_{\beta} \pi \partial_{\nu} \pi \nabla_{\delta} \partial_{\sigma} \pi
\right) \nn
&& + 2  \frac{\partial^2 G_4}{\partial X^2} \partial^{\alpha} \pi 
R_{\rho \alpha \gamma}^{ \ \ \ \ \lambda} \partial_{\lambda} \pi \partial_{\beta} \pi 
\partial_{\nu} \pi \nabla_{\delta} \partial_{\sigma} \pi
+ \frac{\partial^2 G}{\partial \pi \partial X} \partial_{\gamma} \pi 
\partial_{\beta} \pi \partial_{\nu} \pi 
R_{\rho \sigma \delta}^{ \ \ \ \ \lambda} \partial_{\lambda} \pi \nn
&& \left. - \frac{1}{2} \frac{\partial G_4}{\partial X} \partial_{\beta} \pi \partial_{\nu} \pi 
( \nabla_{\rho} R^{\lambda}_{ \ \sigma \gamma \delta}) \partial_{\lambda} \pi 
\right]\, .
\end{eqnarray}
Only the last term in (\ref{V4}) includes the third derivative of the metric. 
This term can be canceled by $\Delta \mathcal{E}^{(G)}_4$, which is given by the variation of $-\frac{1}{8} G_4 \mathcal{L}_6$, 
\begin{eqnarray}
\label{DeltaV4}
\Delta \mathcal{E}^{(G)}_4 &=& 
\epsilon_{\mu} ^{ \ \  \beta \gamma \delta} \epsilon^{\mu \nu \rho \sigma}
\left[
\frac{1}{4} \partial_{\lambda} \frac{\partial G_4}{\partial X} \partial^{\lambda} \pi 
\partial_{\beta} \pi \partial_{\nu} \pi R_{\gamma \delta \rho \sigma}
+ \frac{1}{4} \frac{\partial G_4}{\partial X} \Box \pi \partial_{\beta} \pi \partial_{\nu} \pi
R_{\gamma \delta \rho \sigma}
+ \frac{1}{2} \frac{\partial G_4}{\partial X} \partial^{\lambda} \pi 
\nabla_{\lambda} \partial_{\beta} \pi \partial_{\nu} \pi R_{\gamma \delta \rho \sigma} \right. \nn
&& \left. + \frac{1}{2} \frac{\partial G_4}{\partial X} \partial_{\beta} \pi \partial_{\nu} ( \nabla_{\rho}
R^{\lambda}_{ \ \sigma \gamma \delta}) \partial_{\lambda} \pi
 - \frac{1}{8} \frac{\partial G_4}{\partial \pi} \partial_{\beta} \pi \partial_{\nu} \pi
R_{\gamma \delta \rho \sigma}
+ \frac{1}{4} \partial_{\beta} G_4 \partial_{\nu} \pi R_{\gamma \delta \rho \sigma} 
+ \frac{1}{4} G_4 \nabla_{\beta} \partial_{\nu} \pi R_{\gamma \delta \rho \sigma}
\right]\, .
\end{eqnarray}
The third derivative term in (\ref{V4}) is canceled by the fourth term in (\ref{DeltaV4}). 

We also have
\begin{eqnarray}
\label{V5}
\mathcal{E}^{(G,0)}_5 &=&
\epsilon^{ \alpha  \beta \gamma \delta} \epsilon^{\mu \nu \rho \sigma}
\left[ - \frac{1}{3} \partial_{\lambda} \frac{\partial^2 G_5}{\partial X^2}
\partial^{\lambda} \pi \partial_{\alpha} \pi \partial_{\mu} \pi \nabla_{\beta} \partial_{\nu} \pi
\nabla_{\gamma} \partial_{\rho} \pi \nabla_{\delta} \partial_{\sigma} \pi
\right. \nn 
&& - \frac{1}{3} \frac{\partial^2 G_5}{\partial X^2}
\partial_{\mu} \pi \nabla_{\beta} \partial_{\nu} \pi \nabla_{\gamma} \partial_{\rho} \pi
\nabla_{\delta} \partial_{\sigma} \pi (\Box \pi \partial_{\alpha} \pi + 2 \partial^{\lambda} 
\pi \nabla_{\lambda} \partial_{\alpha} \pi )
+ \frac{1}{6} \frac{\partial^2 G_5}{\partial \pi \partial X} 
\partial_{\alpha} \pi \partial_{\nu} \pi \nabla_{\beta} \partial_{\nu} \pi 
\nabla_{\gamma} \partial_{\rho} \pi \nabla_{\delta} \partial_{\sigma} \pi \nn
&& + \left( 
\partial_{\nu} \frac{\partial^2 G_5}{\partial X^2} \partial_{\lambda} \pi
+ \frac{\partial^2 G_5}{\partial X^2} \nabla_{\nu} \partial_{\lambda} \pi
\right) \nabla_{\beta} \partial^{\lambda} \pi \partial_{\alpha} \pi \partial_{\mu} \pi
 \nabla_{\gamma} \partial_{\rho} \pi \nabla_{\delta} \partial_{\sigma} \pi \nn
&& + \frac{1}{2} \left( 
\partial_{\nu} \frac{\partial^2 G_5}{\partial \pi \partial X} \partial_{\beta} \pi 
+ \frac{\partial^2 G_5}{\partial \pi \partial X} \nabla_{\nu} \partial_{\beta} \pi
\right) \partial_{\alpha} \pi \partial_{\mu} \pi \nabla_{\gamma} \partial_{\rho} \pi
\nabla_{\delta} \partial_{\sigma} \pi \nn
&& \frac{\partial G_5}{\partial X} \left( 
 - \frac{5}{6} \nabla_{\alpha} \partial_{\mu} \pi \nabla_{\beta} \partial_{\nu} \pi 
\nabla_{\gamma} \partial_{\rho} \pi \nabla_{\delta} \partial_{\sigma} \pi
+ \frac{7}{4} R^{\lambda}_{ \ \nu \alpha \beta} \partial_{\mu} \pi \partial_{\lambda} \pi 
\nabla_{\gamma} \partial_{\rho} \pi \nabla_{\delta} \partial_{\sigma} \pi
+ \frac{1}{4} \partial_{\alpha} \pi \partial_{\mu} \pi R^{\lambda}_{ \ \rho \beta \gamma}
\partial_{\lambda} \pi R^{\tau}_{ \ \delta \nu \sigma} \partial_{\tau} \pi \right. \nn
&& \left. - \frac{1}{2} \partial_{\alpha} \pi \partial_{\mu} \pi R^{\lambda}_{ \ \rho \beta \gamma}
\nabla_{\nu} \partial_{\lambda} \pi \nabla_{\delta} \partial_{\sigma} \pi 
\right)
+ \frac{\partial^2 G_5}{\partial X^2} \partial^{\lambda} \pi R_{\nu \lambda \beta}^{ \ \ \ \ \tau}
\partial_{\tau} \pi \partial_{\alpha} \pi \partial_{\mu} \pi \nabla_{\gamma} \partial_{\rho} \pi
\nabla_{\delta} \partial_{\sigma} \pi \nn
&& \left. + \partial_{\beta} \frac{\partial G_5}{\partial X}
\partial_{\alpha} \pi \partial_{\mu} \pi \nabla_{\gamma} \partial_{\rho} \pi
R_{\nu \sigma \delta}^{ \ \ \ \ \lambda} \partial_{\lambda} \pi
 - \frac{1}{2} \frac{\partial G_5}{\partial X} \partial_{\alpha} \pi \partial_{\mu} \pi 
( \nabla_{\nu} R^{\lambda}_{ \  \rho \beta \gamma}) 
\partial_{\lambda} \pi \nabla_{\delta} \partial_{\sigma} \pi \right]\, .
\end{eqnarray}
Only the last term in (\ref{V5}) includes the third derivative term and this term is canceled by the seventh term in $\Delta \mathcal{E}^{(G)}_5$ 
\begin{eqnarray}
\label{DeltaV5}
\Delta \mathcal{E}^{(G)}_5 &=& 
\epsilon^{ \alpha \beta \gamma \delta} \epsilon^{\mu \nu \rho \sigma}
\left[
\frac{1}{4} \left( \partial_{\lambda} \frac{\partial G_5}{\partial X} \partial^{\lambda} \pi
+ \frac{\partial G_5}{\partial X} \Box  \pi \right)
\partial_{\alpha} \pi \partial_{\mu} \pi \nabla_{\beta} \partial_{\nu} \pi R_{\gamma \delta \rho \sigma}
+ \frac{1}{2} \frac{\partial G_5}{\partial X} \partial_{\lambda} \pi 
\nabla_{\lambda} \partial_{\alpha} \pi \partial_{\mu} \pi 
\nabla_{\beta} \partial_{\nu} \pi R_{\gamma \delta \rho \sigma} \right. \nn
&& + \frac{1}{4} \left(
\partial_{\alpha} G_5 \partial_{\mu} \pi + G_5 \nabla_{\alpha} \partial_{\mu} \pi
\right) \nabla_{\beta} \partial_{\nu} \pi R_{\gamma \delta \rho \sigma}
 - \frac{1}{4} \partial_{\beta} G_5 \nabla_{\nu} \partial_{\alpha} \pi \partial_{\mu} \pi
R_{\gamma \delta \rho \sigma}
+ \frac{1}{8} G_5 \nabla_{\nu} \partial_{\alpha} \pi 
\nabla_{\beta} \partial_{\mu} R_{\gamma \delta \rho \sigma} \nn
&& - \frac{1}{4} \frac{\partial G_5}{\partial \pi} \partial_{\alpha} \pi \partial_{\mu} \pi 
\nabla_{\beta} \partial_{\nu} \pi R_{\gamma \delta \rho \sigma}
+ \frac{1}{2} \frac{\partial G_5}{\partial X} \partial_{\alpha} \pi \partial_{\mu} \pi
( \nabla_{\nu} R^{\lambda}_{ \ \rho \beta \gamma}) \partial_{\lambda} \pi 
\nabla_{\delta } \partial_{\sigma} \pi \nn
&& \left.
+ \frac{1}{4} \frac{\partial G_5}{\partial X} \partial_{\lambda} \pi \partial_{\alpha} \pi
\partial_{\mu} \pi R_{\lambda \nu \beta}^{ \ \ \ \ \tau} \partial_{\tau} \pi
R_{ \gamma \delta \rho \sigma}
+ \frac{3}{8} G_5 \partial_{\mu} \pi R_{\alpha \beta \nu}^{ \ \ \ \ \tau} \partial_{\tau} \pi
R_{ \gamma \delta \rho \sigma}
\right]\, .
\end{eqnarray}
It is straightforward to see that the equations of motion given by variation of $\pi$ from 
$G_6 (\pi) \mathcal{L}_6$, $G_7(\pi) \mathcal{L}_7$, and $G_8(\pi) \mathcal{L}_8$ do not include the 
derivative higher than two because 
$G_6(\pi)$, $G_7(\pi)$, and $G_8(\pi)$ do not contain $X$. 


We also investigate 
the equation given by the variation of the metric. 
The equation corresponds to the Einstein equation. 
Since 
\bea
\label{Q4}
\delta g &=& g g^{\mu\nu} \delta g_{\mu\nu}\, , \nn
\delta\Gamma^\kappa_{\mu\nu} &=&\frac{1}{2}g^{\kappa\lambda}\left(
\nabla_\mu \delta g_{\nu\lambda} + \nabla_\nu \delta g_{\mu\lambda}
 - \nabla_\lambda \delta g_{\mu\nu}
\right)\, ,\nn
\delta R^\mu_{\ \nu\lambda\sigma}&=& \nabla_\lambda \delta\Gamma^\mu_{\sigma\nu} 
 - \nabla_\sigma \delta \Gamma^\mu_{\lambda\nu}\, ,\nn
\delta R_{\mu\nu\lambda\sigma}&=&\frac{1}{2}\left[\nabla_\lambda \nabla_\nu \delta g_{\sigma\mu} 
 - \nabla_\lambda \nabla_\mu \delta g_{\sigma\nu} 
 - \nabla_\sigma \nabla_\nu \delta g_{\lambda\mu} 
+ \nabla_\sigma \nabla_\mu \delta g_{\lambda\nu} 
+ \delta g_{\mu\rho} R^\rho_{\ \nu\lambda\sigma} 
 - \delta g_{\nu\rho} R^\rho_{\ \mu\lambda\sigma} \right] \, ,\nn
\delta R_{\mu\nu} &=& \frac{1}{2}\left[\nabla^\rho\left(\nabla_\mu \delta g_{\nu\rho} 
+ \nabla_\nu \delta g_{\mu\rho}\right) - \nabla^2 \delta g_{\mu\nu} 
 - \nabla_\mu \nabla_\nu \left(g^{\rho\lambda}\delta g_{\rho\lambda}\right)\right] \nn
&=& \frac{1}{2}\left[\nabla_\mu\nabla^\rho \delta g_{\nu\rho} 
+ \nabla_\nu \nabla^\rho \delta g_{\mu\rho} - \nabla^2 \delta g_{\mu\nu} 
 - \nabla_\mu \nabla_\nu \left(g^{\rho\lambda}\delta g_{\rho\lambda}\right) 
 - 2R^{\lambda\ \rho}_{\ \nu\ \mu}\delta g_{\lambda\rho} 
+ R^\rho_{\ \mu}\delta g_{\rho\nu} + R^\rho_{\ \nu}\delta g_{\rho\mu} \right]\, ,\nn
\delta R &=& -\delta g_{\mu\nu} R^{\mu\nu} + \nabla^\mu \nabla^\nu \delta g_{\mu\nu} 
 - \nabla^2 \left(g^{\mu\nu}\delta g_{\mu\nu}\right)\, , 
\eea
we find
\be
\label{A1}
0 = \frac{1}{2\kappa^2} \left( \frac{1}{2} g^{\mu\nu} R - R^{\mu\nu} \right) 
+ (\mathcal{H}_2)^{\mu\nu} 
+ (\mathcal{H}_3)^{\mu\nu} + (\mathcal{H}_4)^{\mu\nu} + (\mathcal{H}_5)^{\mu\nu}
+ (\mathcal{H}_6)^{\mu\nu} + (\mathcal{H}_7)^{\mu\nu} + (\mathcal{H}_8)^{\mu\nu} \, .
\ee
Here
\bea
\label{ma23a}
(\mathcal{H}_2)^{\mu\nu} &=& \frac{1}{2}g^{\mu\nu} G_2 (\pi, X) - \frac{\partial G_2 (\pi, X)}{\partial X} 
\partial^\mu \pi \partial^\nu \pi \, , \\
\label{ma23b}
(\mathcal{H}_3)^{\mu\nu} &=& \frac{1}{2}g^{\mu\nu} G_3 (\pi, X) \mathcal{L}_3 
 - \frac{\partial G_3 (\pi, X)}{\partial X} \mathcal{L}_3 
\partial^\mu \pi \partial^\nu \pi \nn
&& - G_3 (\pi, X) \left( \partial^\mu \pi \partial^\nu \pi \Box \pi 
 - \partial^\mu \pi \partial^\rho \pi \nabla^\nu \partial_\rho \pi  
 - \partial^\nu \pi \partial^\rho \pi \nabla^\mu \partial_\rho \pi \right) \nn
&& + \frac{1}{2} \left( -  g^{\mu\nu} \Box \pi +  \partial^\nu \pi \partial^\mu 
+  \partial^\mu \pi \partial^\nu - g^{\mu\nu} \partial^\tau \pi \partial_\tau \right)
\left( G_3 (\pi, X) \partial_\rho \pi \partial^\rho \pi \right) 
 - \frac{1}{2} 
\nabla_\rho \left( G_3 (\pi, X) \partial^\rho \pi \partial^\mu \pi \partial^\nu \pi \right) \, , \\
\label{ma23c}
(\mathcal{H}_4)^{\mu\nu} &=& \frac{1}{2}g^{\mu\nu} \left\{ 
\frac{\partial G_4 (\pi, X)}{\partial X} \mathcal{L}^{(0)}_4 
 - \frac{1}{8} G_4 (\pi, X) \mathcal{L}_6 \right\} 
 - \left\{\frac{\partial^2 G_4 (\pi, X)}{\partial X^2} \mathcal{L}^{(0)}_4 
 - \frac{1}{8} \frac{\partial G_4 (\pi, X)}{\partial X} \mathcal{L}_6 \right\}
\partial^\mu \pi \partial^\nu \pi \nn
&& + \frac{1}{2}\frac{\partial G_4 (\pi, X)}{\partial X} 
\left\{ - \partial^\mu \pi \partial^\nu \pi \left( \left( \Box \pi \right)^2 
- \nabla^\sigma \partial_\rho \pi \nabla^\rho \partial_\sigma \pi \right) 
+ 2 \partial_\rho \pi \partial^\rho \pi \nabla^\mu \partial_\sigma \pi \nabla^\nu \partial^\sigma \pi 
\right. \nn
&& + 2 \partial^\mu \pi \partial^\rho \pi \nabla^\nu \partial_\rho \pi \Box \pi 
+ 2 \partial^\nu \pi \partial^\rho \pi \nabla^\mu \partial_\rho \pi \Box \pi 
 - 2 \partial^\mu \pi \partial^\rho \pi \nabla^\sigma \partial^\nu \pi \nabla_\rho \partial_\sigma \pi 
 - 2 \partial^\nu \pi \partial^\rho \pi \nabla^\sigma \partial^\mu \pi \nabla_\rho \partial_\sigma \pi \nn
&& \left. - 2 \partial^\rho \pi \partial^\sigma \pi \nabla^\mu \partial_\rho \pi \nabla^\nu \partial_\sigma \pi 
\right\} \nn
&& + \frac{1}{2} \left( - g^{\mu\nu} \Box \pi + \partial^\nu \pi \partial^\mu 
+ \partial^\mu \pi \partial^\nu - g^{\mu\nu} \partial^\tau \pi \partial_\tau \right)
\left[ \frac{1}{2}\frac{\partial G_4 (\pi, X)}{\partial X} \left\{ 2 \partial_\rho \pi \partial^\rho \pi \Box \pi 
 - 2 \partial^\rho \pi \partial^\sigma \pi \nabla_\rho \partial_\sigma \pi 
\right\} \right] \nn
&& - \nabla_\lambda \left[ \frac{1}{2} \frac{\partial G_4 (\pi, X)}{\partial X} \left\{
\partial_\rho \pi \partial^\rho \pi \left( \nabla^\lambda \partial^\mu \pi \partial^\nu \pi 
+ \nabla^\lambda \partial^\nu \pi \partial^\mu \pi
 - \partial^\lambda \pi \nabla^\mu \partial^\nu \pi \right) \right. \right. \nn
&& \left. \left. 
+ \partial^\lambda \pi \partial^\mu \pi \partial^\nu \pi \Box \pi 
 - 2 \nabla_\tau \partial^\lambda \pi \partial^\tau \pi \partial^\mu \pi \partial^\nu \pi
\right\} \right] \nn
&& + \frac{1}{2} G_4 (\pi, X) \left\{ \frac{1}{2} \left( R^{\mu\rho} \partial^\nu \pi \partial_\rho \pi 
+ R^{\nu\rho} \partial^\mu \pi \partial_\rho \pi \right) - \frac{1}{2} R \partial^\mu \pi \partial^\nu \pi 
\right\} \nn
&& + \frac{1}{2} \nabla_\rho \nabla^\nu \left( G_4 (\pi, X)  \partial^\rho \pi \partial^\mu \pi \right) 
 - \frac{1}{4} \Box \left( G_4 (\pi, X) \partial^\mu \pi \partial^\nu \pi \right) 
 - \frac{1}{4} \nabla_\rho \nabla_\sigma \left( G_4 (\pi, X) g^{\mu\nu} \partial^\rho \pi \partial^\sigma \pi \right) \nn
&& + \frac{1}{4} G_4 (\pi, X) R^{\mu\nu} \partial_\rho \pi \partial^\rho \pi 
 - \frac{1}{4} \nabla^\mu \nabla^\nu \left(  G_4 (\pi, X) \partial_\rho \pi \partial^\rho \pi \right) 
+ \frac{1}{4} \Box \left( G_4 (\pi, X) g^{\mu\nu}  \partial_\rho \pi \partial^\rho \pi \right) \nn
&& - \frac{1}{2}\partial_\rho \pi \partial_\sigma R^{\mu\sigma\nu\rho} \, , \\
\label{ma23d}
(\mathcal{H}_5)^{\mu\nu} &=& \frac{1}{2}g^{\mu\nu} \left\{ 
\frac{\partial G_5 (\pi, X)}{\partial X} \mathcal{L}^{(0)}_5 - \frac{1}{8} G_5 (\pi, X) \mathcal{L}_7 \right\} 
 - \left\{\frac{\partial^2 G_5 (\pi, X)}{\partial X^2} \mathcal{L}^{(0)}_5 
 - \frac{1}{8} \frac{\partial G_5 (\pi, X)}{\partial X} \mathcal{L}_7 \right\}
\partial^\mu \pi \partial^\nu \pi \nn
&& + \frac{1}{6} \frac{\partial G_5 (\pi , X)}{\partial X} 
\left\{ - \partial^\mu \pi \partial^\nu \pi (\Box \pi)^3
+ 3 \partial^\mu \pi \partial^\nu \pi \Box \pi \nabla^\sigma \partial_\rho \pi \nabla^\rho \partial_\sigma \pi
+ 6 \partial_\rho \pi \partial^\rho \pi \Box \pi \nabla^\mu \partial_\sigma \pi \nabla^\sigma \partial^\nu \pi 
\right. \nn
&& - 2 \partial^\mu \pi \partial^\nu \pi \nabla^\rho \partial_\tau \pi \nabla^\sigma \partial_\rho \pi 
\nabla^\tau \partial_\sigma \pi 
 - 6 \partial_\rho \pi \partial^\rho \pi \nabla^\sigma \partial^\nu \pi \nabla^\tau \partial_\sigma \pi \nabla^\mu
\partial_\tau \pi
+ 3 \partial^\mu \pi \partial^\rho \pi \nabla^\nu \partial_\rho \pi (\Box \pi)^2 \nn
&& + 3 \partial^\nu \pi \partial^\rho \pi \nabla^\mu \partial_\rho \pi (\Box \pi)^2
 - 3 \partial^\mu \pi \partial^\rho \pi \nabla^\nu \partial_\rho \pi \nabla^\sigma \partial_\tau \pi 
\nabla^\tau \partial_\sigma \pi
 - 3 \partial^\nu \pi \partial^\rho \pi \nabla^\mu \partial_\rho \pi \nabla^\sigma \partial_\tau \pi 
\nabla^\tau \partial_\sigma \pi \nn
&& - 6 \partial^\rho \pi \partial^\sigma \pi \nabla_\rho \partial_\sigma \pi \nabla^\mu \partial_\tau \pi 
\nabla^\tau \partial^\nu \pi 
 - 6 \partial^\mu \pi \partial^\rho \pi \nabla^\sigma \partial^\nu \pi \nabla_\rho \partial_\sigma \pi \Box \pi
 - 6 \partial^\nu \pi \partial^\rho \pi \nabla^\sigma \partial^\mu \pi \nabla_\rho \partial_\sigma \pi \Box \pi \nn
&& - 6 \partial ^\rho \pi \partial ^\sigma \pi \nabla ^\mu \partial_\rho \pi \nabla_\sigma \partial^\nu \pi \Box \pi
+ 6 \partial ^\mu \pi \partial ^\rho \pi \nabla^\sigma \partial^\nu \pi \nabla_\rho \partial_\tau \pi \nabla^\tau 
\partial_\sigma \pi
+ 6 \partial ^\nu \pi \partial ^\rho \pi \nabla^\sigma \partial^\mu \pi \nabla_\rho \partial_\tau \pi \nabla^\tau 
\partial_\sigma \pi \nn
&& \left. + 6 \partial^\rho \pi \partial^\sigma \pi \nabla^\mu \partial_\rho \pi \nabla_\sigma 
\partial_\tau \pi \nabla^\tau 
\partial^\nu \pi
+ 6 \partial^\rho \pi \partial^\sigma \pi \nabla^\nu \partial_\rho \pi \nabla_\sigma \partial_\tau \pi \nabla^\tau 
\partial^\mu \pi \right\} \nn
&& + \frac{1}{2} \left( - g^{\mu\nu} \Box \pi + \partial^\nu \pi \partial^\mu 
+ \partial^\mu \pi \partial^\nu - g^{\mu\nu} \partial^\tau \pi \partial_\tau \right)
\left\{ \frac{1}{6} \frac{\partial G_5 (\pi , X)}{\partial X}
\left( 3 \partial _\rho \pi \partial ^\rho \pi (\Box \pi)^2 \right. \right. \nn
&& \left. \left. - 3 \partial_\rho \pi \partial^\rho \pi 
\nabla^\sigma \partial_\tau \pi \nabla^\tau \partial_\sigma \pi
 - 6 \partial^\rho \pi \partial^\sigma \pi \nabla_\rho \partial_\sigma \pi \Box \pi
+ 6 \partial^\rho \pi \partial^\sigma \pi \nabla^\tau \partial_\rho \pi 
\nabla_\sigma \partial_\tau \pi \right) \right\} \nn
&& + \nabla_\lambda \left\{ \frac{1}{6} \frac{\partial G_5 (\pi , X)}{\partial X} 
\left( -3 \partial_\rho \pi \partial^\rho \pi \Box \pi \partial^\nu \pi \nabla^\mu \partial^\lambda \pi
 -3 \partial_\rho \pi \partial^\rho \pi \Box \pi \partial^\mu \pi \nabla^\nu \partial^\lambda \pi
+ 3 \partial_\rho \pi \partial ^\rho \pi \Box \pi \partial^\lambda \pi \nabla^\nu \partial^\mu \pi 
\right. \right. \nn
&& + 3 \partial_\rho \pi \partial ^\rho \pi \nabla^\sigma \partial^\lambda \pi 
\nabla^\mu \partial_\sigma \pi \partial^\nu \pi
 + 3 \partial_\rho \pi \partial ^\rho \pi \nabla^\sigma \partial^\lambda \pi 
\nabla^\nu \partial_\sigma \pi \partial^\mu \pi
 - 3 \partial_\rho \pi \partial^\rho \pi \nabla^\sigma \partial ^\mu \pi 
\nabla^\nu \partial_\sigma \pi \partial^\lambda \pi \nn
&& - \frac{3}{2} \partial^\lambda \pi \partial^\mu \pi \partial^\nu \pi (\Box \pi)^2 
+ \frac{3}{2} \partial^\lambda \pi \partial^\mu \pi \partial^\nu \pi 
\nabla^\rho \partial_\sigma \pi \nabla^\sigma \partial_\rho \pi
+ 3 \partial^\rho \pi \partial^\sigma \pi \nabla_\rho \partial_\sigma \pi 
\nabla^\mu \partial^\lambda \pi \partial ^\nu \pi \nn
&& + 3 \partial^\rho \pi \partial^\sigma \pi \nabla_\rho 
\partial_\sigma \pi \nabla^\nu \partial^\lambda \pi \partial ^\mu \pi 
 - 3 \partial^\rho \pi \partial^\sigma \pi \nabla_\rho \partial_\sigma \pi 
\nabla^\nu \partial^\mu \pi \partial^\lambda \pi 
 - 6 \partial^\rho \pi \partial^\mu \pi \nabla^\sigma \partial_\rho \pi 
\partial^\nu \pi \nabla^\lambda \partial_\sigma \pi \nn
&& + 6 \partial^\mu \pi \partial^\rho \pi \partial^\nu \pi \nabla_\rho \partial^\lambda \pi \Box \pi 
 - 3 \partial^\rho \pi \partial^\sigma \pi \nabla^\mu \partial_\rho \pi 
\nabla_\sigma \partial^\lambda \pi \partial^\nu \pi 
 - 3 \partial^\rho \pi \partial^\sigma \pi \nabla^\nu \partial_\rho \pi 
\nabla_\sigma \partial^\lambda \pi \partial^\mu \pi \nn
&& \left. \left. + 3 \partial^\rho \pi \partial^\sigma \pi \nabla^\nu \partial_\rho \pi 
\nabla_\sigma \partial^\mu \pi \partial^\lambda \pi
\right) \right\} \nn
&& - \frac{1}{8} G_5 (\pi, X) \mathcal{L}_7 \left( - 2 \partial^\mu \pi \partial^\nu \pi \Box \pi R 
+ 4 \partial^\mu \pi \partial^\nu \pi \nabla^\rho \partial^\sigma \pi R_{\rho \sigma} 
+ 4 \partial_\rho \pi \partial^\rho \pi \nabla^\mu \partial^\sigma \pi R^\nu _{\ \sigma} \right. \nn
&& + 4 \partial_\rho \pi \partial^\rho \pi \nabla^\nu \partial^\sigma \pi R^\mu_{\ \sigma}
+ 2 \partial^\mu \pi \partial^\rho \pi \nabla^\nu \partial_\rho \pi R
+ 2 \partial^\nu \pi \partial^\rho \pi \nabla^\mu \partial_\rho \pi R
 - 4 \partial^\mu \pi \partial_\rho \pi \nabla^\nu \partial_\sigma \pi R^{\rho \sigma} \nn
&& - 4 \partial^\nu \pi \partial_\rho \pi \nabla^\mu \partial_\sigma \pi R^{\rho \sigma} 
 - 4 \partial^\rho \pi \partial^\mu \pi \nabla_\rho \partial_\sigma \pi R^{\nu \sigma}
 - 4 \partial^\rho \pi \partial^\nu \pi \nabla_\rho \partial_\sigma \pi R^{\mu \sigma}
 - 4 \partial^\rho \pi \partial_\sigma \pi \nabla_\rho \partial^\mu \pi R^{\sigma \nu} \nn
&& - 4 \partial^\rho \pi \partial_\sigma \pi \nabla_\rho \partial^\nu \pi R^{\sigma \mu} 
+ 4 \partial^\mu \pi \partial^\rho \pi \Box \pi R^\nu _{\ \rho}
+ 4 \partial^\nu \pi \partial^\rho \pi \Box \pi R^\mu _{\ \rho} 
 - 3 \partial^\mu \pi \partial^\rho \pi \nabla_\sigma \partial_\tau R_\rho ^{\ \tau \nu \sigma}
 - 3 \partial^\nu \pi \partial^\rho \pi \nabla_\sigma \partial_\tau R_\rho ^{\ \tau \mu \sigma} \nn
&& - 3 \partial^\rho \pi \partial^\tau \pi \nabla_\sigma \partial^\mu \pi R_{\tau \ \rho}^{\ \nu \ \sigma}
 - 3 \partial^\rho \pi \partial^\tau \pi \nabla_\sigma \partial^\nu \pi R_{\tau \ \rho}^{\ \mu \ \sigma}
 - 2 \partial_\rho \pi \partial^\rho \pi \Box \pi R^{\mu \nu}
+ 2 \partial^\rho \pi \partial^\sigma \pi \nabla_\rho \partial_\sigma \pi R^{\mu \nu}  \Big ) \nn
&& + \frac{1}{2} \left( - g^{\mu\nu} \Box \pi + \partial^\nu \pi \partial^\mu 
+ \partial^\mu \pi \partial^\nu - g^{\mu\nu} \partial^\tau \pi \partial_\tau \right)
\left\{ - \frac{1}{8} G_5 (\pi, X) \mathcal{L}_7 
(2 \partial_\rho \pi \partial^\rho \pi R -4 \partial^\rho \pi \partial^\sigma \pi R_{\rho \sigma}) \right\} \nn
&& + \nabla_\lambda \bigg \{ - \frac{1}{8} G_5 (\pi, X) \mathcal{L}_7 
\left( -2 \partial_\rho \pi \partial^\rho \pi \partial^\nu \pi R^{\lambda \mu}
 - 2 \partial_\rho \pi \partial^\rho \pi \partial^\mu \pi R^{\lambda \nu}
+ 2 \partial_\rho \pi \partial^\rho \pi \partial^\lambda \pi R^{\mu \nu} \right. \nn
&& \left. \left. - \partial^\lambda \pi \partial^\mu \pi \partial^\nu \pi R 
+ 4 \partial^\mu \pi \partial_\rho \pi \partial^\nu \pi R^{\rho \lambda} 
+ 2 \partial^\rho \pi \partial^\sigma \pi \partial^\nu \pi R_{\sigma \ \rho}^{\ \mu \ \lambda} 
+ 2 \partial^\rho \pi \partial^\sigma \pi \partial^\mu \pi R_{\sigma \ \rho}^{\ \nu \ \lambda} 
- 2 \partial^\rho \pi \partial^\sigma \pi \partial^\lambda \pi R_{\sigma \ \rho}^{\ \nu \ \mu} \right)
\right\} \nn
&& + \left( \nabla^\nu \nabla^\mu - g^{\mu \nu} \nabla^2 \right)
\left\{ - \frac{1}{8} G_5 (\pi, X) \mathcal{L}_7 
(2 \partial_\rho \pi \partial^\rho \pi \Box \pi -2 \partial^\rho \pi \partial^\sigma \pi 
\nabla_\rho \partial_\sigma \pi )\right\} \nn
&& + \frac{1}{2} \nabla_\sigma \nabla^\nu \bigg \{ - \frac{1}{8} G_5 (\pi, X) \mathcal{L}_7 
\left( - 8 \partial_\rho \pi \partial^\rho \pi \nabla^\sigma \partial^\mu \pi 
+ 8 \partial^\rho \pi \partial^\sigma \pi \nabla_\rho \partial^\mu \pi 
+ 8 \partial^\rho \pi \partial^\mu \pi \nabla_\rho \partial^\sigma \pi \right. \nn
&& \left. \left. - 8 \partial^\sigma \pi \partial^\mu \pi \Box \pi
\right) \right\} 
+ \frac{1}{2} \nabla_\sigma \nabla^\mu \left\{ - \frac{1}{8} G_5 (\pi, X) \mathcal{L}_7 
\left( - 8 \partial_\rho \pi \partial^\rho \pi \nabla^\sigma \partial^\nu \pi 
+ 8 \partial^\rho \pi \partial^\sigma \pi \nabla_\rho \partial^\nu \pi \right. \right. \nn
&& \left. \left. + 8 \partial^\rho \pi \partial^\nu \pi \nabla_\rho \partial^\sigma \pi
 - 8 \partial^\sigma \pi \partial^\nu \pi \Box \pi
\right) \right\} 
 - \frac{1}{2} \nabla^2 \left\{ - \frac{1}{8} G_5 (\pi, X) \mathcal{L}_7 
\left( - 4 \partial_\rho \pi \partial^\rho \pi \nabla^\mu \partial^\nu \pi \right. \right. \nn
&& \left. \left. + 4 \partial^\rho \pi \partial^\mu \pi \nabla_\rho \partial^\nu \pi
+ 4 \partial^\rho \pi \partial^\nu \pi \nabla_\rho \partial^\mu \pi
 - 4 \partial^\mu \pi \partial^\nu \pi \Box \pi
\right) \right\} - \frac{1}{2} g^{\mu \nu} \nabla_\tau \nabla_\sigma
\left\{ - \frac{1}{8} G_5 (\pi, X) \mathcal{L}_7 \right. \nn
&& \left. \left( - 4 \partial_\rho \pi \partial^\rho \pi \nabla^\sigma \partial^\tau \pi
+ 8 \partial^\rho \pi \partial^\sigma \pi \nabla_\rho \partial^\tau \pi
 - 4 \partial^\sigma \pi \partial^\tau \pi \Box \pi
\right) \right\} \nn
&& + 2 \nabla_\rho \nabla_\sigma \bigg \{ - \frac{1}{8} G_5 (\pi, X) \mathcal{L}_7 
\frac{1}{2} \big ( \partial^\sigma \pi \partial^\nu \pi \nabla^\mu \partial^\rho \pi 
+ \partial^\sigma \pi \partial^\mu \pi \nabla^\nu \partial^\rho \pi \nn
&&- 2 \partial^\sigma \pi \partial^\rho \pi \nabla^\mu \partial^\nu \pi 
- 2 \partial^\mu \pi \partial^\nu \pi \nabla^\sigma \partial^\rho \pi 
+ \partial^\mu \pi \partial^\rho \pi \nabla^\sigma \partial^\nu \pi
+ \partial^\nu \pi \partial^\rho \pi \nabla^\sigma \partial^\mu \pi \big ) \bigg \} \, \\
\label{ma23e}
(\mathcal{H}_6)^{\mu\nu} &=& \frac{1}{2}g^{\mu\nu} G_6 (\pi) \nn
&& - 4 G_6 (\pi) \left\{ - \left( R^{\mu\rho} \partial^\nu \pi \partial_\rho \pi 
+ R^{\nu\rho} \partial^\mu \pi \partial_\rho \pi \right) + \frac{1}{2} R \partial^\mu \pi \partial^\nu \pi 
\right\} \nn
&& - 4 \nabla_\rho \nabla^\nu \left( G_6 (\pi)  \partial^\rho \pi \partial^\mu \pi \right) 
 + 2 \Box \left( G_6 (\pi) \partial^\mu \pi \partial^\nu \pi \right) 
 + 2 \nabla_\rho \nabla_\sigma \left( G_6 (\pi) g^{\mu\nu} \partial^\rho \pi \partial^\sigma \pi \right) \nn
&& - 2 G_6 (\pi) R^{\mu\nu} \partial_\rho \pi \partial^\rho \pi 
 + 2 \nabla^\mu \nabla^\nu \left(  G_6 (\pi) \partial_\rho \pi \partial^\rho \pi \right) 
 - 2 \Box \left( G_6 (\pi) g^{\mu\nu}  \partial_\rho \pi \partial^\rho \pi \right) \, , \\
\label{ma23f}
(\mathcal{H}_7)^{\mu\nu} &=& \frac{1}{2}g^{\mu\nu} G_7 (\pi) \nn
&& + G_7 (\pi) \mathcal{L}_7 
\left( - 2 \partial^\mu \pi \partial^\nu \pi \Box \pi R 
+ 4 \partial^\mu \pi \partial^\nu \pi \nabla^\rho \partial^\sigma \pi R_{\rho \sigma} 
+ 4 \partial_\rho \pi \partial^\rho \pi \nabla^\mu \partial^\sigma \pi R^\nu _{\ \sigma} \right. \nn
&& + 4 \partial_\rho \pi \partial^\rho \pi \nabla^\nu \partial^\sigma \pi R^\mu_{\ \sigma}
+ 2 \partial^\mu \pi \partial^\rho \pi \nabla^\nu \partial_\rho \pi R
+ 2 \partial^\nu \pi \partial^\rho \pi \nabla^\mu \partial_\rho \pi R
 - 4 \partial^\mu \pi \partial_\rho \pi \nabla^\nu \partial_\sigma \pi R^{\rho \sigma} \nn
&& - 4 \partial^\nu \pi \partial_\rho \pi \nabla^\mu \partial_\sigma \pi R^{\rho \sigma} 
 - 4 \partial^\rho \pi \partial^\mu \pi \nabla_\rho \partial_\sigma \pi R^{\nu \sigma}
 - 4 \partial^\rho \pi \partial^\nu \pi \nabla_\rho \partial_\sigma \pi R^{\mu \sigma}
 - 4 \partial^\rho \pi \partial_\sigma \pi \nabla_\rho \partial^\mu \pi R^{\sigma \nu} \nn
&& - 4 \partial^\rho \pi \partial_\sigma \pi \nabla_\rho \partial^\nu \pi R^{\sigma \mu} 
+ 4 \partial^\mu \pi \partial^\rho \pi \Box \pi R^\nu _{\ \rho}
+ 4 \partial^\nu \pi \partial^\rho \pi \Box \pi R^\mu _{\ \rho} 
 - 3 \partial^\mu \pi \partial^\rho \pi \nabla_\sigma \partial_\tau R_\rho ^{\ \tau \nu \sigma}
 - 3 \partial^\nu \pi \partial^\rho \pi \nabla_\sigma \partial_\tau R_\rho ^{\ \tau \mu \sigma} \nn
&& - 3 \partial^\rho \pi \partial^\tau \pi \nabla_\sigma \partial^\mu \pi R_{\tau \ \rho}^{\ \nu \ \sigma}
 - 3 \partial^\rho \pi \partial^\tau \pi \nabla_\sigma \partial^\nu \pi R_{\tau \ \rho}^{\ \mu \ \sigma}
 - 2 \partial_\rho \pi \partial^\rho \pi \Box \pi R^{\mu \nu}
+ 2 \partial^\rho \pi \partial^\sigma \pi \nabla_\rho \partial_\sigma \pi R^{\mu \nu}  \Big ) \nn
&& + \frac{1}{2} \left( - g^{\mu\nu} \Box \pi + \partial^\nu \pi \partial^\mu 
+ \partial^\mu \pi \partial^\nu - g^{\mu\nu} \partial^\tau \pi \partial_\tau \right)
\left\{ G_7 (\pi) \mathcal{L}_7 
(2 \partial_\rho \pi \partial^\rho \pi R -4 \partial^\rho \pi \partial^\sigma \pi R_{\rho \sigma}) \right\} \nn
&& + \nabla_\lambda \bigg \{ G_7 (\pi) \mathcal{L}_7 
\left( -2 \partial_\rho \pi \partial^\rho \pi \partial^\nu \pi R^{\lambda \mu}
 - 2 \partial_\rho \pi \partial^\rho \pi \partial^\mu \pi R^{\lambda \nu}
+ 2 \partial_\rho \pi \partial^\rho \pi \partial^\lambda \pi R^{\mu \nu} \right. \nn
&& \left. \left. - \partial^\lambda \pi \partial^\mu \pi \partial^\nu \pi R 
+ 4 \partial^\mu \pi \partial_\rho \pi \partial^\nu \pi R^{\rho \lambda} 
+ 2 \partial^\rho \pi \partial^\sigma \pi \partial^\nu \pi R_{\sigma \ \rho}^{\ \mu \ \lambda} 
+ 2 \partial^\rho \pi \partial^\sigma \pi \partial^\mu \pi R_{\sigma \ \rho}^{\ \nu \ \lambda} 
- 2 \partial^\rho \pi \partial^\sigma \pi \partial^\lambda \pi R_{\sigma \ \rho}^{\ \nu \ \mu} \right)
\right\} \nn
&& + \left( \nabla^\nu \nabla^\mu - g^{\mu \nu} \nabla^2 \right)
\left\{ G_7 (\pi) \mathcal{L}_7 
(2 \partial_\rho \pi \partial^\rho \pi \Box \pi -2 \partial^\rho \pi \partial^\sigma \pi 
\nabla_\rho \partial_\sigma \pi )\right\} \nn
&& + \frac{1}{2} \nabla_\sigma \nabla^\nu \bigg \{ G_7 (\pi) \mathcal{L}_7 
\left( - 8 \partial_\rho \pi \partial^\rho \pi \nabla^\sigma \partial^\mu \pi 
+ 8 \partial^\rho \pi \partial^\sigma \pi \nabla_\rho \partial^\mu \pi 
+ 8 \partial^\rho \pi \partial^\mu \pi \nabla_\rho \partial^\sigma \pi \right. \nn
&& \left. \left. - 8 \partial^\sigma \pi \partial^\mu \pi \Box \pi
\right) \right\} 
+ \frac{1}{2} \nabla_\sigma \nabla^\mu \left\{ G_7 (\pi) \mathcal{L}_7 
\left( - 8 \partial_\rho \pi \partial^\rho \pi \nabla^\sigma \partial^\nu \pi 
+ 8 \partial^\rho \pi \partial^\sigma \pi \nabla_\rho \partial^\nu \pi \right. \right. \nn
&& \left. \left. + 8 \partial^\rho \pi \partial^\nu \pi \nabla_\rho \partial^\sigma \pi
 - 8 \partial^\sigma \pi \partial^\nu \pi \Box \pi
\right) \right\} 
 - \frac{1}{2} \nabla^2 \left\{ G_7 (\pi) \mathcal{L}_7 
\left( - 4 \partial_\rho \pi \partial^\rho \pi \nabla^\mu \partial^\nu \pi \right. \right. \nn
&& \left. \left. + 4 \partial^\rho \pi \partial^\mu \pi \nabla_\rho \partial^\nu \pi
+ 4 \partial^\rho \pi \partial^\nu \pi \nabla_\rho \partial^\mu \pi
 - 4 \partial^\mu \pi \partial^\nu \pi \Box \pi
\right) \right\} - \frac{1}{2} g^{\mu \nu} \nabla_\tau \nabla_\sigma
\left\{ G_7 (\pi) \mathcal{L}_7 \right. \nn
&& \left. \left( - 4 \partial_\rho \pi \partial^\rho \pi \nabla^\sigma \partial^\tau \pi
+ 8 \partial^\rho \pi \partial^\sigma \pi \nabla_\rho \partial^\tau \pi
 - 4 \partial^\sigma \pi \partial^\tau \pi \Box \pi
\right) \right\} \nn
&&+ 2 \nabla_\rho \nabla_\sigma \bigg \{ 
G_7 (\pi) \mathcal{L}_7 \frac{1}{2} \big ( \partial^\sigma \pi \partial^\nu \pi \nabla^\mu \partial^\rho \pi 
+ \partial^\sigma \pi \partial^\mu \pi \nabla^\nu \partial^\rho \pi \nn
&&- 2 \partial^\sigma \pi \partial^\rho \pi \nabla^\mu \partial^\nu \pi 
- 2 \partial^\mu \pi \partial^\nu \pi \nabla^\sigma \partial^\rho \pi 
+ \partial^\mu \pi \partial^\rho \pi \nabla^\sigma \partial^\nu \pi
+ \partial^\nu \pi \partial^\rho \pi \nabla^\sigma \partial^\mu \pi \big ) \bigg \} \, , \\
\label{ma23g}
(\mathcal{H}_8)^{\mu\nu} &=& 8 \left( \nabla^\mu \nabla^\nu G_8(\pi) \right)R 
 - 8 g^{\mu\nu} \left( \nabla^2 G_8(\pi) \right) R 
 - 16 \left( \nabla_\rho \nabla^\mu G_8(\pi) \right)R^{\nu\rho} 
 - 16 \left( \nabla_\rho \nabla^\nu G_8(\pi) \right)R^{\mu\rho} \nn
&& + 16 \left( \nabla^2 G_8(\pi) \right)R^{\mu\nu}
+ 16 g^{\mu\nu} \left( \nabla_{\rho} \nabla_\sigma G_8(\pi) \right) R^{\rho\sigma}
- 16  \left(\nabla_\rho \nabla_\sigma G_8(\pi) \right) R^{\mu\rho\nu\sigma} \, .
\eea
Thus, 
we have given the explicit expressions of the field equations and 
the equations corresponding to the Einstein equation for the generalized Galileon scalar models. 

\section{FRW dynamics \label{III}}

In this section, we give formulae for the cosmological reconstruction by considering 
the FRW dynamics and we investigate the stability 
of the reconstructed solution. 
Usually, we start from a theory, which is defined by the action, and solve
equations of motion to define the background dynamics.
The reconstruction is the inverse problem, i.e., when 
an arbitrary development of the expansion of the universe is given, we construct 
the explicit form of the action which reproduces the development. 
For the reconstructed action, the solution describing the development of the expansion 
history is not always stable. In this section, we show the conditions so that the solution can become stable or attractor. 

We now take 
the FRW universe with the flat spatial part, 
\be
\label{Gl3}
ds^2 = - dt^2 + a(t)^2 \sum_{i=1,2,3} \left(dx^i\right)^2\, ,
\ee
and we assume $\pi$ depends on only time variable. 

Let us denote the energy density and pressure of the matters by $\rho$ and $p$. Then the first FRW equation 
is given by 
\be
\label{1FRW}
 - \frac{3}{2\kappa^2}H^2 + \sum_{k=2}^8 (\mathcal{H}_k)_{00} = - \frac{\rho}{2} \, .
\ee
We also write $(\mathcal{H}_k)_{ij}$ as $(\mathcal{H}_k)_{ij} = \mathcal{H}_k a^2 \delta_{ij}$. 
Accordingly, 
the second FRW equation is given by
\be
\label{2FRW}
\frac{1}{2\kappa^2}\left( 3 H^2 + 2 \dot H \right) + \sum_{k=2}^8 \mathcal{H}_k = - \frac{p}{2} \, .
\ee
If $\rho$ and $p$ are given by the sums of the contribution of the matters with a constant EoS parameter $w_l$, we find
\be
\label{rhop}
\rho = \sum_l \rho_l a^{-3(1+w_l)}\, , \quad p = \sum_l w_l \rho_l a^{-3(1+w_l)}\, .
\ee
Here $\rho_l$'s are constants.  

We now demonstrate 
the reconstruction of the expansion history of the universe. That is, for arbitrary development of the Hubble rate $H(t)$ or scale factor $a(t)$, we determine the functions $G_2 (\pi, X)$, $G_3 (\pi, X)$, $G_4 (\pi, X)$, 
$G_5 (\pi, X)$, $G_6 (\pi)$, $G_7 (\pi)$, and $G_8 (\pi)$, which gives the development. 
Since the redefinition of the scalar field $\pi$ can be absorbed into the redefinition of $G_i$ $\left(i=2,3, \cdots,8\right)$, we may choose
\be
\label{Grec1}
\pi = t\, .
\ee
Then we find
\be
\label{Grec2} 
X = -1\, ,\quad \dot G_i (\pi, X) = \left. 
\frac{\partial G_i (\pi, X)}{\partial \pi} \right|_{\pi=t,\,X=-1}\, ,
\quad \mbox{etc.}
\ee
We also write $G_i (\pi, X)$ $\left(i=2,3,4,5\right)$ as follows,
\be
\label{Grec3}
G_i (\pi, X) = \sum_{n=0}^\infty G_i^{(n)} (\pi)\left(1+X\right)^n\, .
\ee
In addition, 
we suppose 
the scale factor $a(t)$ can be given by an appropriate function $g(t)$ as 
$a(t)=\e^{g(t)}$, which gives $H(t)=\dot g (t)$. 
Thus 
the first FRW equation (\ref{1FRW}) has the following form: 
\bea
\label{Grec4}
&&  - \frac{3}{2\kappa^2}H^2 - \frac{1}{2} G_2^{(0)}(t) - G_2^{(1)}(t) 
+ \left\{ 3 G_3^{(1)}(t) - \frac{9}{2} G_3^{(0)} (t) \right\} \dot g (t) 
+ \left\{ -6 G_4^{(1)} (t) - 6 G_4^{(2)} (t) + \frac{9}{4} G_4^{(0)} (t) \right\} {\dot g (t)}^2 \nn
&& + \left\{ -5 G_5^{(1)} (t) + 2 G_5^{(2)} (t) + \frac{15}{4} G_5^{(0)} (t) \right\} {\dot g (t)}^3 
 - 18 G_6 (t) {\dot g (t)}^2 - 30 G_7 (t) {\dot g (t)}^3 
+ 48 \dot G_8 (t) {\dot g (t)}^3 \nn 
&=& - \frac{1}{2} \sum_l \rho_l \e^{-3(1+w_l) g(t) }\, .
\eea
On the other hand, the second FRW equation (\ref{2FRW}) has the following 
form: 
\bea
\label{Grec5}
&& \frac{1}{2\kappa^2}\left( 3 H^2 + 2 \dot H \right) + \frac{1}{2} G_2^{(0)}(t) + \frac{1}{2} \dot G_3^{(0)}(t) \nn
&& - \frac{3}{2} G_4^{(1)}(t) {\dot g (t)}^2 - \dot G_4^{(1)}(t) \dot g (t) - G_4^{(1)}(t) \ddot g (t)
+ \frac{3}{4} G_4^{(0)}(t) {\dot g (t)}^2 + \frac{1}{2} \dot G_4^{(0)}(t) \dot g (t) 
+ \frac{1}{2} G_4^{(0)}(t) \ddot g (t) \nn
&& + G_5^{(1)}(t) {\dot g (t)}^3 + \frac{1}{2} \dot G_5^{(1)}(t) {\dot g (t)}^2 
+ G_5^{(1)}(t) \dot g (t) \ddot g (t) 
 - \frac{3}{2} G_5^{(0)}(t) {\dot g (t)}^3 - \frac{3}{4} \dot G_5^{(0)}(t) {\dot g (t)}^2 
 - \frac{3}{2} G_5^{(0)}(t) \dot g (t) \ddot g (t) \nn
&& - 6 G_6 (t) {\dot g (t)}^2 - 4 \dot G_6 (t) \dot g (t) - 4 G_6 (t) \ddot g (t) 
+ 12 G_7 (t) {\dot g (t)}^3 + 6 \dot G_7 (t) {\dot g (t)}^2 + 12 G_7 (t) \dot g (t) \ddot g (t) \nn
&& - 32 \dot G_8 (t) {\dot g (t)}^3 - 16 \ddot G_8 (t) {\dot g (t)}^2 
 - 32 \dot G_8 (t) \dot g (t) \ddot g (t) = - \frac{1}{2}\sum_l w_l \rho_l \e^{-3(1+w_l) g(t)}\, .
\eea
Then if we choose $G_2^{(0)}(t)$, $G_2^{(1)}(t)$, $G_3^{(0)}(t)$, $G_3^{(1)}(t)$, $G_4^{(0)}(t)$, 
$G_4^{(1)}(t)$, $G_4^{(2)}(t)$, $G_5^{(0)}(t)$, $G_5^{(1)}(t)$, 
$G_5^{(2)}(t)$, $G_6 (t)$, $G_7 (t)$, and $G_8 (t)$ so that the FRW equations 
(\ref{Grec4}) and (\ref{Grec5}) 
can be satisfied, 
we have the following solution
\be
\label{Grec6}
H (t) = \dot g(t)\, ,\quad \pi = t\, .
\ee
We should note that $G_2^{(i)}(t)$ $\left(i=2,3,\cdots\right)$, 
$G_3^{(i)}(t)$ $\left(i=2,3,\cdots\right)$, 
$G_4^{(i)}(t)$ $\left(i=3,4,\cdots\right)$, and 
$G_5^{(i)}(t)$ $\left(i=3,4,\cdots\right)$ are irrelevant for the expansion of the universe. 
Then we can completely and explicitly separate $G_n$ to the parts relevant for the expansion 
and the irrelevant parts. We should note, however, that some of the irrelevant are related 
with the stability of the reconstructed solution as we see in the following. 

Just for the simplicity, we neglect $G_{4}\left(\pi,X\right)$, $G_{5}\left(\pi,X\right)$, $G_{6}\left(\pi\right)$, 
$G_{7}\left(t\right)$, and $G_{8}\left(t\right)$ in the rest of this section 
and consider the examples of the reconstruction and the (in)stability. 
Then, we acquire
\begin{eqnarray}
\label{FRW23_4}
\frac{3}{\kappa^{2}}H^{2} & = & \frac{1}{\kappa^{2}}g''\left(\pi\right)\left(1+X\right)
+\frac{3}{\kappa^{2}}g'^{2}\left(\pi\right)-\dot{G}_{3}^{\left(0\right)}\left(\pi\right)\nonumber \\
\, & \, & +\left(X-1\right)\left\{ -\frac{1}{2}\dot{G}_{3}^{\left(0\right)}\left(\pi\right)
+3G_{3}^{\left(1\right)}\left(\pi\right)g'\left(\pi\right)
 -\frac{9}{2}G_{3}^{\left(0\right)}\left(\pi\right)g'\left(\pi\right)\right\} \nonumber \\
\, & \, & -\underset{n=2}{\sum}G_{2}^{\left(n\right)}\left(\pi\right)\left(1+X\right)^{n}
 -2\underset{n=2}{\sum}G_{2}^{\left(n\right)}\left(\pi\right)n\left(1+X\right)^{n-1}\dot{\pi}^{2}
+6\frac{\partial G_{3}\left(\pi,X\right)}{\partial X}H\dot{\pi}^{5}\nonumber \\
&& -9G_{3}\left(\pi,X\right)H\dot{\pi}^{3} \, , \\
\label{FRW23_5}
-\frac{1}{\kappa^{2}}\left(3H^{2}+2\dot{H}\right) 
& = & \frac{1}{\kappa^{2}}g''\left(\pi\right)\left(-1+X\right)-\frac{3}{\kappa^{2}}g'^{2}\left(\pi\right)
+\dot{G}_{3}^{\left(0\right)}\left(\pi\right)\nonumber \\
\, & \, & +\left(X+1\right)\left\{ -\frac{1}{2}\dot{G}_{3}^{\left(0\right)}\left(\pi\right)
+3G_{3}^{\left(1\right)}\left(\pi\right)g'\left(\pi\right)
 -\frac{9}{2}G_{3}^{\left(0\right)}\left(\pi\right)g'\left(\pi\right)\right\} \nonumber \\
\, & \, & +\underset{n=2}{\sum}G_{2}^{\left(n\right)}\left(\pi\right)\left(1+X\right)^{n}
 -\dot{G}_{3}\left(\pi,X\right)\dot{\pi}^{3}-3G_{3}\left(\pi,X\right)\dot{\pi}^{2}\ddot{\pi}\, .
\end{eqnarray}
By combining the equation of motion for $\pi$ with (\ref{FRW23_5}), we obtain
\begin{eqnarray}
\label{FRW23_7}
\ddot{\pi}
&=& A^{-1} \left [  
-18 \sum_{n=0}^{\infty} G_3^{(n)}(\pi) n (1+X)^{n-1} H^2 \dot{\pi}^4 -6
\sum_{n=0}^{\infty} G_3'^{(n)}(\pi) n (1+X)^{n-1} H \dot{\pi}^5 \right. \nonumber \\
&& +\left ( -6 \sum_{n=0}^{\infty} G_3^{(n)} n(1+X)^{n-1} \dot{\pi}^4
+9 \sum_{n=0}^{\infty} G_3^{(n)} (1+X)^n \dot{\pi}^2
 \right ) \nonumber \\
&& \times \left \{ - \frac{1}{2} g''(-1+X) + \frac{3}{2} g'^2 - \frac{1}{2} \kappa^2
G_3'^{(0)}(\pi) - \frac{\kappa^2}{2} (X+1) 
\left ( - \frac{1}{2} G_3'^{(0)}(\pi) + 3 G_3^{(1)} g'(\pi) - \frac{9}{2}
G_3^{(0)} g'(\pi) \right ) \right .\nonumber  \\
&& \left . - \frac{\kappa^2}{2} \sum_{n=2}^{\infty} G_2^{(n)} (1+X)^n 
+ \frac{\kappa^2}{2} \sum_{n=0}^{\infty} G_3'^{(n)} (1+X)^n \dot{\pi}^4
- \frac{3}{2} H^2 \right  \}  \nonumber \\
&& + 27  \sum_{n=0}^{\infty} G_3^{(n)}(1+X)^n H^2 \dot{\pi}^2
+ 9  \sum_{n=0}^{\infty} G_3'^{(n)} (1+X)^n H \dot{\pi}^3
 -3  \sum_{n=0}^{\infty} G_3'^{(n)} (1+X)^n H \dot{\pi}^3 \nonumber \\
&& -6 \left \{ \frac{1}{\kappa^2} g''(\pi) - \frac{1}{2} G_3'^{(0)}(\pi) +3 G_3^{(1)}(\pi)
g'(\pi) - \frac{9}{2} G_3^{(0)}(\pi) g'(\pi) +  \sum_{n=2}^{\infty} G_2^{(n)}(\pi) n (1+X)^{n-1}
 \right \} H \dot{\pi} \nonumber \\
&& -2 \left  \{
\frac{1}{\kappa^2} g'''(\pi) - \frac{1}{2} G_3''^{(0)}(\pi) +3 G_3'^{(1)}(\pi) g'(\pi)
+ 3 G_3^{(1)}(\pi) g''(\pi) - \frac{9}{2} G_3'^{(0)}(\pi) g'(\pi)
 - \frac{9}{2} G_3^{(0)}(\pi) g''(\pi) \right . \nonumber \\
&& \left . +  \sum_{n=2}^{\infty} G_2'^{(n)} n(1+X)^n
 \right  \} \nonumber \\
&& - \left \{ - \frac{6}{\kappa^2} g'(\pi) g''(\pi) - \frac{2}{\kappa^2} g'''(\pi)
+ \left ( \frac{1}{\kappa^2} g'''(\pi) - \frac{1}{2} G_3''^{(0)} + 3
G_3^{(1)}(\pi) g'(\pi) + 3 G_3^{(1)}(\pi) g''(\pi) \right. \right. \nonumber \\
&& \left.  \left.\left. - \frac{9}{2} G_3^{(0)}(\pi) g''(\pi) 
 \right ) (1+X) + G_3''^{(0)}+  \sum_{n=2}^{\infty} G_2'^{(n)} (1+X)^n
\right \} \right ] \, ,\\
\label{FRW23_8}
\dot{H} &=&  -\frac{1}{2} g''(\pi) (-1+X) + \frac{3}{2} g'^2(\pi) - \frac{1}{2}
\kappa^2 G_3'^{(0)}(\pi) \nn
&& - \frac{\kappa^2}{2} (X+1) \left (
- \frac{1}{2} G_3'^{(0)}(\pi) + 3 G_3^{(1)}(\pi) g'(\pi) - \frac{9}{2} G_3^{(0)}(\pi) g'(\pi)
 \right ) \nonumber \\
&& - \frac{\kappa^2}{2} \sum_{n=2}^{\infty} G_2^{(n)}(1+X)^n 
+ \frac{\kappa^2}{2}  \sum_{n=0}^{\infty} G_3'^{(n)}(\pi) (1+X)^n \dot{\pi}^4
 - \kappa^2  \sum_{n=0}^{\infty} G_3^{(n)}(\pi) n (1+X)^{n-1} \dot{\pi}^4
\ddot{\pi} \nonumber \\
&& + \frac{3}{2} \kappa^2  \sum_{n=0}^{\infty} G_3^{(n)}(\pi) (1+X)^n
\dot{\pi}^2 \ddot{\pi} - \frac{3}{2} H^2 \, .
\end{eqnarray}
We can eliminate $\ddot{\pi}$ by substituting (\ref{FRW23_7}) into
(\ref{FRW23_8}). 
Here
\begin{eqnarray}
\label{FRW23_9}
A & \equiv & \left (
\kappa^2 \sum_{n=0}^{\infty} G_3^{(n)}(\pi) n (1+X)^{n-1} \dot{\pi}^4
- \frac{3}{2} \kappa^2 \sum_{n=0}^{\infty} G_3^{(n)}(\pi) (1+X)^n
\dot{\pi}^2 \right ) \nonumber \\
&&  \times \left ( -6 \sum_{n=0}^{\infty} G_3^{(n)}(\pi) n(1+X)^{n-1} \dot{\pi}^4
+9 \sum_{n=0}^{\infty} G_3^{(n)}(\pi) (1+X)^n \dot{\pi}^2 \right )  \nonumber \\
&& + 42 \sum_{n=0}^{\infty} G_3^{(n)}(\pi) n (1+X)^{n-1}H \dot{\pi}^3
 -18  \sum_{n=0}^{\infty} G_3^{(n)}(\pi) (1+X)^n H \dot{\pi} \nn
&& -12  \sum_{n=0}^{\infty} G_3^{(n)}(\pi) n(n-1) (1+X)^{n-2} H \dot{\pi}^5
+ \frac{2}{\kappa^2} g''(\pi) - G_3'^{(0)}(\pi)+ 6 G_3^{(1)}(\pi) g'(\pi) 
 -9  G_3^{(0)}(\pi) g'(\pi) \nonumber \\
&& +2  \sum_{n=2}^{\infty} G_2^{(n)}(\pi) n (1+X)^{n-1} 
 -4  \sum_{n=2}^{\infty} G_2^{(n)}(\pi) n(n-1) (1+X)^{n-2} \, ,
\end{eqnarray}
We now define variables $x$ and $y$ by 
$x\equiv\dot{\pi}$ and $y\equiv g'\left(\pi\right)/H$. 
Then for the reconstructed solution (\ref{Grec6}), we find $x=1$ and $y=1$. 
Since 
\be
\label{FRW23_13}
\frac{dx}{dN} = \frac{\ddot{\pi}y}{g'\left(\pi\right)} \, ,\quad 
\frac{dy}{dN} = \frac{g''\left(\pi\right)}{g'^{2}\left(\pi\right)}y^{2}x
 -\frac{y^{3}}{g'^{2}\left(\pi\right)}\dot{H}\, ,\quad 
\frac{d\pi}{dN} = \frac{yx}{g'\left(\pi\right)} \, ,
\ee
we can examine the perturbation from the reconstructed solution $x=y=1$ and $\pi=t$ 
as $x=1+\delta x$, $y=1+\delta y$, and $\pi=t+\delta\pi$ by using $x$ and $y$.  
Eqs.~(\ref{FRW23_7}) and (\ref{FRW23_8}) as follows: 
\begin{equation}
\label{FRW23_16}
\frac{d}{dN}\begin{pmatrix}\delta x\\
\vphantom{M_{12}}\delta y\\
\vphantom{M_{12}}\delta\pi
\end{pmatrix}
= M \begin{pmatrix}\delta x\\
\vphantom{M_{12}}\delta y\\
\vphantom{M_{12}}\delta\pi
\end{pmatrix} \, ,\quad 
M \equiv \begin{pmatrix}M_{11} & M_{12} & M_{13}\\
M_{21} & M_{22} & M_{23}\\
M_{31} & M_{32} & M_{33}
\end{pmatrix} \, .
\end{equation}
Here
\begin{eqnarray}
\label{FRW23_17}
 M_{11}
&=& A_0^{-1} g'^{-1}(t) \left [ 
72 G_3^{(2)}(t)g'^2(t)-72 G_3^{(1)}(t) g'^2(t) +24 \dot{G}_3^{(2)}(t)
g'(t) -30 \dot{G}_3^{(1)}(t) g'(t) \right . \nonumber \\
&& + \left (g''(t)+ 3g'^2(t) - \kappa^2 \dot{G}_3^{(0)}(t) \right ) \left (
24 G_3^{(2)}(t)-42 G_3^{(1)}(t) +18 G_3^{(0)}(t)  \right ) \nonumber \\
&& + \left (-6G_3^{(1)}(t)+9G_3^{(0)}(t)   \right )
\left \{ g''(t) + \kappa^2 \left ( \frac{3}{2} \dot{G}_3^{(0)}(t) 
+3 G_3^{(1)}(t) g'(t) - \frac{9}{2} G_3^{(0)}(t) g'(t)- \dot{G}_3^{(1)} \right )
\right  \} \nonumber \\
&&  +54 G_3^{(1)}(t) g'^2(t)+54 G_3^{(0)}(t) g'^2(t)
 -12 \dot{G}_3^{(1)}(t) g'(t) +18 \dot{G}_3^{(0)}(t) g'(t)
+ 24 G_2^{(2)}(t)g'(t) \nonumber \\
&& -6 \left (
\frac{1}{\kappa^2}g'(t) g''(t) - \frac{1}{2} \dot{G}_3^{(0)}(t) g'(t) +3
G_3^{(1)}(t) g'^2(t) - \frac{9}{2} G_3^{(0)}(t) g'^2(t) \right )
+ 8 \dot{G}_2^{(2)} \nonumber \\
&& \left . +2 \left (
\frac{1}{\kappa^2} g'''(t)- \frac{1}{2} \ddot{G}_3^{(0)}(t)
+3 \dot{G}_3^{(1)}(t) g'(t) +3 G_3^{(1)}(t) g''(t)
 - \frac{9}{2} \dot{G}_3^{(0)}(t) g'(t) - \frac{9}{2} G_3^{(0)}(t) g''(t)
\right  ) \right ] \, , \\
\label{FRW23_18}
M_{12} &=& A_0^{-1} g'^{-1}(t)
\left [ 36 G_3^{(1)}(t) g'^2(t)+6 \dot{G}_3^{(1)}(t) g'(t)
+3 \left ( -6 G_3^{(1)}(t)+9 G_3^{(0)}(t)
\right )g'^2(t)  -54 G_3^{(0)}(t) g'^2(t) \right . \nonumber \\
&& \left . -9 \dot{G}_3^{(0)}(t) g'(t)
+3 \dot{G}_3^{(0)}(t) g'(t) +6 g'(t) \left ( \frac{1}{\kappa^2} g''(t)
 - \frac{1}{2} \dot{G}_3^{(0)}(t) +3 G_3^{(1)}(t) g'(t)
 - \frac{9}{2} G_3^{(0)}(t) g'(t) \right ) \right ]\, , \\
\label{FRW23_19}
M_{13} &=& A_0^{-1} g'^{-1}(t)
\left [ -18 \dot{G}_3^{(1)}(t) g'^2(t) -36 G_3^{(1)}(t) g'(t) g''(t)
 -6 \ddot{G}_3^{(1)}(t) g'(t) -6 G_3^{(1)}(t) g''(t) \right . \nonumber \\
&& \left (
g''(t)+ \frac{3}{2} g'^2(t)- \kappa^2 \dot{G}_3^{(0)}+ \frac{3}{2} g'^2(t) \right )
\left ( -6 \dot{G}_3^{(1)}(t)+9 \dot{G}_3^{(0)} \right ) 
+\left ( -6 G_3^{(1)}(t)+9 G_3^{(0)}(t) \right ) g'''(t) \nonumber \\
&& +27 \dot{G}_3^{(0)}(t) g'^2(t) +54 G_3^{(0)}(t) g'(t)g''(t)
+6 \ddot{G}_3^{(0)}(t) g'(t) +6 \dot{G}_3^{(0)}(t) g''(t) \nonumber \\
&& -6 g'(t) \left ( \frac{1}{\kappa^2} g'''(t) - \frac{1}{2} \ddot{G}_3^{(0)}(t)
+3 \dot{G}_3^{(1)}(t) g'(t) +3 G_3^{(1)}(t) g''(t) - \frac{9}{2} \dot{G}_3^{(0)}(t) g'(t)
 - \frac{9}{2} G_3^{(0)}(t) g''(t)  \right ) \nonumber \\
&& -6 g''(t) \left ( \frac{1}{\kappa^2} g''(t) - \frac{1}{2} \dot{G}_3^{(0)}(t)
+3 G_3^{(1)}(t) g'(t) - \frac{9}{2} G_3^{(0)}(t) g'(t) \right ) \nonumber \\
&& -2 \left ( \frac{1}{\kappa^2} g''''(t) - \frac{1}{2} \dddot{G}_3^{(0)}(t)
+3 \ddot{G}_3^{(1)}(t) g'(t) +6 \dot{G}_3^{(1)}(t) g''(t) +3 G_3^{(1)}(t) g'''(t)
 - \frac{9}{2} \ddot{G}_3^{(0)}(t) g'(t) -9 \dot{G}_3^{(0)}(t) g''(t) 
\right ) \nonumber \\ 
&& \left . + \frac{6}{\kappa^2} g''^2(t) + \frac{6}{\kappa^2} g'(t) g'''(t) 
 - \dddot{G}_3^{(0)} \right ]\, , \\
\label{FRW23_20}
M_{21} &=& \frac{g''(t)}{g'^2(t)} - g'^{-2}(t) \left \{ g''(t)
+\kappa^2 \left ( \frac{3}{2} \dot{G}_3^{(0)}(t)
+3 G_3^{(1)}(t) g'(t)- \frac{9}{2} G_3^{(0)}(t) g'(t)- \dot{G}_3^{(1)}(t)
 \right )  \right \} \nonumber  \\
&& - g'^{-1}(t) \left ( - \kappa^2 G_3^{(1)}(t)  + \frac{3}{2} \kappa^2 G_3^{(0)}(t)
 \right ) M_{11} \, , \\
\label{FRW23_21}
M_{22} &=& - g'^{-2}(t) g''(t) - 3 - g'^{-1}(t)
\left ( -\kappa^2 G_3^{(1)}(t) + \frac{3}{2} \kappa^2 G_3^{(0)}(t)  \right ) M_{12} \, ,\\
\label{FRW23_21B}
M_{23} &=& - g'(t)^{-1}
\left ( -\kappa^2 G_3^{(1)}(t)+ \frac{3}{2} \kappa^2 G_3^{(0)}(t) \right ) M_{13}
+ 3 g'^{-1}(t) g''(t) -g'^{-2}(t) g'''(t)  \, ,
\end{eqnarray}
where 
\begin{eqnarray}
\label{FRW23_21C}
A_0 &=&- \frac{\kappa^2}{6} 
\left ( 6 G_3^{(1)}(t) - 9 G_3^{(0)}(t)  \right )^2
+42 G_3^{(1)}(t) g'(t) -27 G_3^{(0)}(t)g'(t) -24 G_3^{(2)}(t) g'(t)
+ \frac{2}{\kappa^2} g''(t)- \dot{G}_3^{(0)}(t) \nonumber \\
&& +6 G_3^{(1)}(t) g'(t) -8 G_2^{(2)}(t)\, ,
\end{eqnarray}
and
\be
\label{FRW23_22}
M_{31} = g'^{-1}\left(t\right)\, , \quad 
M_{32} = g'^{-1}\left(t\right)\, , \quad 
M_{33} = -g'^{-2}\left(t\right)g''\left(t\right)\, .
\ee
The eigenvalue equation for the matrix $M$ in (\ref{FRW23_16}) has the following form:
\begin{equation}
\label{FRW23_25}
\lambda^{3}+\alpha\lambda^{2}+\beta\lambda+\gamma=0\, , 
\end{equation}
where 
\begin{eqnarray}
\label{FRW23_26}
\alpha & = & -M_{11} - M_{22}+g'^{-2}\left(t\right)g''\left(t\right) \, , \\
\label{FRW23_27}
\beta & = & -(M_{11}+M_{22}) g'^{-2}(t)g''(t) -(M_{23}+M_{13})g'^{-1}(t)
+M_{11}M_{22}-M_{12}M_{21} \, , \\
\label{FRW23_28}
\gamma & = &
(M_{11}M_{22}-M_{12}M_{21})g'^{-2}(t)g''(t)
+(M_{11}M_{23}-M_{13}M_{21}-M_{12}M_{23}+M_{13}M_{22})g'^{-1}(t) \, .
\end{eqnarray}
When all the eigenvalues $\lambda$ are negative, the reconstructed solution becomes stable. 
The Hurwitz theorem tells the condition that all the eigenvalues are negative is given by
\begin{equation}
\label{FRW23_29}
\left(\mathrm{i}\right)\ \alpha>0\, ,\quad \left(\mathrm{ii}\right)\ \alpha\beta>\gamma\, , \quad 
\left(\mathrm{iii}\right)\ \alpha\beta\gamma>\gamma^{2}\, .
\end{equation}
For the simplicity, we further put $G_{3}^{\left(0\right)}\left(t\right)=G_{3}^{\left(1\right)}\left(t\right)=0$. 
Thus we find 
\begin{eqnarray}
\label{FRW23_30}
&& M_{11} = A_0^{-1}g'^{-1}(t)
\left \{ 72 G_3^{(2)}(t)g'^2(t)+24 \dot{G}_3^{(2)}(t) g'(t)
+ 24 G_3^{(2)}  \left ( g''(t)+ 3 g'^2(t) \right  )
+24 G_2^{(2)} (t) g'(t) \right .  \nonumber  \\
&& \left . \qquad \qquad - \frac{6}{\kappa^2} g'(t) g''(t) 
+ 8 \dot{G}_2^{(2)} + \frac{2}{\kappa^2} g'''(t) \right \} \, , \nn
&& M_{12} =
6 A_0^{-1}
\frac{1}{\kappa^2} g''(t)
\, , \nn
&& M_{13} =
-2 A_0^{-1} g'^{-1}(t) 
 \frac{1}{\kappa^2} g''''(t)
\, , \nn
&& M_{21} = 0 \, , \quad 
M_{22} = - g'^{-2}\left(t\right)g''\left(t\right)-3 \, , \quad 
M_{23} = 3 g'^{-1}(t) g''(t)- g'^{-2}(t) g'''(t) \, ,
\nonumber  \\
&& M_{31} = M_{32}=g'^{-1}\left(t\right) \, , \quad 
M_{33} = -g'^{-2}\left(t\right)g''\left(t\right)\, .
\end{eqnarray}
Here $A_0$ has the following form:
\begin{equation}
\label{FRW23_36}
A_0 = -24 G_3^{(2)}(t) g'(t) + \frac{2}{\kappa^2} g''(t) - 8 G_2^{(2)} \, .
\end{equation}
The eigenvalue equation becomes 
\begin{equation}
\label{FRW23_37}
\left(M_{22}-\lambda\right)\left\{ \lambda^{2}
 -\left(M_{11}+M_{33}\right)\lambda+M_{11}M_{33}-M_{13}M_{31}\right\} =0 \, .
\end{equation}
Therefore we obtain
\begin{equation}
\label{FRW23_38A}
\lambda=M_{22}\, ,\quad 
\frac{1}{2}\left[M_{11}+M_{33}\pm\sqrt{\left(M_{11}+M_{33}\right)^{2}
 -4\left(M_{11}M_{33}-M_{13}M_{31}\right)}\right]\, .
\end{equation}
Consequently, the condition of the stability is given by
\begin{equation}
\label{FRW23_38}
\left(\mathrm{i}\right)\ M_{22}<0\, , \quad 
\left(\mathrm{ii}\right)\ M_{11}+M_{33}<0\, , \quad 
\left(\mathrm{iii}\right)\ M_{11}M_{33}-M_{13}M_{31}>0\, .
\end{equation}
 From the first condition $\left(\mathrm{i}\right)$ in (\ref{FRW23_38}), we find 
\begin{equation}
\label{FRW23_39}
0 > -g'^{-2}\left(t\right)g''\left(t\right)-3\, ,
\end{equation}
and from the second condition $\left(\mathrm{ii}\right)$, 
\begin{eqnarray}
\label{FRW23_40}
0 & > & 
A_0^{-1}
\left \{
- g'^{-2}(t)g''(t)A_0
+144 G_3^{(2)}(t)g'(t)+24 \dot{G}_3^{(2)}(t) +24
G_3^{(2)}(t) g'^{-1}(t) g''(t) \right . \nonumber \\
&&
\left .
+24 G_2^{(2)}(t)
- \frac{6}{\kappa^2} g''(t)
+8 \dot{G}_2^{(2)}(t) g'^{-1}(t)
+ \frac{2}{\kappa^2} g'''(t) g'^{-1}(t)
 \right \}
\, .
\end{eqnarray}
The third condition $\left(\mathrm{iii}\right)$ in (\ref{FRW23_38}) presents
\begin{eqnarray}
\label{FRW23_41}
0 & < & 
A_0^{-1} g'^{-1}(t)
\left [ -g'^{-2}(t) g''(t) \left \{
72 G_3^{(2)}(t) g'^2(t) +24 \dot{G}_3^{(2)}(t) g'(t) +24 G_3^{(2)}
(g''(t)+3 g'^2(t)) \right . \right . \nonumber \\
&&
\left . \left .
+24 G_2^{(2)}(t) g'(t) - \frac{6}{\kappa^2} g'(t) g''(t) +8
\dot{G}_2^{(2)}(t) + \frac{2}{\kappa^2} g'''(t) \right \}
+2 g'^{-1}(t) \frac{1}{\kappa^2} g''''(t) \right ] \, .
\end{eqnarray}

Before going to non-trivial case, 
in order to check the above formulation could work, we consider the de Sitter space-time, 
where $g'\left(t\right)$ is constant. Then we find 
\be
\label{FRW23_42}
M=\left(\begin{array}{ccc}
M_{11} & 0 & 0\\
0 & -3 & 0\\
g'^{-1}\left(t\right) & g'^{-1}\left(t\right) & 0
\end{array}\right)\, ,\quad 
M_{11}
= \frac{144G_3^{(2)}(t) g'^2(t)+24
\dot{G}_3^{(2)}(t)g'(t)+24G_2^{(2)}(t)g'(t)+8
\dot{G}_2^{(2)}}{-24G_3^{(2)}(t) g'^2(t) + \frac{2}{\kappa^2} g'(t) g''(t) -8
G_2^{(2)}(t) g'(t)} \, .
\ee
Then by solving the eigenvalue equation (\ref{FRW23_37}), we find the eigenvalues: 
\begin{equation}
\label{FRW23_44}
\lambda=-3\, ,\quad 0\, , \quad \frac{144G_3^{(2)}(t) g'^2(t)
+24 \dot{G}_3^{(2)}(t)g'(t)+24G_2^{(2)}(t)g'(t)+8
\dot{G}_2^{(2)}}{-24G_3^{(2)}(t) g'^2(t) + \frac{2}{\kappa^2} g'(t) g''(t)
 -8 G_2^{(2)}(t) g'(t)}\, .
\end{equation}
Note that the eigenvector corresponding to the eigenvalue $0$ is given by 
\begin{equation}
\label{FRW23_45}
\left(\begin{array}{c}
\delta x\\
\delta y\\
\delta\pi
\end{array}\right)=s\left(\begin{array}{c}
0\\
0\\
1
\end{array}\right)\, ,
\end{equation}
which tells that the eigenvector with the eigenvalue $0$ 
corresponds to the shift of $\pi$ or the origin of time 
and therefore the eigenvalue $0$ does not corresponds to any instability. 
Especially if we consider the case that 
$G_{3}^{\left(2\right)}\left(t\right)=0$, the condition that the last eigenvalue in (\ref{FRW23_44}) is given by
\begin{equation}
\label{FRW23_46}
 -\frac{\dot{G}_{2}^{\left(2\right)}\left(t\right)g'^{-1}\left(t\right)}
{G_{2}^{\left(2\right)}\left(t\right)}-3<0\, ,
\end{equation}
can be satisfied if
\begin{equation}
\label{FRW23_47}
G_{2}^{\left(2\right)}\left(t\right)=D\exp\left(-f(t)\right)\, ,
\end{equation}
where 
$D$ is a constant and $f(t)$ is an arbitrary function satisfying the condition 
$0<f'(t)<3g'\left(t\right)$. 
We should also note that Eq.~(\ref{FRW23_47}) can be satisfied 
if $G_{2}^{\left(2\right)}\left(t\right)$ is a constant. 

As a little bit non-trivial example, we study 
the case that $g'\left(t\right)$ is a solution of the $\Lambda$CDM model:
\be
\label{FRW23_48}
a\left(t\right) = A\sinh^{\frac{2}{3}}\left(bt\right) \, , 
\ee
where $A$ and $b$ are positive constants. 
Eq.~(\ref{FRW23_48}) gives
\be
\label{FRW23_49}
g'\left(t\right) = \frac{2}{3}b\coth\left(bt\right)\, ,\quad 
g''\left(t\right) = - \frac{2b^2}{3\sinh^{2}\left(bt\right)}\, .
\ee
Then the first condition $\left(\mathrm{i}\right)$ in (\ref{FRW23_38}) gives
\be
\label{FRW23_50}
0 > \frac{3}{2\cosh^{2}\left(bt\right)}-3\, ,
\ee
which is trivially satisfied, 
and the second condition $\left(\mathrm{ii}\right)$ presents 
\begin{eqnarray}
\label{FRW23_51}
0 & > &
A_0^{-1} \left \{
24 \dot{G}_3^{(2)} - \frac{48 b}{\cosh(bt) \sinh(bt)} G_3^{(2)} 
+ \left(24 - \frac{12}{\cosh^2 (bt)} \right) G_2^{(2)}(t)
\right. \nonumber \\
&&
\left.
+\frac{12}{b} \tanh (bt) \dot{G}_2^{(2)}  + \frac{b^2}{\kappa^2} \left(
\frac{8}{ \sinh^2(bt)} - \frac{2}{\cosh^2(bt) \sinh^2(bt)} \right)
 \right \}
\, .
\end{eqnarray}
Here
\begin{equation}
\label{FRW23_52}
A_0 = -16 G_3^{(2)} b \coth(bt) - \frac{4b^2}{3 \kappa^2} 
\frac{1}{\sinh^2(bt)} -8 G_2^{(2)} \, .
\end{equation}
By using the third condition $\left(\mathrm{iii}\right)$ in (\ref{FRW23_38}), we find
\begin{eqnarray}
\label{FRW23_53}
0 & < &  A_0^{-1}
\left \{ 64 b^2 \coth^2 (bt) G_3^{(2)} + 16 b \coth (bt) \dot{G}_3^{(2)}
 - \frac{16 b^2}{\sinh^2(bt)} G_3^{(2)} +16 b \coth(bt) G_2^{(2)} \right. \nonumber \\
&& \left .
+8 \dot{G}_2^{(2)}  - \frac{b^3}{\kappa^2} \left( \frac{16}{3} \coth (bt) 
+ \frac{8\cosh (bt)}{3\sinh^3 (bt)} \right) \right \} \, .
\end{eqnarray}
Although the above expressions (\ref{FRW23_51}) and (\ref{FRW23_52}) are very complicated, 
these conditions can be satisfied. 
The simplest example is given by 
\be
\label{FRW23_53B}
G_{2}^{\left(2\right)}\left(t\right) = C
\, , \quad G_{3}^{\left(2\right)}\left(t\right) = 0 \, . 
\ee 
Here $C$ is a constant. 
By substituting (\ref{FRW23_53B}) into (\ref{FRW23_51}), (\ref{FRW23_52}), and (\ref{FRW23_53}), 
we find
\bea
\label{sn1}
A_0 &=& - \frac{4b^2}{3 \kappa^2} \frac{1}{\sinh^2(bt)} -8 C\, , \\
\label{sn2}
0 &>& A_0^{-1} \left\{ \left( 24 - \frac{12}{\cosh^2 (bt) } \right) C 
+ \frac{b^2}{\kappa^2} \left(
\frac{8}{ \sinh^2(bt)} - \frac{2}{\cosh^2(bt) \sinh^2(bt)} \right) \right \}
\, , \\
\label{sn3}
0 & < & 
A_0^{-1}
\left \{
16 b \left( C - \frac{b^2}{3\kappa^2} \right)\coth(bt) 
 - \frac{8b^3\cosh (bt)}{3\kappa^2\sinh^3 (bt)}
\right \}\, .
\eea
If $C>0$, we find $A_0<0$ in (\ref{sn1}) and the inequality (\ref{sn2}) is 
satisfied. The inequality (\ref{sn3}) is also satisfied if
\be
\label{sn4}
0 < C < \frac{b^2}{3\kappa^2}\, .
\ee
Then stability can be realized. 
Note that even for the de Sitter space-time, if $G_{2}^{\left(2\right)}\left(t\right) $ and 
$G_{3}^{\left(2\right)}\left(t\right)$ are given by (\ref{FRW23_53B}), the stability is 
realized as discussed after (\ref{FRW23_47}). 

As a result, in this section, by exploring 
the FRW dynamics, we have given the explicit formulae for the reconstruction and we have 
investigated the condition of the stability for the reconstructed solution. 


\section{Vainshtein mechanism \label{IV}}

In this section, in order to investigate if the Vainshtein mechanism could work, 
we consider the behavior of the Galileon scalar field $\pi$ in the spherically symmetric space-time, 
especially in the Schwarzschild background. 

The Schwarzschild space-time has a spherically symmetric and static metric. 
The general spherically symmetric and static space-time has the metric of the 
following form:
\begin{eqnarray}
\label{Sch1}
ds^2 = - \e^{2 \Phi} dt^2 + \e^{2 \Lambda} dr^2 
+ r^2 (d \theta^2 + \sin^2{\theta} d \phi^2)\, , 
\end{eqnarray}
where 
$\Phi$ and $\Lambda$ only depend on the radial coordinate $r$.
In the Schwarzschild background, the equation derived 
by the variation of the Galileon scalar field $\pi$ is given by
\begin{eqnarray}
\label{Sch5}
&& 
\frac{1}{\sqrt{-g}} \frac{\delta \sqrt{-g} G_2(\pi,X)}{\delta \pi}
+ \frac{1}{\sqrt{-g}} \frac{\delta \sqrt{-g} G_3(\pi,X) \mathcal{L}_3}{\delta \pi}
+ \frac{1}{\sqrt{-g}} \frac{\delta \sqrt{-g} \frac{\partial G_4(\pi,X)}{\partial X} \mathcal{L}_4^{(0)}}{\delta \pi}
 - \frac{1}{8} \frac{1}{\sqrt{-g}} \frac \delta{ \sqrt{-g} G_4(\pi,X) \mathcal{L}_6}{\delta \pi} \nonumber \\
&&
+ \frac{1}{\sqrt{-g}} \frac{\delta \sqrt{-g} \frac{\partial G_5(\pi,X)}{\partial X} \mathcal{L}_5^{(0)}}{\delta \pi}
+ \frac{1}{\sqrt{-g}} \frac{\delta \sqrt{-g} G_5(\pi,X) \mathcal{L}_7}{\delta \pi}
+ \frac{1}{\sqrt{-g}} \frac{\delta \sqrt{-g} G_6(\pi) \mathcal{L}_6}{\delta \pi}
+ \frac{1}{\sqrt{-g}} \frac{\delta \sqrt{-g} G_7(\pi) \mathcal{L}_7}{\delta \pi}
\nonumber \\ &&
+ \frac{1}{\sqrt{-g}} \frac{\delta \sqrt{-g} G_8(\pi) \mathcal{L}_8}{\delta \pi} =0\, , 
\end{eqnarray}
where we have used the notation $\dfrac{\delta A}{\delta \pi} 
\equiv \dfrac{\partial A}{\partial \pi} - \partial_{\mu} \dfrac{\partial A}{ \partial (\partial_{\mu} \pi)}$. 
The explicit forms of $\mathcal{E}_k$ and also 
$(\mathcal{H}_k)_{\mu\nu}$ are given in Appendix \ref{Ap2}. 

We also note that the Einstein tensor is given by
\begin{eqnarray}
\label{Sch25}
G_{tt} &=& \frac{1}{r^2} \e^{2 \Phi} \frac{d}{dr} 
\left[ r(1- \e^{-2 \Lambda})  \right]\, ,\nn
G_{rr} &=& - \frac{1}{r^2} \e^{2 \Lambda} (1 - \e^{-2 \Lambda})+
\frac{2}{r} \frac{d \Phi}{dr}\, , \nn
G_{\theta \theta} &=& r^2 \e^{-2 \Lambda} \left[ \frac{d^2 \Phi}{dr^2} 
+ \left( \frac{d \Phi}{dr} \right)^2 
+ \frac{1}{r} \frac{d \Phi}{dr} - \frac{d \Phi}{dr} \frac{d \Lambda}{dr} 
 - \frac{1}{r} \frac{d \Lambda}{dr} \right]\, ,\nn
G_{\phi \phi} &=& \sin^2{\theta} G_{\theta \theta}\, .
\end{eqnarray}
The Einstein equation (\ref{A1}) tells that in order that 
the modification of the Schwarzschild geometry 
due to the Galileon scalar could be small, we find 
\be
\label{Sch36}
\sum_{i=2}^8 (\mathcal{H}_i)_{\mu\nu} \sim 0\, .
\ee

We now investigate how the Vainshtein mechanism works for the generalized Galileon scalar model. 
In the limit of $r\gg GM$, the Schwarzschild metric 
\be
\label{Sch12}
ds^2 = - \left( 1-\frac{2GM}{r}  \right)dr^2 
+ \frac{dr^2}{1- \frac{2GM}{r}} + r^2(d \theta^2 + \sin^2{\theta} d \phi^2)\, ,
\ee
for general 
spherically symmetric and static space-time in (\ref{Sch1}) behaves as
\be
\label{Sch37}
\Phi(r) \simeq - \frac{GM}{r}\, ,\quad \Lambda(r) \simeq \frac{GM}{r}\, .
\ee
If the Vainshtein mechanism works, the gravity with the Galileon scalar behaves as the usual Einstein gravity 
in the short distance compared with the cosmological scale and $\Phi(r)$ and $\Lambda(r)$ 
in (\ref{Sch1}) behaves as those in (\ref{Sch37}). 

Before examining 
the Vainshtein mechanism in the covariant and generalized Galileon model, 
we review how the Vainshtein mechanism works when we only include $\mathcal{L}_4$ in (\ref{Gl2}). 
In this case the energy momentum tensors of the Galileon scalar behave as 
\be
\label{Sch38}
T_{00}^\pi = 8 c_4 \frac{(\pi')^3 \pi''}{r} + c_4 \frac{(\pi')^4}{r^2} + \cdots\, ,\quad 
T_{11}^\pi = - 5 c_4 \frac{(\pi')^4}{r^2} + \cdots\, .
\ee
If we write $c_4 = {\tilde c}_4 \frac{M_\mathrm{Pl}^2}{H_0^4}$, we find ${\tilde c}_4\sim \mathcal{O}(1)$. 

We now study 
the generalized Galileon model in (\ref{DeltaGG8}) but just for the 
simplicity, we only include $\mathcal{L}_2$ and $\mathcal{L}_3$:
\be
\label{Sch39}
\mathcal{L}_{\pi} =  - \frac{1}{2} c_2 X + G_3(\pi ,X) \mathcal{L}_3 \, .
\ee
We may assume the first term in (\ref{Sch39}) dominates in the short distance but the second term does in the long distance. 
Since the field equation has a form like 
\begin{eqnarray}
\label{Sch40}
\pi''(r) = f(\pi(r),\pi'(r)) \frac{\pi'(r)}{r}\, ,
\end{eqnarray}
when $r\ll r_V$, we find
\be
\label{Sch41}
\pi'(r) = \pi'(r_V) \exp \left( \int_{r_V}^{r} \frac{f(\pi(r'),\pi'(r'))}{r'} dr'  \right) 
= \pi'(r_V) \frac{e^{F(r)}}{e^{F(r_V)}}\, , 
\ee
where
\be
\label{Sch42}
F(r) \equiv \int^{r} \frac{f(\pi(r'),\pi'(r'))}{r'} dr \, .
\ee
If we provide 
$\pi'(r_V) \sim \Phi'_N(r_V)$, $\pi'(r)$ becomes small when we choose 
$f(\pi(r),\pi'(r))$ so that we can have $F(r)\ll F(r_V)$. 
If $\pi'(r)\ll \pi'(r_V)$ as required so that the Vainshtein mechanism could work, we may find 
$\pi(r)\ll \pi(r_V)$. Then if we can choose $G_3(\pi ,X)$ so that $f(\pi(r),\pi'(r))>0$ 
when $\pi'(r)<\pi'(r_V)$ and $\pi(r)<\pi(r_V)$, we may surely obtain 
$\pi'(r)\ll \pi'(r_V)$, consistently. 
We now have 
\begin{eqnarray}
\label{Sch43}
f(\pi(r),\pi'(r)) =  
\frac{-2 r^2 -2 \frac{\partial G_3}{\partial \pi} (\pi')^2 r^2 - 3 G_3 \pi' r
 -2 \frac{\partial G_3}{\partial X} (\pi')^3 r  -2 \frac{\partial^2 G_3}{\partial \pi \partial X} (\pi')^4 r^2}
{ r^2 +6 G_3 \pi' r +14 \frac{\partial G_3}{\partial X} (\pi')^3 r +4 \frac{\partial^2 G_3}{\partial X^2} 
(\pi')^5 r}\, .
\end{eqnarray}
It is easy to find that there exists $G_3(\pi ,X)$ which satisfies the condition that $f(\pi(r),\pi'(r))>0$ 
when $\pi'(r)<\pi'(r_V)$ and $\pi(r)<\pi(r_V)$ but $G_3(\pi ,X)$ cannot be uniquely determined. 

Finally we show that any spherically symmetric and static geometry given by arbitrary $\Phi(r)$ and 
$\Lambda(r)$ can be realized by properly choosing $G_i$'s. 
Since the redefinition of the scalar field can be absorbed into the redefinition of $G_i$'s, here we may 
identify the Galileon scalar field with the radial coordinate, $\pi=r$. 
Then we find
\begin{eqnarray}
\label{Sch37_00}
0 &= & \frac{1}{2\kappa^2 r^2} \e^{2 \Phi} \frac{d}{dr} 
\left[ r(1- \e^{-2 \Lambda})  \right] - \frac{1}{2}\e^{2\Phi} G_2 (r, \e^{-2\Lambda}) 
+ \frac{\e^{2\Phi - 2\Lambda}}{2} \frac{d}{dr} \left( G_3 (r, \e^{-2\Lambda}) \e^{-2\Lambda} \right) \nn
&& -5 \frac{\partial G_4 (r, \e^{-2\Lambda})}{\partial X} \frac{d \Lambda}{dr}
\frac{\e^{2 \Phi - 6 \Lambda}}{r} +  \frac{d}{dr}  
\left( \frac{\partial G_4 (r, \e^{-2\Lambda})}{\partial X}  \right)
\frac{\e^{2 \Phi -6 \Lambda}}{r} 
+ \frac{1}{2} \frac{\partial G_4 (r, \e^{-2\Lambda})}{\partial X}  
\frac{\e^{2 \Phi -6 \Lambda}}{r^2} \nn
&& - \frac{3}{2}  G_4 (r, \e^{-2\Lambda}) \frac{d \Lambda}{dr} \frac{\e^{2 \Phi -4 \Lambda}}{r}
+ \frac{1}{2} \frac{d G_4 (r, \e^{-2\Lambda})}{dr} \frac{\e^{2 \Phi-4 \Lambda}}{r} 
+ \frac{1}{4} G_4 (r, \e^{-2\Lambda}) \frac{\e^{2 \pi - 4 \Lambda}}{r^2} 
+ \frac{1}{4} G_4 (r, \e^{-2\Lambda}) \frac{\e^{-4 \Phi -2 \Lambda}}{r^2} \nn
&& + \frac{1}{2} \frac{d}{dr} \left( \frac{\partial G_5 (r, \e^{-2\Lambda})}{\partial X}
\right) \frac{\e^{2\Phi -8 \Lambda}}{r^2} 
 - \frac{7}{2} \frac{\partial G_5 (r, \e^{-2\Lambda})}{\partial X} 
\frac{d \Lambda}{dr} \frac{\e^{2 \Phi -8 \Lambda}}{r^2} 
+\frac{3}{4} G_5 (r, \e^{-2\Lambda}) \frac{d \Lambda}{dr} \frac{\e^{2 \Phi -4 \Lambda}}{r^2} \nn
&& - \frac{15}{4} G_5 (r, \e^{-2\Lambda}) \frac{d \Lambda}{dr} \frac{\e^{2 \Phi -6 \Lambda}}{r^2} 
 - \frac{1}{4} \frac{d G_5 (r, \e^{-2\Lambda})}{dr} \frac{\e^{2 \Phi -4 \Lambda}}{r^2} 
+ \frac{3}{4} \frac{d G_5 (r, \e^{-2\Lambda})}{dr} \frac{\e^{2 \Phi -6 \Lambda}}{r^2} \nn
&& + 12 G_6 (r) \frac{d \Lambda}{dr} \frac{\e^{2 \Phi -4 \Lambda}}{r}
 - 4 \frac{d G_6 (r)}{dr} \frac{\e^{2 \Phi-4 \Lambda}}{r} 
 -2 G_6 (r) \frac{\e^{2 \Phi - 4 \Lambda}}{r^2} 
 -2 G_6 (r) \frac{\e^{-4 \Phi -2 \Lambda}}{r^2} \nn
&& -6  G_7 (r) \frac{d \Lambda}{dr} \frac{\e^{2 \Phi -4 \Lambda}}{r^2} 
+ 30 G_7 (r) \frac{d \Lambda}{dr} \frac{\e^{2 \Phi -6 \Lambda}}{r^2} 
+ 2 \frac{d G_7 (r)}{dr} \frac{\e^{2 \Phi -4 \Lambda}}{r^2} 
 -6 \frac{d G_7 (r)}{dr} \frac{\e^{2 \Phi -6 \Lambda}}{r^2} \nn
&& -192 \frac{d G_8 (r)}{dr} \frac{\e^{2 \Phi -2 \Lambda}}{r^2} 
\frac{d \Phi}{dr} + 16 \frac{d G_8 (r)}{dr} \frac{\e^{2 \Phi -2 \Lambda}}{r^2} 
\frac{d \Lambda}{dr}
 -16 \frac{d^2 G_8 (r)}{dr^2} \frac{\e^{2 \Phi -2 \Lambda}}{r^2}
 -48 \frac{d G_8 (r)}{dr} \frac{\e^{2 \Phi-4 \Lambda}}{r^2} \frac{d
\Lambda}{dr} \nn
&& +16 \frac{d^2 G_8 (r)}{dr^2} \frac{\e^{2 \Phi -4 \Lambda}}{r^2}\, , \\
\label{Sch38-00}
0 &= & \frac{1}{2\kappa^2}\left\{- \frac{1}{r^2} \e^{2 \Lambda} (1 - \e^{-2 \Lambda})+
\frac{2}{r} \frac{d \Phi}{dr}\right\} \nn
&& + \frac{1}{2} \e^{2\Lambda} G_2 (r, \e^{-2\Lambda}) 
 - \frac{\partial G_2 (r, \e^{-2\Lambda})}{\partial X} 
+ \frac{1}{2}\e^{2\Lambda} G_3 (r, \e^{-2\Lambda}) 
\left( \frac{\e^{-3\Lambda}}{r^2} \left( \frac{d}{dr} \left(r^2 
\e^{\Phi - \Lambda} \right)\right)
+ \e^{-4\Lambda} \frac{d\Lambda}{dr} \right) \nn
&& - \frac{\partial G_3 (r, \e^{-2\Lambda})}{\partial X} 
\left( \frac{\e^{-3\Lambda}}{r^2} \left( \frac{d}{dr} \left(r^2 
\e^{\Phi - \Lambda} \right)\right)
+ \e^{-4\Lambda} \frac{d\Lambda}{dr} \right) \nn
&& - G_3 (r, \e^{-2\Lambda}) \left( \left( \frac{\e^{- \Phi - \Lambda}}{r^2} 
\left( \frac{d}{dr} \left( r^2 \e^{\Phi - \Lambda} \right) \right) \right)
+ 2 \e^{-4\Lambda} \frac{d\Lambda}{dr} \pi' \right) \nn
&& + \frac{1}{2} \left( -  \frac{\e^{- \Phi + \Lambda}}{r^2} \left( \frac{d}{dr} 
\left( r^2 \e^{\Phi - \Lambda} \right) \right)
+ \frac{d}{dr} \right) \left( G_3 (r, \e^{-2\Lambda}) \e^{-2\Lambda} \right) 
 - \frac{1}{2} \e^{-2\Lambda} \frac{d}{dr} \left( G_3 (r, \e^{-2\Lambda}) \right) \nn 
&& -5 \frac{\partial G_4 (r, \e^{-2\Lambda})}{\partial X} \frac{d \Phi}{dr} \frac{\e^{-4 \Lambda}}{r} 
 - \frac{5}{2} \frac{\partial G_4 (r, \e^{-2\Lambda})}{\partial X} \frac{\e^{-4 \Lambda}}{r^2} 
 - 2 \frac{\partial^2 G_4 (r, \e^{-2\Lambda})}{\partial X^2} \frac{d \Phi}{dr} \frac{\e^{-6 \Lambda}}{r} 
 - \frac{\partial^2 G_4 (r, \e^{-2\Lambda})}{\partial X^2} \frac{\e^{-6 \Lambda}}{r^2} \nonumber \\
&& - \frac{3}{2} G_4 (r, \e^{-2\Lambda}) \frac{d \Phi}{dr} \frac{\e^{-2 \Lambda}}{r} 
 - \frac{3}{4} G_4 (r, \e^{-2\Lambda}) \frac{\e^{-2 \Lambda}}{r^2} 
+ \frac{1}{4} \frac{(G_4 (r, \e^{-2\Lambda})}{r^2} 
 -\frac{\partial G_4 (r, \e^{-2\Lambda})}{\partial X} \frac{d \Phi}{dr}
\frac{\e^{-4 \Lambda}}{r} \nn
&& + \frac{1}{2} \frac{\partial G_4 (r, \e^{-2\Lambda})}{\partial X}
\frac{\e^{-2 \Lambda}}{r^2} 
 -\frac{1}{2} \frac{\partial G_4 (r, \e^{-2\Lambda})}{\partial X} \frac{\e^{-2 \Lambda}}{r^2} 
 - \frac{7}{2} \frac{\partial G_5 (r, \e^{-2\Lambda})}{\partial X} \frac{\e^{-6 \Lambda}}{r^2} 
\frac{d \Phi}{dr} - \frac{\partial^2 G_5 (r, \e^{-2\Lambda})}{\partial X^2}
\frac{\e^{-8 \Lambda}}{r^2}  \frac{d \Phi}{dr} \nonumber \\
&& + \frac{3}{4} G_5 (r, \e^{-2\Lambda}) \frac{d \Phi}{dr} 
\frac{\e^{-2 \Lambda}}{r^2} - \frac{15}{4} G_5 (r, \e^{-2\Lambda}) \frac{d \Phi}{dr} \frac{\e^{-4 \Lambda}}{r^2} 
+ \frac{1}{2} \frac{\partial G_5 (r, \e^{-2\Lambda})}{\partial X} \frac{d \Phi}{dr} \frac{\e^{-4 \Lambda}}{r^2}
 - \frac{3}{2} \frac{\partial G_5 (r, \e^{-2\Lambda})}{\partial X} \frac{d \Phi}{dr} \frac{\e^{-6 \Lambda}}{r^2} \nn
&& + 12 G_6 (r) \frac{d \Phi}{dr} \frac{\e^{-2 \Lambda}}{r} 
+ 6 G_6 (r) \frac{\e^{-2 \Lambda}}{r^2} - \frac{2 G_6 (r)}{r^2} 
 - 6 G_7 (r) \frac{d \Phi}{dr} \frac{\e^{-2 \Lambda}}{r^2} 
+ 30 G_7 (r) \frac{d \Phi}{dr} \frac{\e^{-4 \Lambda}}{r^2} \nn
&& + 16 \frac{d G_8 (r)}{dr} \frac{1}{r^2} \frac{d \Phi}{dr} -48 \frac{dG_8 (r)}{dr}
\frac{1}{r^2} \frac{d \Phi}{dr} \e^{-2 \Lambda}\, . 
\end{eqnarray}
As in (\ref{Grec3}), if we wrote 
\be
\label{SchR1}
G_2 (\pi, X) = \sum_{n=0}^\infty {\tilde G}_2^{(n)} (\pi)\left(1 - \e^{2\Lambda(r=\phi)}X\right)^n\, .
\ee
we find
\be
\label{SchR2}
G_2 (r, \e^{-2\Lambda}) = {\tilde G}_2^{(0)} (r)\, ,\quad 
\frac{\partial G_2 (r, \e^{-2\Lambda})}{\partial X} 
= {\tilde G}_2^{(1)} (r)\, .
\ee
Then Eqs.~(\ref{Sch37_00}) and (\ref{Sch38-00}) can be rewritten as 
\begin{eqnarray}
\label{Sch39-00}
{\tilde G}_2^{(0)} (r) 
& = & F^{(0)}(r) \nn 
&\equiv& 2\e^{-2\Phi} \left[ \frac{1}{2\kappa^2 r^2} \e^{2 \Phi} \frac{d}{dr} 
\left[ r(1- \e^{-2 \Lambda})  \right]
\frac{\e^{2\Phi - 2\Lambda}}{2} 
\frac{d}{dr} \left( G_3 (r, \e^{-2\Lambda}) \e^{-2\Lambda} \right) \right. \nn
&& -5 \frac{\partial G_4 (r, \e^{-2\Lambda})}{\partial X} \frac{d \Lambda}{dr}
\frac{\e^{2 \Phi - 6 \Lambda}}{r} +  \frac{d}{dr}  
\left( \frac{\partial G_4 (r, \e^{-2\Lambda})}{\partial X}  \right)
\frac{\e^{2 \Phi -6 \Lambda}}{r} 
+ \frac{1}{2} \frac{\partial G_4 (r, \e^{-2\Lambda})}{\partial X}  
\frac{\e^{2 \Phi -6 \Lambda}}{r^2} \nn
&& - \frac{3}{2}  G_4 (r, \e^{-2\Lambda}) \frac{d \Lambda}{dr} \frac{\e^{2 \Phi -4 \Lambda}}{r}
+ \frac{1}{2} \frac{d G_4 (r, \e^{-2\Lambda})}{dr} \frac{\e^{2 \Phi-4 \Lambda}}{r} 
+ \frac{1}{4} G_4 (r, \e^{-2\Lambda}) \frac{\e^{2 \pi - 4 \Lambda}}{r^2} 
+ \frac{1}{4} G_4 (r, \e^{-2\Lambda}) \frac{\e^{-4 \Phi -2 \Lambda}}{r^2} \nn
&& + \frac{1}{2} \frac{d}{dr} \left( \frac{\partial G_5 (r, \e^{-2\Lambda})}{\partial X}
\right) \frac{\e^{2\Phi -8 \Lambda}}{r^2} 
 - \frac{7}{2} \frac{\partial G_5 (r, \e^{-2\Lambda})}{\partial X} 
\frac{d \Lambda}{dr} \frac{\e^{2 \Phi -8 \Lambda}}{r^2} 
+\frac{3}{4} G_5 (r, \e^{-2\Lambda}) \frac{d \Lambda}{dr} \frac{\e^{2 \Phi -4 \Lambda}}{r^2} \nn
&& - \frac{15}{4} G_5 (r, \e^{-2\Lambda}) \frac{d \Lambda}{dr} \frac{\e^{2 \Phi -6 \Lambda}}{r^2} 
 - \frac{1}{4} \frac{d G_5 (r, \e^{-2\Lambda})}{dr} \frac{\e^{2 \Phi -4 \Lambda}}{r^2} 
+ \frac{3}{4} \frac{d G_5 (r, \e^{-2\Lambda})}{dr} \frac{\e^{2 \Phi -6 \Lambda}}{r^2} \nn
&& + 12 G_6 (r) \frac{d \Lambda}{dr} \frac{\e^{2 \Phi -4 \Lambda}}{r}
 - 4 \frac{d G_6 (r)}{dr} \frac{\e^{2 \Phi-4 \Lambda}}{r} 
 -2 G_6 (r) \frac{\e^{2 \Phi - 4 \Lambda}}{r^2} 
 -2 G_6 (r) \frac{\e^{-4 \Phi -2 \Lambda}}{r^2} \nn
&& -6  G_7 (r) \frac{d \Lambda}{dr} \frac{\e^{2 \Phi -4 \Lambda}}{r^2} 
+ 30 G_7 (r) \frac{d \Lambda}{dr} \frac{\e^{2 \Phi -6 \Lambda}}{r^2} 
+ 2 \frac{d G_7 (r)}{dr} \frac{\e^{2 \Phi -4 \Lambda}}{r^2} 
 -6 \frac{d G_7 (r)}{dr} \frac{\e^{2 \Phi -6 \Lambda}}{r^2} \nn
&& -192 \frac{d G_8 (r)}{dr} \frac{\e^{2 \Phi -2 \Lambda}}{r^2} 
\frac{d \Phi}{dr} + 16 \frac{d G_8 (r)}{dr} \frac{\e^{2 \Phi -2 \Lambda}}{r^2} 
\frac{d \Lambda}{dr}
 -16 \frac{d^2 G_8 (r)}{dr^2} \frac{\e^{2 \Phi -2 \Lambda}}{r^2}
 -48 \frac{d G_8 (r)}{dr} \frac{\e^{2 \Phi-4 \Lambda}}{r^2} \frac{d
\Lambda}{dr} \nn
&& \left. +16 \frac{d^2 G_8 (r)}{dr^2} \frac{\e^{2 \Phi -4 \Lambda}}{r^2} \right] \, , \\
\label{Sch40-00}
{\tilde G}_2^{(1)} (r) & = & F^{(1)}(r) \nn
& \equiv & \frac{1}{2} \e^{2\Lambda} F^{(0)}(r) + \frac{1}{2\kappa^2}\left\{
- \frac{1}{r^2} \e^{2 \Lambda} (1 - \e^{-2 \Lambda}) +
\frac{2}{r} \frac{d \Phi}{dr}\right\} \nn
&& + \frac{1}{2}\e^{2\Lambda} G_3 (r, \e^{-2\Lambda}) 
\left( \frac{\e^{-3\Lambda}}{r^2} \left( \frac{d}{dr} \left(r^2 
\e^{\Phi - \Lambda} \right)\right)
+ \e^{-4\Lambda} \frac{d\Lambda}{dr} \right) \nn
&& - \frac{\partial G_3 (r, \e^{-2\Lambda})}{\partial X} 
\left( \frac{\e^{-3\Lambda}}{r^2} \left( \frac{d}{dr} \left(r^2 
\e^{\Phi - \Lambda} \right)\right)
+ \e^{-4\Lambda} \frac{d\Lambda}{dr} \right) \nn
&& - G_3 (r, \e^{-2\Lambda}) \left( \left( \frac{\e^{- \Phi - \Lambda}}{r^2} 
\left( \frac{d}{dr} \left( r^2 \e^{\Phi - \Lambda} \right) \right) \right)
+ 2 \e^{-4\Lambda} \frac{d\Lambda}{dr} \pi' \right) \nn
&& + \frac{1}{2} \left( -  \frac{\e^{- \Phi + \Lambda}}{r^2} \left( \frac{d}{dr} 
\left( r^2 \e^{\Phi - \Lambda} \right) \right)
+ \frac{d}{dr} \right) \left( G_3 (r, \e^{-2\Lambda}) \e^{-2\Lambda} \right) 
 - \frac{1}{2} \e^{-2\Lambda} \frac{d}{dr} \left( G_3 (r, \e^{-2\Lambda}) \right) \nn 
&& -5 \frac{\partial G_4 (r, \e^{-2\Lambda})}{\partial X} \frac{d \Phi}{dr} \frac{\e^{-4 \Lambda}}{r} 
 - \frac{5}{2} \frac{\partial G_4 (r, \e^{-2\Lambda})}{\partial X} \frac{\e^{-4 \Lambda}}{r^2} 
 - 2 \frac{\partial^2 G_4 (r, \e^{-2\Lambda})}{\partial X^2} \frac{d \Phi}{dr} \frac{\e^{-6 \Lambda}}{r} 
\nonumber \\
&& - \frac{\partial^2 G_4 (r, \e^{-2\Lambda})}{\partial X^2} \frac{\e^{-6 \Lambda}}{r^2} 
 - \frac{3}{2} G_4 (r, \e^{-2\Lambda}) \frac{d \Phi}{dr} \frac{\e^{-2 \Lambda}}{r} 
 - \frac{3}{4} G_4 (r, \e^{-2\Lambda}) \frac{\e^{-2 \Lambda}}{r^2} 
+ \frac{1}{4} \frac{(G_4 (r, \e^{-2\Lambda})}{r^2}  \nn
&& -\frac{\partial G_4 (r, \e^{-2\Lambda})}{\partial X} \frac{d \Phi}{dr} \frac{\e^{-4 \Lambda}}{r} 
+ \frac{1}{2} \frac{\partial G_4 (r, \e^{-2\Lambda})}{\partial X}
\frac{\e^{-2 \Lambda}}{r^2} 
 -\frac{1}{2} \frac{\partial G_4 (r, \e^{-2\Lambda})}{\partial X} \frac{\e^{-2 \Lambda}}{r^2} 
\nonumber \\
&&  - \frac{7}{2} \frac{\partial G_5 (r, \e^{-2\Lambda})}{\partial X} \frac{\e^{-6 \Lambda}}{r^2} 
\frac{d \Phi}{dr} - \frac{\partial^2 G_5 (r, \e^{-2\Lambda})}{\partial X^2}
\frac{\e^{-8 \Lambda}}{r^2}  \frac{d \Phi}{dr} 
+ \frac{3}{4} G_5 (r, \e^{-2\Lambda}) \frac{d \Phi}{dr} 
\frac{\e^{-2 \Lambda}}{r^2} \nn
&& - \frac{15}{4} G_5 (r, \e^{-2\Lambda}) \frac{d \Phi}{dr} \frac{\e^{-4 \Lambda}}{r^2} 
+ \frac{1}{2} \frac{\partial G_5 (r, \e^{-2\Lambda})}{\partial X} \frac{d \Phi}{dr} \frac{\e^{-4 \Lambda}}{r^2}
 - \frac{3}{2} \frac{\partial G_5 (r, \e^{-2\Lambda})}{\partial X} \frac{d \Phi}{dr} \frac{\e^{-6 \Lambda}}{r^2} \nn
&& + 12 G_6 (r) \frac{d \Phi}{dr} \frac{\e^{-2 \Lambda}}{r} 
+ 6 G_6 (r) \frac{\e^{-2 \Lambda}}{r^2} - \frac{2 G_6 (r)}{r^2} 
 - 6 G_7 (r) \frac{d \Phi}{dr} \frac{\e^{-2 \Lambda}}{r^2} 
+ 30 G_7 (r) \frac{d \Phi}{dr} \frac{\e^{-4 \Lambda}}{r^2} \nn
&& + 16 \frac{d G_8 (r)}{dr} \frac{1}{r^2} \frac{d \Phi}{dr} -48 \frac{dG_8 (r)}{dr}
\frac{1}{r^2} \frac{d \Phi}{dr} \e^{-2 \Lambda}\, . 
\end{eqnarray}
Then arbitrary geometry given by $\Phi(r)$ and $\Lambda(r)$ can be realized 
for arbitrary functions $G_i$ $\left(i=3,4,\cdots,8\right)$ 
by choosing ${\tilde G}_2^{(0)} (r)$ and ${\tilde G}_2^{(1)} (r)$ as in 
(\ref{Sch39-00}) and (\ref{Sch40-00}). 
This may also tell that the spherical symmetric solution could have fourth hair corresponding 
to the scalar field in addition to the usual three hairs corresponding to mass, angular momentum, 
and electric charge. 
Then only ${\tilde G}_2^{(0)}$ and ${\tilde G}_2^{(1)}$ in $G_2 (\pi, X) $ are relevant for the reconstruction 
of the spherically symmetric and static solution and ${\tilde G}_2^{(n)} (\pi)$, $n=2,3,4,\cdots $ are irrelevant 
for the reconstruction although they may be related with the stability of the reconstructed solution. 

As we have succeeded in the reconstruction of arbitrary spherically symmetric and static 
geometry, we may show the reconstruction is compatible with the cosmological reconstruction 
in the last section. 
Let us assume $G_i^{(n)}$, $n=0,1,\cdots, N_i$ in (\ref{Grec3}) are relevant for the cosmological 
reconstruction and $G_i^{(n)}$, $n=N_i+1,N_i+2,\cdots$ are irrelevant. 
We also assume ${\tilde G}_i^{(n)}$, $n=0,1,\cdots {\tilde N}_i$ as in (\ref{SchR1}) 
\be
\label{SchRi}
G_i (\pi, X) = \sum_{n=0}^\infty {\tilde G}_i^{(n)} (\pi)
\left(1 - \e^{2\Lambda(\phi)}X\right)^n\, .
\ee
are relevant for the reconstruction of spherically symmetric and static geometry 
and ${\tilde G}_i^{(n)}$, $n={\tilde N}_i + 1, {\tilde N}_i +2,\cdots$ are irrelevant. 
By expanding (\ref{Grec3}) as a power series of $1 - \e^{2\Lambda(\phi)}X$ by 
using $1-X=1+\e^{-2\Lambda(\phi)} - \e^{-2\Lambda(\phi)}\left(1 - \e^{2\Lambda(\phi)}X\right)$, 
we find
\bea
\label{SchRi1}
G_i (\pi, X) &=& \sum_{n=0}^\infty \sum_{k=0}^n
G_i^{(n)}(\pi) \, _n C_k \left( 1+\e^{-2\Lambda(\phi)} \right)^{n-k} 
(-1)^k \e^{-2k\Lambda(\phi)}\left(1 - \e^{2\Lambda(\phi)}X\right)^k \nn
&=& \sum_{n=0}^\infty (-1)^n \e^{-2n\Lambda(\phi)} \sum_{k=n}^\infty G_i^{(k)}(\pi) 
\, _k C_n \left( 1+\e^{-2\Lambda(\phi)} \right)^{k-n} 
\left(1 - \e^{2\Lambda(\phi)}X\right)^n\, .
\eea
By comparing the expression in (\ref{SchRi1}) and (\ref{SchRi}), we find
\be
\label{SchRi2}
{\tilde G}_i^{(n)} (\pi)= (-1)^n \e^{-2n\Lambda(\phi)}
\sum_{k=n}^\infty\, _k C_n \left( 1+\e^{-2\Lambda(\phi)} \right)^{k-n}  G_i^{(k)}(\pi)\, .
\ee
Then when $n<N_i$, we find 
\bea
\label{SchRi3}
&& {\tilde G}_i^{(n)} (\pi) - (-1)^n \e^{-2n\Lambda(\phi)}
\sum_{k=n}^N\, _k C_n \left( 1+\e^{-2\Lambda(\phi)} \right)^{k-n}  G_i^{(k)}(\pi) \nn
&& = (-1)^n \e^{-2n\Lambda(\phi)}
\sum_{k=N_i+1}^\infty\, _k C_n \left( 1+\e^{-2\Lambda(\phi)} \right)^{k-n}  G_i^{(k)}(\pi)\, , 
\eea
and when $n\geq N_i$
\be
\label{SchRi4}
{\tilde G}_i^{(n)} (\pi) 
= (-1)^n \e^{-2n\Lambda(\phi)}
\sum_{k=N_i+1}^\infty\, _k C_n \left( 1+\e^{-2\Lambda(\phi)} \right)^{k-n}  G_i^{(k)}(\pi)\, .
\ee
For a set given by $G_i^{(n)}(\pi)$, $n=0,1,\cdots N_i$ and 
${\tilde G}_i^{(n)}(\pi)$, $n=0,1,\cdots {\tilde N}_i$, 
we can always choose $G_i^{(n)}(\pi)$, $n=N_i,N_i+1,\cdots, N_i + {\tilde N}_i$ to satisfy (\ref{SchRi2}) 
(we may set $G_i^{(n)}(\pi)=0$, $n=N_i + {\tilde N}_i,N_i + {\tilde N}_i+1,\cdots$ and 
${\tilde G}_i^{(n)}(\pi)=0$, $n={\tilde N}_i+1,{\tilde N}_i+2,\cdots$). 
Therefore we can obtain an action generating both an arbitrary spherically symmetric and static 
geometry and an arbitrary expansion history of the universe, simultaneously. 

\section{Summary \label{V}}

In this paper, we gave explicit formulae of the equations in the generalized Galileon models. 
Even in generalized Galileon models, the derivatives higher than two do not appear 
in the field equation of motion nor the Einstein equation. These structures are clearly given by using the Levi-Civita symbol 
$\epsilon^{\mu\nu\rho\sigma}$. 
We have also developed the formulation of the reconstruction. 
As a consequence, we can construct an explicit action, 
which reproduces an arbitrary development of the expansion of the universe. 
We should note that the functions $G_n$'s can be completely and explicitly separate 
to the parts relevant for the expansion and the irrelevant parts. 
Some of the irrelevant parts are related with the stability of the reconstructed solution and 
we can also identify which parts in $G_n$'s are relevant for the stability.  
Then we have succeeded to show the conditions how the reconstructed solution 
becomes stable and therefore it becomes an attractor solution. 
Working in the spherically symmetric and static space-time, we 
have investigated how the Vainshtein mechanism works. 
It has been also shown that arbitrary spherically symmetric and static geometry 
can be realized by properly choosing $G_i$'s, which may tell that the solution could have fourth hair corresponding 
to the scalar field. We again identified the parts in in $G_2 (\pi, X) $ relevant for the reconstruction. 
We should note that by choosing $G_i\left(\pi,X\right)$ appropriately, we can obtain an action which has both 
the solution corresponding to an arbitrarily given spherically symmetric and static 
geometry and the solution an arbitrarily given expansion history of the universe. 

The generalized Galileon model contains many functions denoted by $G_i$, which cannot 
be determined by the expansion history of the universe, the stability of the reconstructed solution, nor 
the condition that the Vainshtein mechanism works. 
These function could be restricted if we consider the developments of the several kinds 
of fluctuations by density perturbation, etc. The analysis by the perturbation could be very complicated in the models but 
it might become possible by the development of the technologies. 

\section*{Acknowledgments}

We are grateful to S.~D.~Odintsov for the discussion when he stayed in 
Nagoya University. 
K.B. would like to sincerely acknowledge very kind and warm hospitality 
at Eurasian National University very much, where this work has been completed. 
This research has been supported in part
by Global COE Program of Nagoya University (G07)
provided by the Ministry of Education, Culture, Sports, Science \&
Technology and by the JSPS Grant-in-Aid for Scientific Research (S) \# 22224003
and (C) \# 23540296 (SN).

\appendix

\section{Explicit forms of $\mathcal{E}_k$ and $(\mathcal{H}_k)_{\mu\nu}$ in the FRW universe \label{Ap1}}

In this appendix, by assuming the FRW universe (\ref{Gl3}), we give explicit forms of $\mathcal{E}_k$ 
in (\ref{GG4}), (\ref{Delta5}), and (\ref{GG7}), 
and $(\mathcal{H}_k)_{\mu\nu}$ in (\ref{ma23a}). 

In the FRW universe (\ref{Gl3}), we have
\bea
\label{GG9}
&& \Gamma^t_{ij}=a^2 H \delta_{ij}\, ,\quad \Gamma^i_{jt}=\Gamma^i_{tj}=H\delta^i_{\ j}\, ,\nn
&& R_{itjt} = - \left(\dot H + H^2\right)a^2\delta_{ij}\, ,\quad 
R_{ijkl} = a^4 H^2\left(\delta_{ik} \delta_{lj} - \delta_{il} \delta_{kj}\right)\, .
\eea
Then we find
\bea
\label{GG10}
\mathcal{E}_1 &=& 1\, , \nn
\mathcal{E}_2 &=&
2 \ddot{\pi} + 6 H \dot{\pi}\, , \nn
\mathcal{E}_3 &=&
-18 H \dot{\pi} \ddot{\pi} -27 H^2 \dot{\pi}^2 -9 \dot{H} \dot{\pi}^2\, , \nn
\mathcal{E}_4^{(0)} &=&
 36 H^2 \dot{\pi}^2 \ddot{\pi} + 36 H^3 \dot{\pi}^3 + 30 H \dot{H}
\dot{\pi}^3 + \mbox{(third derivative terms)}\, , \nn
\Delta \mathcal{E}_4 &=&
18 H^2 \dot{\pi}^2 \ddot{\pi} + 18 H^3 \dot{\pi}^3 + 6 H \dot{H}
\dot{\pi}^3 + \mbox{(third derivative terms)}\, , \nn
\mathcal{E}_5^{(0)} &=&
-20 H^3 \dot{\pi}^3 \ddot{\pi} -15 H^4 \dot{\pi}^4 -21 H^2 \dot{H} \dot{\pi}^4 + \mbox{(third derivative terms)}\, , \nn
\Delta \mathcal{E}_5 &=&
-30 H^3 \dot{\pi}^3 \ddot{\pi} - \frac{45}{2} H^4 \dot{\pi}^4 - \frac{33}{2} H^2 \dot{H} \dot{\pi}^4 
+ \mbox{(third derivative terms)}\, , \nn
\mathcal{E}_6 &=&
24 H^2 \ddot{\pi} +72 H^3 \dot{\pi}+48 H \dot{H} \dot{\pi}\, , \nn
\mathcal{E}_7 &=&
-72 H^3 \dot{\pi} \ddot{\pi} -108 H^4 \dot{\pi}^2 -108 H^2 \dot{H} \dot{\pi}^2\, .
\eea
We also obtain
\bea
\label{HH0}
(\mathcal{H}_2)_{00} &=& - \frac{1}{2} G_2(\pi,X) - \frac{\partial G_2(\pi,X)}{\partial X} \dot{\pi}^2 \, ,\nn
(\mathcal{H}_2)_{ij} &=& \frac{1}{2} G_2(\pi,X) a^2 \delta_{ij} \, , \\
\label{HH1}
(\mathcal{H}_3)_{00} &=& 3 \frac{\partial G_3(\pi,X)}{\partial X} H \dot{\pi}^5 - \frac{9}{2} G_3(\pi,X) H
\dot{\pi}^3\, ,\nn
(\mathcal{H}_3)_{ij} &=& \left[ - \frac{1}{2} \frac{dG_3(\pi,X)}{dt} 
\dot{\pi}^3 - \frac{3}{2} G_3(\pi,X) \dot{\pi}^2 \ddot{\pi} \right] a^2 \delta_{ij} \, , \\
\label{HH3}
(\mathcal{H}_4)_{00} &=& - 6 \frac{\partial G_4(\pi,X)}{\partial X} H^2 \dot{\pi}^4 
 - 3 \frac{\partial^2 G_4(\pi,X)}{\partial X^2} H^2 \dot{\pi}^6
+\frac{9}{4} G_4(\pi,X) H^2 \dot{\pi}^2 \, , \nn
(\mathcal{H}_4)_{ij} &=& \left[ - \frac{3}{2} \frac{\partial G_4(\pi,X)}{\partial X} H^2 \dot{\pi}^4
 - \frac{d}{dt} \left( \frac{\partial G_4(\pi,X)}{\partial X}  \right) H
\dot{\pi}^4 - \frac{\partial G_4(\pi,X)}{\partial X} \dot{H} \dot{\pi}^4 - 4
\frac{\partial G_4(\pi,X)}{\partial X} H \dot{\pi}^3 \ddot{\pi} \right. \nn
&& \left. +\frac{3}{4} G_4(\pi,X) H^2 \dot{\pi}^2 + \frac{1}{2} \frac{d G_4(\pi,X)}{dt} 
H \dot{\pi}^2 +\frac{1}{2} G_4(\pi,X) \dot{H} \dot{\pi}^2
+ G_4(\pi,X) H \dot{\pi} \ddot{\pi} \right] a^2 \delta_{ij}\, , \\
(\mathcal{H}_5)_{00} &=&  - 5 \frac{\partial G_5(\pi,X)}{\partial X} H^3 \dot{\pi}^5
+ \frac{\partial^2 G_5(\pi,X)}{\partial X^2} H^3 \dot{\pi}^7 
+ \frac{15}{4} G_5(\pi,X) H^3 \dot{\pi}^3 \, , \nn
(\mathcal{H}_5)_{ij} &=& \left[ \frac{\partial G_5(\pi,X)}{\partial X} H^3 \dot{\pi}^5 +\frac{1}{2}
\frac{d}{dt} \left( \frac{\partial G_5(\pi,X)}{\partial X}  \right) H^2 \dot{\pi}^5
+ \frac{\partial G_5(\pi,X)}{\partial X} H \dot{H} \dot{\pi}^5 
+ \frac{5}{2} \frac{\partial G_5(\pi,X)}{\partial X} H^2 \dot{\pi}^4 
\ddot{\pi} \right. \nn
&& \left. - \frac{3}{2} G_5(\pi,X) H^3 \dot{\pi}^3 - \frac{3}{4} \frac{d G_5(\pi,X)}{dt} H^2
\dot{\pi}^3 - \frac{3}{2} G_5(\pi,X) H \dot{H} \dot{\pi}^3 
 - \frac{9}{4} G_5(\pi,X) H^2 \dot{\pi}^2 \ddot{\pi} \right] a^2 \delta_{ij}\, , \\
\label{HH2}
\label{HH4}
(\mathcal{H}_6)_{00} &=& - 18 G_6(\pi) H^2 \dot{\pi}^2\, , \nn  
(\mathcal{H}_6)_{ij} &=& \left[ - 6 G_6(\pi)H^2 \dot{\pi}^2 
 - 4 \frac{d G_6(\pi)}{dt} H \dot{\pi}^2 - 4 G_6(\pi) \dot{H} \dot{\pi}^2
 - 8 G_6(\pi) H \dot{\pi} \ddot{\pi} \right] a^2 \delta_{ij} \, , \\
\label{HH5}
(\mathcal{H}_7)_{00} &=& - 30 G_7(\pi) H^3 \dot{\pi}^3\, , \nn
(\mathcal{H}_7)_{ij} &=& \left[ 12 G_7(\pi) H^3 \dot{\pi}^3 + 6 \frac{d G_7(\pi)}{dt} H^2 \dot{\pi}^3 
+ 12 G_7(\pi) H \dot{H} \dot{\pi}^3 + 18 G_7(\pi) H^2 \dot{\pi}^2 \ddot{\pi} \right] a^2 \delta_{ij}\, , \\
\label{HH6}
(\mathcal{H}_8)_{00} &=& 48 \frac{d G_8(\pi)}{dt} H^3\, , \nn
(\mathcal{H}_8)_{ij} &=&  \left[ - 32 \frac{d G_8(\pi)}{dt} H^3 
 -16 \frac{d^2 G_8(\pi)}{dt^2} H^2 - 32 \frac{dG_8(\pi)}{dt} \dot{H} H \right] a^2 \delta_{ij}\, .
\eea

\section{Explicit forms of $\mathcal{E}_k$ and $(\mathcal{H}_k)_{\mu\nu}$ 
in the static and spherically symmetric space-time \label{Ap2}}

In this appendix, we give explicit forms of $\mathcal{E}_k$ and $(\mathcal{H}_k)_{\mu\nu}$ in the static and spherically 
symmetric space-time, especially in the Schwarzschild space-time. 

In the static and spherically symmetric space-time in (\ref{Sch1}), we find 
\begin{eqnarray}
\label{Sch6}
\mathcal{E}_5^{(0)} &=&
7 \frac{(\pi')^4 \e^{-8 \Lambda}}{r^2} \frac{d^2 \Phi}{dr^2} 
+ 5 \frac{(\pi')^4 \e^{-8 \Lambda}}{r^2} \left( \frac{d \Phi}{dr}
\right)^2
 -41 \frac{(\pi')^4 \e^{-8 \Lambda}}{r^2} \frac{d \Lambda}{dr} 
\frac{d \Phi}{dr}
+2 \frac{(\pi')^4 \e^{-6 \Lambda}}{r^3} \frac{d \Phi}{dr} \nn
&& +20 \frac{(\pi')^3 (\pi'') \e^{-8 \Lambda}}{r^2} \frac{d \Phi}{dr}
 - 4 \frac{(\pi')^4 \e^{-8 \Lambda}}{r^3} \frac{d \Phi}{dr}
+ \mbox{(third derivative terms)}\, , \\
\label{Sch7}
\mathcal{E}_4^{(0)} &=& 10 \frac{ (\pi')^3 \e^{-6 \Lambda}}{r} \frac{d^2
\Phi}{dr^2}
+ 8 \frac{(\pi')^3 \e^{-6 \Lambda}}{r} \left( \frac{d \Phi}{dr} \right)^2
 -44 \frac{(\pi')^3 \e^{-6 \Lambda}}{r} \frac{d \Lambda}{dr} \frac{d
\Phi}{dr}
+24 \frac{(\pi')^2 (\pi'') \e^{-6 \Lambda}}{r} \frac{d \Phi}{dr} \nn
&& +10 \frac{(\pi')^3 \e^{-6 \Lambda}}{r^2} \frac{d \Phi}{dr}
 -22 \frac{(\pi')^3 \e^{-6 \Lambda}}{r^2} \frac{d \Lambda}{dr} 
+ 2 \frac{(\pi')^3 \e^{-4 \Lambda}}{r^3} 
+12 \frac{(\pi')^2 (\pi'') \e^{-6 \Lambda}}{r^2}
 -2 \frac{(\pi')^3 \e^{-6 \Lambda}}{r^3} \nn
&& + \mbox{(third derivative terms)} \, , \\ 
\label{Sch8}
\mathcal{E}_3 &=&
2 (\pi')^2 \e^{-4 \Lambda} \frac{d^2 \Phi}{dr^2} 
+ 2 (\pi')^2 \e^{-4 \Lambda} \left( \frac{d \Phi}{dr}  \right)^2
 -8 (\pi')^2 \e^{-4 \Lambda} \frac{d \Lambda}{dr} \frac{d \Phi}{dr} 
+ 12 \frac{(\pi')^2 \e^{-4 \Lambda}}{r} \frac{d \Phi}{dr} \nn
&& +6 (\pi')(\pi'') \e^{-4 \Lambda} \frac{d \Phi}{dr}
-16 \frac{(\pi')^2 \e^{-4 \Lambda}}{r} \frac{d \Lambda}{dr}
+12 \frac{(\pi')(\pi'') \e^{-4 \Lambda}}{r} 
+ 6 \frac{(\pi')^2 \e^{-4 \Lambda}}{r^2}\, ,\\
\label{Sch9}
\mathcal{E}_2 &=&
 -2 (\pi'') \e^{-2 \Lambda}
 -2 (\pi') \e^{-2 \Lambda} \frac{d \Phi}{dr} +2 (\pi') \e^{-2 \Lambda}
\frac{d \Lambda}{dr} - 4 \frac{(\pi') \e^{-2 \Lambda}}{r}\, ,\\
\label{Sch10}
\mathcal{E}_6 &=& 
 - 16 \frac{(\pi') \e^{-4 \Lambda}}{r} \frac{d^2 \Phi}{dr^2} 
 -16 \frac{(\pi') \e^{-4 \Lambda}}{r} \left ( \frac{d \Phi}{dr} \right)^2
+ 48 \frac{(\pi') \e^{-4 \Lambda}}{r} \frac{d \Lambda}{dr} 
\frac{d\Phi}{dr}
+ 8 \frac{(\pi') \e^{-2 \Lambda}}{r^2} \frac{d \Phi}{dr} \nn
&& - 16 \frac{(\pi'') \e^{-4 \Lambda}}{r} \frac{d \Phi}{dr}
 - 24 \frac{(\pi') \e^{-4 \Lambda}}{r^2} \frac{d \Phi}{dr}
 -8 \frac{(\pi') \e^{-2 \Lambda}}{r^2} \frac{d \Lambda}{dr}
+24 \frac{(\pi') \e^{-4 \Lambda}}{r^2} \frac{d \Lambda}{dr} \nn 
&& +8 \frac{(\pi'') \e^{-2 \Lambda}}{r^2}
 -8 \frac{(\pi'') \e^{-4 \Lambda}}{r^2}\, , \\
\label{Sch11}
\mathcal{E}_7 &=&
12 \frac{(\pi')^2 \e^{-4 \Lambda}}{r^2} \frac{d^2 \Phi}{dr^2}
 -36 \frac{(\pi')^2 \e^{-6 \Lambda}}{r^2} \frac{d^2 \Phi}{dr^2}
+12 \frac{(\pi')^2 \e^{-4 \Lambda}}{r^2} \left( \frac{d \Phi}{dr} \right)^2
-36 \frac{(\pi')^2 \e^{-6 \Lambda}}{r^2} \left( \frac{d \Phi}{dr} \right)^2 \nn 
&& -36 \frac{(\pi')^2 \e^{-4 \Lambda}}{r^2} \frac{d \Lambda}{dr} 
\frac{d\Phi}{dr} +180 \frac{(\pi')^2 \e^{-6 \Lambda}}{r^2} \frac{d \Lambda}{dr} 
\frac{d\Phi}{dr} 
+24 \frac{(\pi') (\pi'') \e^{-4 \Lambda}}{r^2} \frac{d \Phi}{dr}
 -72 \frac{(\pi') (\pi'') \e^{-6 \Lambda}}{r^2} \frac{d \Phi}{dr}\, .
\end{eqnarray}
Here $\pi'\equiv \dfrac{d \pi}{dr}$ and $\pi''\equiv \dfrac{d^2 \pi}{dr^2}$.
Especially for the Schwarzschild metric (\ref{Sch12}), we have 
\begin{eqnarray}
\label{Sch13}
\mathcal{E}_5^{(0)} &=&
\frac{16 (\pi')^4 GM (2GM -r)^2(3GM -r)}{r^8} - \frac{20 (\pi')^3 GM
(2 GM -r)^2 (2 \pi'' r GM -\pi' GM -\pi'' r^2)}{r^8} \nn
&& + \mbox{(third derivative terms)}\, , \\
\label{Sch14}
\Delta \mathcal{E}_5 &=&
 - \frac{96 (\pi')^3 \pi'' G^4 M^4}{r^7} + \frac{138 (\pi')^4 G^4 M^4}{r^8}
+ \frac{200 (\pi')^3 \pi'' G^3 M^3}{r^6} - \frac{182 (\pi')^4 G^3 M^3}{r^7}
 - \frac{116 (\pi')^3 \pi'' G^2M^2}{r^5} \nn
&& + \frac{117 (\pi')^4G^2 M^2}{2x^6}
+ \frac{20 (\pi')^3 \pi'' GM}{r^4} - \frac{4 (\pi')^4 GM}{r^5} \nn
&& + \mbox{(third derivative terms)}\, , \\
\label{Sch15}
\mathcal{E}_7  &=&
 - \frac{288 \pi' \pi'' G^3 M^3}{r^6} + \frac{576 (\pi')^2 G^3 M^3}{r^7}
+ \frac{240 \pi' \pi'' G^2 M^2}{r^5} - \frac{360 (\pi')^2 G^2 M^2}{r^6} \nn
&& - \frac{48 \pi' \pi'' GM}{r^4} + \frac{48 (\pi')^2GM}{r^5}\, , \\
\label{Sch16}
\mathcal{E}_4^{(0)} &=&
\frac{12 (\pi')^2 (2GM-r)(2 \pi' G^2 M^2 +2 \pi'' r^2 GM-2 \pi' rGM
 -\pi'' r^3)}{r^6} - \frac{8 (\pi')^3 GM (2GM -r)(3GM-r)}{r^6} \nn
&& + \mbox{(third derivative terms)}\, , \\
\label{Sch17}
\Delta \mathcal{E}_4 &=& 0 + \mbox{(third derivative terms)}\, , \\
\label{Sch18}
\mathcal{E}_6 &=&0\, , \\
\label{Sch19}
\mathcal{E}_3 &=& 
\frac{36 \pi' \pi'' G^2 M^2}{r^3} - \frac{18 (\pi')^2 G^2M^2}{r^4} 
 - \frac{42 \pi' \pi'' GM}{r^2} +12 \frac{\pi' \pi'' }{r} 
+ \frac{6 (\pi')^2}{r^2}\, ,\\
\label{Sch20}
\mathcal{E}_2 &=&
 -2 \left ( 1- 2 \frac{GM}{r}  \right) \pi'' + \frac{4}{r^2} (GM-r)
\pi' \, .
\end{eqnarray}
Since 
\begin{eqnarray}
\label{Sch21}
\Delta \mathcal{E}_4 &=&
\frac{2 (\pi')^3 \e^{-6 \Lambda}}{r} \frac{d^2 \Phi}{dr^2} 
+ \frac{4 (\pi')^3 \e^{-6 \Lambda}}{r} \left( \frac{d \Phi}{dr} \right)^2
 - \frac{16 (\pi')^3 \e^{-6 \Lambda}}{r} \frac{d \Lambda}{dr} \frac{d \Phi}{dr}
 - 2 \frac{(\pi')^3 \e^{-4 \Lambda}}{r^2} \frac{d \Phi}{dr} \nn
&& + \frac{12 (\pi')^2 (\pi'') \e^{-6 \Lambda}}{r} \frac{d \Phi}{dr}
+ \frac{8 (\pi')^3 \e^{-6 \Lambda}}{r^2} \frac{d \Phi}{dr} 
+ \frac{6 (\pi')^3 \e^{-4 \Lambda}}{r^2} \frac{d \Lambda}{dr} 
 - \frac{8 (\pi')^3 \e^{-6 \Lambda}}{r^2} \frac{d \Lambda}{dr} \nn
&& - \frac{6 (\pi')^2 (\pi'') \e^{-4 \Lambda}}{r^2}
 - \frac{2 (\pi')^3 \e^{-4 \Lambda}}{r^3} 
+ \frac{6 (\pi')^2 (\pi'') \e^{-6 \Lambda}}{r^2} \nn
&& + \mbox{(third derivative terms)}\, , \\
\label{Sch22}
\Delta \mathcal{E}_5 &=&
 - \frac{7}{4} \frac{(\pi')^4 \e^{-6 \Lambda}}{r^2} \frac{d^2 \Phi}{dr^2} 
+ \frac{19}{4} \frac{(\pi')^4 \e^{-8 \Lambda}}{r^2} \frac{d^2 \Phi}{dr^2}
 - \frac{7}{4} \frac{(\pi')^4 \e^{-6 \Lambda}}{r^2} \left( 
\frac{d \Phi}{dr} \right)^2
+ \frac{27}{4} \frac{(\pi')^4 \e^{-8 \Lambda}}{r^2} \left( 
\frac{d \Phi}{dr} \right)^2 \nn
&& + \frac{6 (\pi')^4 \e^{-4 \Lambda}}{r^2} \frac{d \Lambda}{dr} 
\frac{d \Phi}{dr}
 - \frac{49}{4} \frac{(\pi')^4 \e^{-6 \Lambda}}{r^2} \frac{d \Lambda}{dr}
\frac{d \Phi}{dr} 
 - \frac{105}{4} \frac{(\pi')^4 \e^{-8 \Lambda}}{r^2} \frac{d \Lambda}{dr}
\frac{d \Phi}{dr}
 - \frac{6 (\pi')^3 (\pi'') \e^{-4 \Lambda}}{r^2} \frac{d \Phi}{dr} \nn
&& + \frac{14 (\pi')^3 (\pi'') \e^{-6 \Lambda}}{r^2} \frac{d \Phi}{dr}
 - \frac{2 (\pi')^4 \e^{-6 \Lambda}}{r^3} \frac{d \Phi}{dr}
+ \frac{12 (\pi')^3 (\pi'') \e^{-8 \Lambda}}{r^2} \frac{d \Phi}{dr}
+ \frac{4 (\pi')^4 \e^{-8 \Lambda}}{r^3} \frac{d \Phi}{dr} \nn
&& + \mbox{(third derivative terms)}\, , \ .
\end{eqnarray}
we acquire 
\begin{eqnarray}
\label{Sch23}
\mathcal{E}_5 &=& 
 - \frac{7}{4} \frac{(\pi')^4 \e^{-6 \Lambda}}{r^2} \frac{d^2 \Phi}{dr^2}
+ \frac{47}{4} \frac{(\pi')^4 \e^{-8 \Lambda}}{r^2} \frac{d^2 \Phi}{dr^2}
 - \frac{7}{4} \frac{(\pi')^4 \e^{-6 \Lambda}}{r^2} \left( 
\frac{d \Phi}{dr}  \right)^2
+ \frac{47}{4} \frac{(\pi')^4 \e^{-8 \Lambda}}{r^2} \left( 
\frac{d \Phi}{dr} \right)^2 \nn
&& + \frac{6 (\pi')^4 \e^{- 4 \Lambda}}{r^2} \frac{d \Lambda}{dr} 
\frac{d \Phi}{dr}
 - \frac{49}{4} \frac{(\pi')^4 \e^{-6 \Lambda}}{r^2}
 - \frac{269}{4} \frac{(\pi')^4 \e^{-8 \Lambda}}{r^2} \frac{d \Lambda}{dr}
\frac{d \Phi}{dr}
 - \frac{6 (\pi')^3 (\pi'') \e^{-4 \Lambda}}{r^2} \frac{d \Phi}{dr} \nn
&& + \frac{14 (\pi')^3 (\pi'') \e^{-6 \Lambda}}{r^3} \frac{d \Phi}{dr}
+ \frac{32 (\pi')^3 (\pi'') \e^{-8 \Lambda}}{r^2} \frac{d \Phi}{dr}\, , \\
\label{Sch24}
\mathcal{E}_4 &=&
\frac{12 (\pi')^3 \e^{-6 \Lambda}}{r} \frac{d^2 \Phi}{dr^2} 
+ \frac{12 (\pi')^3 \e^{-6 \Lambda}}{r} \left( \frac{d \Phi}{dr}  \right)^2
 - \frac{60 (\pi')^3 \e^{-6 \Lambda}}{r} \frac{d \Lambda}{dr} 
\frac{d\Phi}{dr}
 - \frac{2 (\pi')^3 \e^{-4 \Lambda}}{r^2} \frac{d \Phi}{dr}  \nn
&& + \frac{36 (\pi')^2(\pi'') \e^{-6 \Lambda}}{r} \frac{d \Phi}{dr}
+ \frac{18 (\pi')^3 \e^{-6 \Lambda}}{r^2} \frac{d \Phi}{dr}
+ \frac{6 (\pi')^3 \e^{-4 \Lambda}}{r^2} \frac{d \Lambda}{dr}
- \frac{30 (\pi')^3 \e^{-6 \Lambda}}{r^2} \frac{d \Lambda}{dr} \nn
&& - \frac{6 (\pi')^2 (\pi'') \e^{-4 \Lambda}}{r^2}
+ \frac{18 (\pi')^2 (\pi'') \e^{-6 \Lambda}}{r^2} \, . 
\end{eqnarray}
Thus we obtain
\begin{eqnarray}
\label{VV2}
\mathcal{E}^{(G)}_2 &=& \frac{\partial G_2(\pi,X)}{\partial \pi} 
 -2 \e^{-2 \Lambda} \left( \frac{d \Phi}{dr} - \frac{d \Lambda}{dr}  \right) 
\frac{\partial G_2(\pi,X)}{\partial X} \pi'\nn
&& -4 \frac{\e^{-2 \Lambda}}{r} \frac{\partial G_2(\pi,X)}{\partial X} \pi' -2 \e^{-2 \Lambda} 
\frac{d}{dr} \left( \frac{\partial G_2(\pi,X)}{\partial X} \right) \pi'
 -2 \e^{-2 \Lambda} \frac{\partial G_2(\pi,X)}{\partial X} \pi''\, , \\
\label{VV3}
\mathcal{E}^{(G)}_3 &=&
\frac{\partial G_3(\pi,X)}{\partial \pi} \e^{-4 \Lambda} (\pi')^3 \left( \frac{d \Phi}{dr} + \frac{2}{r} \right)
 -3 G_3(\pi,X) \e^{-4 \Lambda} \left( \frac{d \Phi}{dr} -3 \frac{d \Lambda}{dr}  \right) 
\left( \frac{d \Phi}{dr} + \frac{2}{r}  \right) (\pi')^2 \nn
&& -6 G_3(\pi,X) \frac{\e^{-4 \Lambda}}{r}(\pi')^2 \left( \frac{d \Phi}{dr} + \frac{2}{r}  \right) 
 -6 G_3(\pi,X) \e^{-4 \Lambda} \pi' \pi'' \left( \frac{d \Phi}{dr} + \frac{2}{r}  \right) \nn
&& -3 G_3(\pi,X) \e^{-4 \Lambda} (\pi')^2 \left( \frac{d^2 \Phi}{dr^2} - \frac{2}{r^2}  \right)
 -3 \frac{d G_3(\pi,X)}{dr} \e^{-4 \Lambda} (\pi')^2 \left( \frac{d \Phi}{dr} + \frac{2}{r}  \right) \nn 
&& -2 \frac{\partial G_3(\pi,X)}{\partial X} \e^{-6 \Lambda} (\pi')^4 \left( \frac{d \Phi}{dr} 
 -5 \frac{d \Lambda}{dr}  \right) \left( \frac{d \Phi}{dr} + \frac{2}{r}  \right)
 -4 \frac{\partial G_3(\pi,X)}{\partial X} \frac{\e^{-6 \Lambda}}{r} (\pi')^4 \left( \frac{d \Phi}{dr}
+ \frac{2}{r}  \right) \nn
&& -8 \frac{\partial G_3(\pi,X)}{\partial X} \e^{- 6 \Lambda} (\pi')^3 \pi'' 
\left( \frac{d \Phi}{dr} + \frac{2}{r}  \right) 
 - 2 \frac{\partial G_3(\pi,X)}{\partial X} \e^{-6 \Lambda} (\pi')^4 
\left( \frac{d^2 \Phi}{dr^2} - \frac{2}{r^2}  \right) \nn
&& - 2  \frac{d}{dr} \left( \frac{\partial G_3(\pi,X)}{\partial X}  \right) \e^{-6 \Lambda}
 (\pi')^4 \left( \frac{d \Phi}{dr} + \frac{2}{r}  \right)\, ,\\
\label{VV4}
\mathcal{E}^{(G,0)}_4 &=&
 - \frac{\partial^2 G_4(\pi,X)}{\partial \pi \partial X} \frac{\e^{-6 \Lambda} }{r} (\pi')^4
\left( 2 \frac{d \Phi}{dr} + \frac{1}{r}  \right) \nn
&& + \left( 2 \frac{d^2 \Phi}{dr^2} - \frac{1}{r^2}  \right)
\left( 4 \frac{\partial G_4(\pi,X)}{\partial X} \frac{\e^{-6 \Lambda}}{r} (\pi')^3 
+ 2 \frac{\partial^2 G_4(\pi,X)}{\partial X^2} \frac{\e^{-8 \Lambda}}{r} (\pi')^5
\right) \nn
&& + \left( 2 \frac{d \Phi}{dr} + \frac{1}{r}  \right)
\left[ 
4 \frac{\partial G_4(\pi,X)}{\partial X} \frac{\e^{-6 \Lambda}}{r^2} (\pi')^3
+ 12 \frac{\partial G_4(\pi,X)}{\partial X}  \frac{\e^{-6 \Lambda}}{r} (\pi')^2 \pi'' 
\right. \nn
&& \left. +4 \frac{d}{dr } \left( \frac{\partial G_4(\pi,X)}{\partial X} 
\right) \frac{\e^{-6 \Lambda}}{r} (\pi')^3
+ 4 \frac{\partial G_4(\pi,X)}{\partial X} \frac{\e^{-6 \Lambda}}{r} (\pi')^3
 \left( \frac{d \Phi}{dr} -5 \frac{d \Lambda}{dr}  \right)
\right] \nn
&& + \left( 2 \frac{d \Phi}{dr} + \frac{1}{r}  \right)
\left[ 
2 \frac{\partial^2 G_4(\pi,X)}{\partial X^2} \frac{\e^{-8 \Lambda}}{r^2} (\pi')^5
+10 \frac{\partial^2 G_4(\pi,X)}{\partial X^2} \frac{\e^{-8 \Lambda}}{r} (\pi')^4 \pi''
\right. \nn
&& \left. + 2 \frac{d}{dr} \left( \frac{\partial^2 G_4(\pi,X)}{\partial X^2}  
\right) \frac{\e^{-8 \Lambda}}{r}(\pi')^5 
+ 2 \frac{\partial^2 G_4(\pi,X)}{\partial X^2} \frac{\e^{-8 \Lambda}}{r} (\pi')^5
\left( \frac{d \Phi}{dr} -7 \frac{d \Lambda}{dr}  \right) \right]\, , \\
\label{VV5}
\mathcal{E}^{(G,0)}_5 &=&
 - \frac{\partial^2 G_5(\pi,X)}{\partial \pi \partial X} \frac{\e^{-8 \Lambda}}{r^2} ( \pi')^5 \frac{d \Phi}{dr}
+ 5 \frac{\partial G_5(\pi,X)}{\partial X} \frac{\e^{-8 \Lambda}}{r^2} (\pi')^4 \left( \frac{d \Phi}{dr}
 - 7 \frac{d \Lambda}{dr} \right) \frac{d \Phi}{dr} \nn
&& + 20 \frac{\partial G_5(\pi,X)}{\partial X} \frac{\e^{-8 \Lambda}}{r^2} (\pi')^3 \pi'' \frac{d \Phi}{dr}
+5 \frac{\partial G_5(\pi,X)}{\partial X} \frac{\e^{-8 \Lambda}}{r^2} (\pi')^4 \frac{d^2 \Phi}{dr^2} \nn
&& +5 \frac{d}{dr} \left( \frac{\partial G_5(\pi,X)}{\partial X}  \right)
\frac{\e^{-8 \Lambda}}{r^2} (\pi')^4 \frac{d \Phi}{dr}
+ 2 \frac{\partial^2 G_5(\pi,X)}{\partial X^2} \frac{\e^{-10 \Lambda}}{r^2} (\pi')^6
 \left( \frac{d \Phi}{dr} -9 \frac{d \Lambda}{dr}  \right) \frac{d \Phi}{dr} \nn
&& +12 \frac{\partial^2 G_5(\pi,X)}{\partial X^2} \frac{\e^{-10 \Lambda}}{r^2} (\pi')^5 \pi''
\frac{d \Phi}{dr} + 2 \frac{\partial^2 G_5(\pi,X)}{\partial X^2}
\frac{\e^{-10 \Lambda}}{r^2} (\pi')^6 \frac{d^2 \Phi}{d r^2} \nn
&& + 2 \frac{d}{dr} \left( \frac{\partial^2 G_5(\pi,X)}{\partial X^2}  \right)
\frac{\e^{-10 \Lambda}}{r^2} ( \pi')^6 \frac{d \Phi}{dr}\, , \\
\label{DeltaVV4}
\Delta \mathcal{E}^{(G)}_4 &=&
 - \frac{\partial G_4(\pi,X)}{\partial \pi} \frac{(\pi')^2}{r} \left( 
\e^{-4 \Lambda} \frac{d \Phi}{dr} - \frac{1}{2} \frac{\e^{-2 \Lambda}}{r} + \frac{1}{2} \frac{\e^{-4 \Lambda}}{r}
\right) \nn
&& + 2 \left[ 
\frac{\e^{-4 \Lambda}}{r^2} \frac{d \Phi}{dr} + \frac{ \e^{-4 \Lambda}}{r} 
\left( \frac{d \Phi}{dr} -3 \frac{d \Lambda}{dr}  \right) \frac{d \Phi}{dr}
+ \frac{\e^{-4 \Lambda}}{r} \frac{d^2 \Phi}{dr^2}
 - \frac{1}{2} \frac{\e^{-2 \Lambda}}{r^2} \left( \frac{d \Phi}{dr} - \frac{d \Lambda}{dr}  \right)
\right. \nn
&& \left. + \frac{1}{2} \frac{\e^{-4 \Lambda}}{r^2} 
\left( \frac{d \Phi}{dr} -3 \frac{d \Lambda}{dr} \right) \right] 
\left( \frac{\partial G_4(\pi,X)}{\partial X} \e^{-2 \Lambda} (\pi')^3 + G_4(\pi,X) \pi' \right) \nn
&& +2 \left[ 
\frac{\e^{-4 \Lambda}}{r} \frac{d \Phi}{dr} - \frac{1}{2} \frac{\e^{-2 \Lambda}}{r^2}
+ \frac{1}{2} \frac{\e^{-4 \Lambda}}{r^2}
\right] \left[
\frac{d}{dr} \left( \frac{\partial G_4(\pi,X)}{\partial X}  \right) \e^{-2 \Lambda} (\pi')^3 \right. \nn
&& \left. -2 \frac{\partial G_4(\pi,X)}{\partial X} \e^{-2 \Lambda} \frac{d \Lambda}{dr} (\pi')^3
+ 3 \frac{\partial G_4(\pi,X)}{\partial X} \e^{-2 \Lambda} (\pi')^2 \pi'' 
+ \frac{d G_4(\pi,X)}{dr} \pi' + G_4(\pi,X) \pi''\right]\, , \\
\label{DeltaVV5}
\Delta \mathcal{E}^{(G)}_5 &=&
\frac{\partial G_5(\pi,X)}{\partial \pi}
\frac{(\pi')^3}{2 r^2} (\e^{-4 \Lambda} -3 \e^{-6 \Lambda}) \frac{d \Phi}{dr} \nn
&& -3 \pi' \pi'' G_5(\pi,X) (\e^{-4 \Lambda}-3 \e^{-6 \Lambda}) \frac{d \Phi}{dr}
 - \frac{3}{2} (\pi')^2 \frac{d G_5(\pi,X)}{dr} (\e^{-4 \Lambda}-3 \e^{-6 \Lambda}) \frac{d \Phi}{dr} \nn
&& - \frac{3}{2} (\pi')^2 G_5(\pi,X) \left[ 
\e^{-4 \Lambda} \left( \frac{d \Phi}{dr} -3 \frac{d \Lambda}{dr}  \right)
 -3 \e^{-6 \Lambda} \left( \frac{d \Phi}{dr} -5 \frac{d \Lambda}{dr}  \right)
\right] \frac{d \Phi}{dr} \nn
&& - \frac{3}{2} (\pi')^2 G_5(\pi,X)  (\e^{-4 \Lambda}-3 \e^{-6 \Lambda}) 
\frac{d^2 \Phi}{dr^2} \nn
&& -4 \frac{\partial G_5(\pi,X)}{\partial X}  (\pi')^3 \pi'' ( \e^{-6 \Lambda} -3 \e^{-8 \Lambda}) \frac{d \Phi}{dr}
 - \frac{d}{dr} \left( \frac{\partial G_5(\pi,X)}{\partial X} \right)(\pi')^4 ( \e^{-6 \Lambda} -3 \e^{-8 \Lambda})
\frac{d \Phi}{dr} \nn
&& \frac{\partial G_5(\pi,X)}{\partial X} (\pi')^4
\left[ \e^{-6 \Lambda} \left( \frac{d \Phi}{dr} -5 \frac{d \Lambda}{dr}  \right)
-3 \e^{-8 \Lambda} \left( \frac{d \Phi}{dr} -7 \frac{d \Lambda}{dr}  \right)
\right] \frac{d \Phi}{dr} \nn
&& - \frac{\partial G_5(\pi,X)}{\partial X} (\pi')^4 ( \e^{-6 \Lambda} -3 \e^{-8 \Lambda})\frac{d^2 \Phi}{dr^2}\, ,\\
\label{VV6}
\mathcal{E}^{(G)}_6 &=&
8 \frac{\partial G_6(\pi)}{\partial \pi} \frac{(\pi')^2}{r} \left( 
\e^{-4 \Lambda} \frac{d \Phi}{dr} - \frac{1}{2} \frac{\e^{-2 \Lambda}}{r} + \frac{1}{2} \frac{\e^{-4 \Lambda}}{r}
\right) \nn
&& -16 \left[ 
\frac{\e^{-4 \Lambda}}{r^2} \frac{d \Phi}{dr} + \frac{ \e^{-4 \Lambda}}{r} 
\left( \frac{d \Phi}{dr} -3 \frac{d \Lambda}{dr}  \right) \frac{d \Phi}{dr}
+ \frac{\e^{-4 \Lambda}}{r} \frac{d^2 \Phi}{dr^2}
 - \frac{1}{2} \frac{\e^{-2 \Lambda}}{r^2} \left( \frac{d \Phi}{dr} - \frac{d \Lambda}{dr}  \right)
\right. \nn
&& \left. + \frac{1}{2} \frac{\e^{-4 \Lambda}}{r^2} \left( \frac{d \Phi}{dr} -3 \frac{d \Lambda}{dr} \right)
\right] G_6(\pi) \pi' \nn
&& -16 \left[ 
\frac{\e^{-4 \Lambda}}{r} \frac{d \Phi}{dr} - \frac{1}{2} \frac{\e^{-2 \Lambda}}{r^2}
+ \frac{1}{2} \frac{\e^{-4 \Lambda}}{r^2}
\right] \left( \frac{d G_6(\pi)}{dr} \pi' + G_6(\pi) \pi'' \right) \, , \\
\label{VV7}
\mathcal{E}^{(G)}_7 &=& -8 \frac{\partial G_7(\pi)}{\partial \pi}
\frac{(\pi')^3}{2 r^2} (\e^{-4 \Lambda} -3 \e^{-6 \Lambda}) \frac{d \Phi}{dr} \nn
&& +24 \pi' \pi'' G_7(\pi) (\e^{-4 \Lambda}-3 \e^{-6 \Lambda}) \frac{d \Phi}{dr}
+ 12 (\pi')^2 \frac{d G_7(\pi)}{dr} (\e^{-4 \Lambda}-3 \e^{-6 \Lambda}) \frac{d \Phi}{dr} \nn
&& + 12 (\pi')^2 G_7(\pi) \left[ 
\e^{-4 \Lambda} \left( \frac{d \Phi}{dr} -3 \frac{d \Lambda}{dr}  \right)
 -3 \e^{-6 \Lambda} \left( \frac{d \Phi}{dr} -5 \frac{d \Lambda}{dr}  \right) \right] \frac{d \Phi}{dr} \nn
&& +12  (\pi')^2 G_7(\pi)  (\e^{-4 \Lambda}-3 \e^{-6 \Lambda}) \frac{d^2 \Phi}{dr^2}\, , \\
\label{V8}
\mathcal{E}^{(G)}_8 &=&
- \frac{32 \e^{-4 \Lambda}}{r^2} \frac{\partial G_8(\pi)}{\partial \pi}
\left[ 
\e^{2 \Lambda} \frac{d^2 \Phi}{dr^2} - \frac{d^2 \Phi}{dr^2}
+ \e^{2 \Lambda} \left( \frac{d \Phi}{dr}  \right)^2
 - \left( \frac{d \Phi}{dr}  \right)^2
 - \e^{2 \Lambda} \frac{d \Lambda}{dr} \frac{d \Phi}{dr}
+3 \frac{d \Lambda}{dr} \frac{d \Phi}{dr} \right]\, .
\end{eqnarray}

In the static and spherically symmetric space-time in (\ref{Sch1}), we obtain
\begin{eqnarray}
\label{Schma23at}
(\mathcal{H}_2)_{tt} &=& - \frac{1}{2}\e^{2\Phi} G_2 (\pi, X) \, , \\
\label{Schma23ar}
(\mathcal{H}_2)_{rr} &=& \frac{1}{2} \e^{2\Lambda} G_2 (\pi, X) 
 - \frac{\partial G_2 (\pi, X)}{\partial X} \left( \pi' \right)^2\, , \\
\label{Scma23bt}
(\mathcal{H}_3)_{tt} &=& \frac{1}{2} G_3 (\pi, X) 
\e^{-4\Lambda} \left( \pi' \right)^2 \left(\pi'' - \frac{d\Lambda}{dr} \pi' \right) 
+ \frac{1}{2} \e^{2\Phi - 2\Lambda} \pi' \frac{d}{dr} 
\left( G_3 (\pi, X) \e^{-2\Lambda} \left( \pi' \right)^2 \right) \, , \\
\label{Scma23br}
(\mathcal{H}_3)_{rr} &=& \frac{1}{2}\e^{2\Lambda} G_3 (\pi, X) 
\left( \frac{\e^{-3\Lambda}}{r^2} \left( \pi' \right)^2 \left( \frac{d}{dr} \left(r^2 
\e^{\Phi - \Lambda} \pi' \right)\right)
 - \e^{-4\Lambda} \left( \pi' \right)^2 \left(\pi'' - \frac{d\Lambda}{dr} \pi' \right) \right) \nn
&& - \frac{\partial G_3 (\pi, X)}{\partial X} 
\left( \frac{\e^{-3\Lambda}}{r^2} \left( \pi' \right)^2 \left( \frac{d}{dr} \left(r^2 
\e^{\Phi - \Lambda} \pi' \right)\right)
 - \e^{-4\Lambda} \left( \pi' \right)^2 \left(\pi'' - \frac{d\Lambda}{dr} \pi' \right) \right)
\left( \pi' \right)^2 \nn
&& - G_3 (\pi, X) \left( \left( \pi' \right)^2 \left( \frac{\e^{- \Phi - \Lambda}}{r^2} 
\left( \frac{d}{dr} \left( r^2 \e^{\Phi - \Lambda} \pi' \right) \right) \right)
 - 2 \e^{-4\Lambda} \left( \pi' \right)^2 \left( \pi'' - \frac{d\Lambda}{dr} \pi' \right) \right) \nn
&& + \frac{1}{2} \left( -  \frac{\e^{- \Phi + \Lambda}}{r^2} \left( \frac{d}{dr} 
\left( r^2 \e^{\Phi - \Lambda} \pi' \right) \right)
+ \pi' \frac{d}{dr} \right) 
\left( G_3 (\pi, X) \e^{-2\Lambda} \left( \pi' \right)^2 \right) \nn
&& - \frac{1}{2} \e^{-2\Lambda} \frac{d}{dr} \left( G_3 (\pi, X) \left( \pi' \right)^3 \right) \, , \\
\label{Sch28}
(\mathcal{H}_4)_{tt} &=&
 -5 \frac{\partial G_4 (\pi, X)}{\partial X} \frac{(\pi')^4}{r} \frac{d \Lambda}{dr}
\e^{2 \Phi - 6 \Lambda} +  \frac{d}{dr} 
\left( \frac{\partial G_4 (\pi, X)}{\partial X}  \right)
\frac{(\pi')^4}{r} \e^{2 \Phi -6 \Lambda}
+ 4 \frac{\partial G_4 (\pi, X)}{\partial X} \frac{(\pi')^4 \pi''}{r} 
\e^{2 \Phi -6 \Lambda} \nn
&& + \frac{1}{2} \frac{\partial G_4 (\pi, X)}{\partial X} \frac{(\pi')^4}{r^2} 
\e^{2 \Phi -6 \Lambda} 
 - \frac{3}{2}  G_4 (\pi, X) \frac{(\pi')^2}{r}
\frac{d \Lambda}{dr} \e^{2 \Phi -4 \Lambda}
+ \frac{1}{2} \frac{d G_4 (\pi, X)}{dr}
\frac{(\pi')^2}{r} \e^{2 \Phi-4 \Lambda} \nn
&& + G_4 (\pi, X) \frac{\pi \pi''}{ r} \e^{2 \Phi - 4 \Lambda}
+ \frac{1}{4} G_4 (\pi, X) \frac{(\pi')^2}{r^2} \e^{2 \pi - 4 \Lambda} 
+ \frac{1}{4} G_4 (\pi, X) \frac{(\pi')^2}{r^2} 
\e^{-4 \Phi -2 \Lambda}\, , \\
\label{Sch29}
(\mathcal{H}_4)_{rr} &=&
 -5 \frac{\partial G_4 (\pi, X)}{\partial X} \frac{(\pi')^4}{r} \frac{d \Phi}{dr}
\e^{-4 \Lambda} - \frac{5}{2} \frac{\partial G_4 (\pi, X)}{\partial X}
\frac{(\pi')^4}{r^2} \e^{-4 \Lambda}
 -2 \frac{\partial^2 G_4 (\pi, X)}{\partial X^2} \frac{(\pi')^6}{r} 
\frac{d \Phi}{dr} \e^{-6 \Lambda} \nn 
&& - \frac{\partial^2 G_4 (\pi, X)}{\partial X^2}
\frac{(\pi')^6}{r^2} \e^{-6 \Lambda} 
 - \frac{3}{2} G_4 (\pi, X) \frac{(\pi')^2}{r} \frac{d \Phi}{dr} \e^{-2 \Lambda}
 - \frac{3}{4} G_4 (\pi, X) \frac{(\pi')^2}{r^2} \e^{-2 \Lambda}
+ \frac{1}{4} G_4 (\pi, X) \frac{(\pi')^2}{r^2} \nn
&& -\frac{\partial G_4 (\pi, X)}{\partial X} \frac{(\pi')^4}{r} \frac{d \Phi}{dr}
\e^{-4 \Lambda} + \frac{1}{2} \frac{\partial G_4 (\pi, X)}{\partial X}
\frac{(\pi')^4}{r^2} \e^{-2 \Lambda} 
 -\frac{1}{2} \frac{\partial G_4 (\pi, X)}{\partial X} \frac{(\pi')^4}{r^2} 
\e^{-2 \Lambda}\, , \\
\label{Sch26}
(\mathcal{H}_5)_{tt} &=& 
\frac{1}{2} \frac{d}{dr} \left( \frac{\partial G_5 (\pi, X)}{\partial X}
\right) \frac{(\pi')^5}{r^2} \e^{2\Phi -8 \Lambda}
+ \frac{5}{2} \frac{\partial G_5 (\pi, X)}{\partial X} \frac{(\pi')^4
(\pi'')}{r^2} \e^{2\Phi -8 \Lambda}
 - \frac{7}{2} \frac{\partial G_5 (\pi, X)}{\partial X} \frac{(\pi')^5}{r^2}
\frac{d \Lambda}{dr} \e^{2 \Phi -8 \Lambda} \nonumber \\ 
&& + \frac{3}{4}  G_5 (\pi, X) \frac{(\pi')^3}{r^2}
\frac{d \Lambda}{dr} \e^{2 \Phi -4 \Lambda} 
 - \frac{15}{4} G_5 (\pi, X) \frac{(\pi')^3}{r^2} \frac{d \Lambda}{dr} 
\e^{2 \Phi -6 \Lambda} - \frac{1}{4} \frac{d G_5 (\pi, X)}{dr}
\frac{(\pi')^3}{r^2} \e^{2 \Phi -4 \Lambda} \nn
&& - \frac{3}{4} G_5 (\pi, X) \frac{(\pi')^2 \pi''}{r^2} \e^{2 \Phi -4 \Lambda} 
+ \frac{3}{4} \frac{d G_5 (\pi, X)}{dr} 
\frac{(\pi')^3}{r^2} \e^{2 \Phi -6 \Lambda} 
+ \frac{9}{4} G_5 (\pi, X) \frac{(\pi')^2 \pi''}{r^2} 
\e^{2 \Phi -6 \Lambda}\, , \\
\label{Sch27}
(\mathcal{H}_5)_{rr}
&=& - \frac{7}{2} \frac{\partial G_5 (\pi, X)}{\partial X} \frac{(\pi')^5}{r^2} 
\e^{-6 \Lambda} \frac{d \Phi}{dr} - \frac{\partial^2 G_5 (\pi, X)}{\partial X^2}
\frac{(\pi')^7}{r^2} \e^{-8 \Lambda} \frac{d \Phi}{dr} 
+ \frac{3}{4} G_5 (\pi, X) \frac{(\pi')^3}{r^2} \frac{d \Phi}{dr} \e^{-2 \Lambda} \nn
&& - \frac{15}{4} G_5 (\pi, X) \frac{(\pi')^3}{r^2} \frac{d \Phi}{dr} \e^{-4 \Lambda} 
+ \frac{1}{2} \frac{\partial G_5 (\pi, X)}{\partial X} \frac{(\pi')^5}{r^2}
\frac{d \Phi}{dr} \e^{-4 \Lambda} - \frac{3}{2} \frac{\partial
G_5 (\pi, X)}{\partial X} \frac{(\pi')^5}{r^2} \frac{d \Phi}{dr} \e^{-6 \Lambda}\, , \\
\label{Sch30}
(\mathcal{H}_6)_{tt} 
&=& 12 G_6 (\pi) \frac{(\pi')^2}{r}
\frac{d \Lambda}{dr} \e^{2 \Phi -4 \Lambda}
 - 4 \frac{d G_6 (\pi)}{dr} \frac{(\pi')^2}{r} \e^{2 \Phi-4 \Lambda}
 -8 G_6 (\pi) \frac{\pi \pi''}{ r} \e^{2 \Phi - 4 \Lambda}
 -2 G_6 (\pi) \frac{(\pi')^2}{r^2} \e^{2 \Phi - 4 \Lambda} \nonumber \\
&& -2 G_6 (\pi) \frac{(\pi')^2}{r^2} \e^{-4 \Phi -2 \Lambda}\, , \\
\label{Sch31}
(\mathcal{H}_6)_{rr} &=&
12 G_6 (\pi) \frac{(\pi')^2}{r} \frac{d \Phi}{dr} \e^{-2 \Lambda}
+ 6 G_6 (\pi) \frac{(\pi')^2}{r^2} \e^{-2 \Lambda}
-2 G_6 (\pi) \frac{(\pi')^2}{r^2}\, ,  \\
\label{Sch32}
(\mathcal{H}_7)_{tt}  &=& -6 G_7 (\pi) \frac{(\pi')^3}{r^2}
\frac{d \Lambda}{dr} \e^{2 \Phi -4 \Lambda} 
+ 30 G_7 (\pi) \frac{(\pi')^3}{r^2} \frac{d \Lambda}{dr} 
\e^{2 \Phi -6 \Lambda} + 2 \frac{d G_7 (\pi)}{dr}
\frac{(\pi')^3}{r^2} \e^{2 \Phi -4 \Lambda} \nn
&& + 6 G_7 (\pi) \frac{(\pi')^2 \pi''}{r^2} \e^{2 \Phi -4 \Lambda} 
 -6 \frac{d G_7 (\pi)}{dr} \frac{(\pi')^3}{r^2} \e^{2 \Phi -6 \Lambda} 
 - 18 G_7 (\pi) \frac{(\pi')^2 \pi''}{r^2} \e^{2 \Phi -6 \Lambda} \, , \\
\label{Sch33}
(\mathcal{H}_7)_{rr} &=&
 - 6 G_7 (\pi) \frac{(\pi')^3}{r^2} \frac{d \Phi}{dr} \e^{-2 \Lambda}
+ 30 G_7 (\pi) \frac{(\pi')^3}{r^2} \frac{d \Phi}{dr} \e^{-4 \Lambda}\, ,\\ 
\label{Sch34}
(\mathcal{H}_8)_{tt} &=&
 -192 \frac{d G_8 (\pi)}{dr} \frac{\e^{2 \Phi -2 \Lambda}}{r^2} 
\frac{d \Phi}{dr} + 16 \frac{d G_8 (\pi)}{dr} \frac{\e^{2 \Phi -2 \Lambda}}{r^2} 
\frac{d \Lambda}{dr}
 -16 \frac{d^2 G_8 (\pi)}{dr^2} \frac{\e^{2 \Phi -2 \Lambda}}{r^2} \nn
&& -48 \frac{d G_8 (\pi)}{dr} \frac{\e^{2 \Phi-4 \Lambda}}{r^2} \frac{d \Lambda}{dr} 
+16 \frac{d^2 G_8 (\pi)}{dr^2} \frac{\e^{2 \Phi -4 \Lambda}}{r^2}\, , \\
\label{Sch35}
(\mathcal{H}_8)_{rr} &=&
16 \frac{d G_8 (\pi)}{dr} \frac{1}{r^2} \frac{d \Phi}{dr} -48 \frac{dG_8 (\pi)}{dr}
\frac{1}{r^2} \frac{d \Phi}{dr} \e^{-2 \Lambda}\, . 
\end{eqnarray}

\end{document}